\documentclass[reprint,showpacs,preprintnumbers,amssymb,nofootinbib,aps,prd]{revtex4-1}
\usepackage{epsfig}
\usepackage[usenames,dvipsnames]{color}
\usepackage{hyperref}
\usepackage{amsmath}
\usepackage{graphicx}
\usepackage{amssymb}
\usepackage{bm}% bold math
\usepackage{dcolumn,multirow,subfig}% Align table columns on decimal point

\usepackage[normalem]{ulem}

\usepackage{array}
\newcommand{\ba}{\begin{array}}
\newcommand{\ea}{\end{array}}
\def\beq{\begin{equation}}
\def\eeq{\end{equation}}
\def\bea{\begin{eqnarray}}
\def\eea{\end{eqnarray}}
\def\nn{\nonumber}

\def\roughly#1{\mathrel{\raise.3ex\hbox
{$#1$\kern-.75em\lower1ex\hbox{$\sim$}}}}

\def\sla#1{\raise.15ex\hbox{$/$}\kern-.57em #1}% Feynman slash

\def\bd{B_d^0}

\def\order{\lower 1.8ex \hbox{\LARGE\~{}}}

\def\bdtau{B\to D^{(\ast)}\tau\nu_{\tau}}

\def\bdell{B\to D^{(\ast)}\ell\nu_{\ell}}
\def\bdstell{B\to D^\ast\ell\nu_{\ell}}
\def\bdnstell{B\to D \ell\nu_{\ell}}
\def\bd0tau{B\to D \tau\nu_{\tau}}
\def\bdasttau{B\to D^{\ast}\tau\nu_{\tau}}
\def\be {\begin{equation}}
\def\ee {\end{equation}}
\def\diff{d\Gamma /{dq^2}}

\usepackage{relsize}
\def\Babar{{\mbox{\slshape B\kern-0.1em{\smaller A}\kern-0.1em B\kern-0.1em{\smaller A\kern-0.2em R}}}}
\begin{document}

%opening
\title{Looking for possible new physics in $\bdtau$ in light of recent data}%Analysis of possible new physics contributions 
%in the decay $\bdtau$ in the light of recent data

\author{Srimoy Bhattacharya}
\email{bhattacharyasrimoy@gmail.com}
\affiliation{Indian Institute of Technology, North Guwahati, Guwahati 781039, Assam, India }

\author{Soumitra Nandi}
\email{soumitra.nandi@gmail.com}
\affiliation{Indian Institute of Technology, North Guwahati, Guwahati 781039, Assam, India }

\author{Sunando K. Patra}
\email{sunando.patra@gmail.com}
\affiliation{Indian Institute of Technology, North Guwahati, Guwahati 781039, Assam, India }

\begin{abstract}
We study the decays $\bdtau$ in light of the available data from \Babar,~Belle and LHCb. 
We divide our analysis into two parts: in one part we fit the form-factors in these decays directly from the 
data without adding any additional new physics (NP) contributions and compare our fit results with those 
available from the decays $\bdell$. We find that the $q^2$-distributions of the form-factors associated with the pseudo-vector
current, obtained from $\bdtau$ and $\bdell$ respectively, do not agree with each other, whereas the other form-factors are consistent 
with each other. In the next part of our analysis, we look for possible new effective operators of dimension 6 amongst new vector, scalar, 
and tensor-type that can best explain the current data in the 
decays $\bdtau$. We use the information-theoretic approaches, especially of `Second-order Akaike Information Criterion' (AIC$_c$) 
in the analysis of empirical data. 
%Multi-parameter goodness-of-fit tests 
Normality tests for the distribution of residuals are done after selecting the best possible scenarios, for cross-validation. 
We find that it is the contribution from the operator involving left or right-handed vector current 
%$(V+A)$ type interaction 
that passes all the selection criteria defined for the best-fit scenario and can successfully accommodate all the available data set.
%  Multi-model goodness-of-fit analysis of $\bdtngen$ observables $R(D^{(*)})$ as well as bin-by-bin data of
%  $q^2$ distribution of number of events. As parameters we take combination of Wilson coefficients associated 
%  with all possible new physics(NP) operators. We use \Babar \cite{babarexp}, Belle\cite{bellexp} and LHCb\cite{lhcbexp} 
%  results as our experimental inputs.
\end{abstract}

\maketitle

\section{Introduction}
\begin{figure}[!hbt]
\centering
 \includegraphics[scale=0.4]{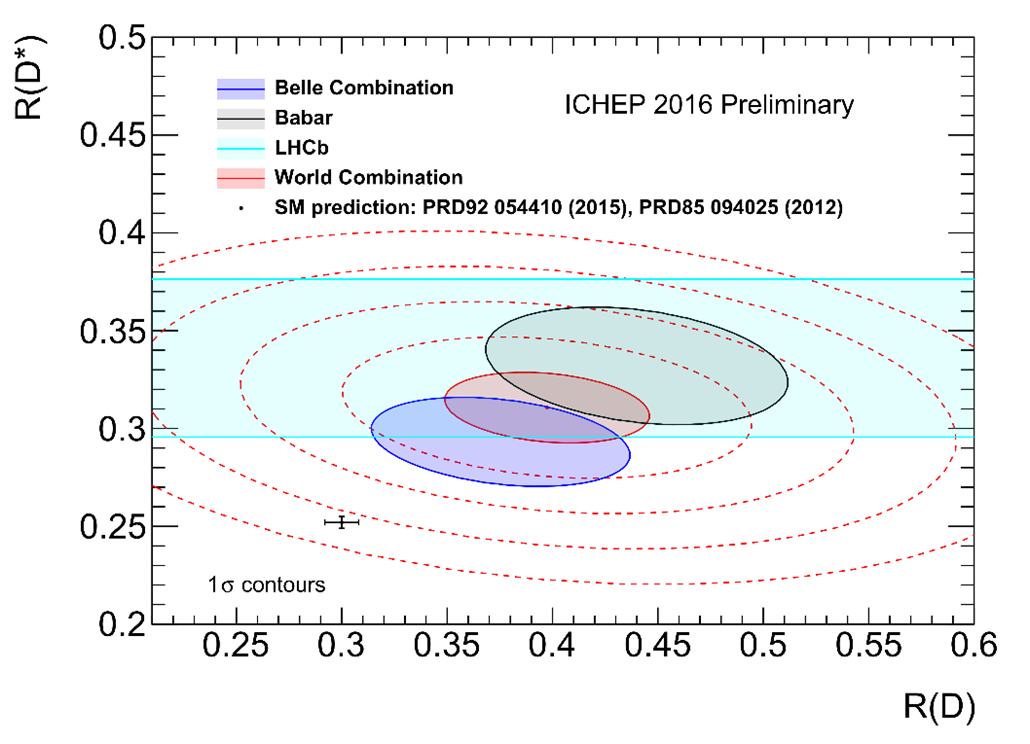}
 \caption{Current experimental status in the measurements of $R(D)$ and $R(D^*)$.}
 \label{fig_belle}
\end{figure}

The semitaunic decays $\bdtau$ have drawn a lot of attention in recent years as sensitive probes of NP 
\cite{Hou,Chen:2006nua,Nierste,Tanaka:2010,Fajfer:2012vx,Bhattacharya:2015}. The present experimental status is summarized in 
Fig. \ref{fig_belle} \cite{belle_talk}. 
Here, $R(D)$ and $R(D^*)$ are defined as
\begin{align}
\nn \mathcal{R}(D) &= \frac{\mathcal{B}\left(\overline{B} \to D \tau^- \overline{\nu}_{\tau}\right)}
{\mathcal{B}\left(\overline{B} \to D l^- \overline{\nu}_{l}\right)}~~ {\rm and}\\
\mathcal{R}(D^*) &=
\frac{\mathcal{B}\left(\overline{B} \to D^* \tau^- \overline{\nu}_{\tau}\right)}{\mathcal{B}
\left(\overline{B} \to D^* l^- \overline{\nu}_{l}\right)}\,.
\end{align}

%ASII
The Standard Model (SM) predictions for $R(D^{(*)})$ are taken from \cite{Fajfer:2012vx}and \cite{Na:2015kha}, 
respectively. The theory uncertainties in these observables are only a few percent, and
independent of the CKM element $|V_{cb}|$. In the figure, the contours show the correlation 
between the measured values of $R(D)$ and $R(D^{*})$ from different experimental collaborations.
We note that the contour obtained after averaging the Belle measurements \cite{Huschle:2015rga,bellexp,Abdesselam:2016xqt}, 
which is more than 3$\sigma$ away from the SM prediction, lies in between the SM expectation and the
\Babar~measurement \cite{babarexp}. LHCb results on $\mathcal{R}(D^{*})$ \cite{lhcbexp} are $2.1 \sigma$ larger than the value 
expected in SM. Although the Belle average is slightly smaller than the LHCb and \Babar~results, it is still considerably larger 
than the SM prediction.

One can explain this excess by considering the contribution from some NP model of one's choice, {\it e.g.} 
\cite{model,Sakai:2013}. On the other hand, one may write down the most general relevant effective NP operators which may include new 
scalar, vector and tensor currents other than the SM, and then try to estimate the size of the NP Wilson coefficients from the 
excess \cite{modelind} in a model independent analysis. 

We observe that with passing time and increasing statistics, the measured value of $R(D^*)$ is becoming closer to that of the SM.
However, we have to wait for more precise measurements on $R(D)$. This is important, since the NP sensitivity of $R(D)$ and $R(D^*)$ 
are not the same \cite{Bhattacharya:2015}. 

Also, the sensitivity to a particular type of interaction is more apparent 
in the binned data, compared to that from the integrated observables like $R(D^{(*)})$ \cite{Bhattacharya:2015}.
On the other hand, as the measured values of $R(D^{(*)})$ are highly model sensitive due to the model-dependence of 
the kinetic distribution, one may get different signal yields per bin from fits using different models. 
Consequently, the measured values obtained from fits assuming only the SM background should not be used to fit the NP parameters. 
Although we use the background-subtracted and normalized binned data for most of our analysis, we compensate for any 
systematic errors coming from such assumption by doing a separate study with over-estimated errors and their correlations.
%However, it is always safe to use the normalized binned data for NP analysis. 

In this article, we systematically divide our analysis into two parts. In the first part of our analysis (section \ref{sec:form-factors}), 
we will assume that there is no NP in $\bdtau$, just as in $\bdell$ ($\ell = e$ or $\mu$), and will fit the form-factors. 
Different experimental collaborations have already fitted the form-factor parameters \cite{CLN} from the data 
collected for the decays $\bdell$, {\it e.g.} \cite{Aubert:2009,Dungel:2010}. Using the present data on $\bdtau$, 
we can check whether the fitted form-factors are in good agreement with those obtained from the decay $\bdell$. 
Any discrepancy between the two will indicate a possible new effect in $\bdtau$, which is absent in $\bdell$. 
It will help us to pinpoint the possible type(s) of new interaction which could be responsible for such deviations. 

In the second part of the analysis (section \ref{sec:np}), we will consider the contributions from different NP interactions in $\bdtau$, 
but not in $\bdell$. 
Our goal will be the search for new interactions most compatible with and best elucidates the present data. Throughout our analysis we 
will use the $q^2$-binned data on the decay rate as well the data on $R(D^{(*)})$. 

Detailed discussion on our methodology can be found 
in sections \ref{latparfit}, \ref{sec:goodfit}, \ref{ffresults} and \ref{sec:npmethod}.

\section{Form-factors from $\bdtau$ }\label{sec:form-factors}

\subsection{Formalism}\label{fftheory}

%\subsection{Theory}\label{latTheo}
The amplitudes of semileptonic $B$ meson decays can be factorized in the product of the matrix elements of leptonic and 
hadronic currents. The matrix elements of the hadronic currents are non-perturbative objects called form-factors. 
For a precise determination of the form-factors, we have to rely either on lattice QCD calculations or on the light cone sum 
rule approaches (LCSR). The uncertainties in the form-factors is one of the major sources of uncertainties in the predictions 
of the decay rates. 

In the SM, the differential decay rates for the decay $B \to D^{(*)} \ell \nu_{\ell}$, where $\ell = e,~\mu$ or $\tau$, 
are given by \cite{Korner:1989}

\begin{align}
  \nn &\frac{d\Gamma \left(\overline{B} \rightarrow D \ell \overline{\nu}_{\ell}\right)}{d q^2} = \frac{G^2_F 
 \left|V_{cb}\right|^2}{192 \pi^3 m^3_B} q^2 \sqrt{\lambda_D(q^2)} \\
  & \left(1 - \frac{m^2_{\ell}}{q^2}\right)^2 \left[ \left(1 + \frac{m^2_{\ell}}{2 q^2}\right) H^{s 2}_{V,0} + 
 \frac{3}{2} \frac{m^2_{\ell}}{q^2} H^{s 2}_{V,t}\right] \,, 
\label{dgamd}
\end{align}
\begin{align}
 \nn &\frac{d\Gamma \left(\overline{B} \rightarrow D^* \ell \overline{\nu}_{\ell}\right)}{d q^2} = 
 \frac{G^2_F \left|V_{cb}\right|^2}{192 \pi^3 m^3_B} q^2 \sqrt{\lambda_D^*(q^2)} \left(1 - \frac{m^2_{\ell}}{q^2}\right)^2 \\ 
 &\left[\left(1 + \frac{m^2_{\ell}}{2 q^2}\right)
 \left(H^2_{V,+} + H^2_{V,-} + H^2_{V,0}\right) + \frac{3}{2} \frac{m^2_{\ell}}{q^2} H^{2}_{V,t}\right] \,, 
\label{dgamdst}
\end{align}

where $ \lambda_D^{(*)}(q^2) = ((m_B - m_D^{(*)})^2 - q^2)((m_B + m_D^{(*)})^2 - q^2)$. 
Here, the helicity amplitudes $H^{\lambda_M}_{i,\lambda }$'s are defined through the hadronic matrix elements 
\begin{equation}
 H^{\lambda_M}_{i,\lambda } = \epsilon^*_{\mu} \langle M (\lambda_M) |{ \bar c}\gamma^{\mu}(1 - \gamma_5) b |{\bar B} \rangle,
\end{equation}
where $\lambda_M$  and $\lambda$ are the helicities of the final
state meson $M$ and the virtual intermediate boson in the $B$ meson rest frame respectively. Also note that whereas for $D$ meson 
$\lambda_M = s$, for $D^*$ meson $\lambda_M = \pm 1,~0$ and $\lambda = 0,~\pm 1$ and $t$. These helicity amplitudes are related
to the form-factors
\begin{align}
 H^s_{V,0}(q^2) &= \sqrt{\frac{\lambda_D(q^2)}{q^2}}F_1(q^2), \nn \\
 H^s_{V,t}(q^2) &= {\frac{m_B^2 - m_D^2}{\sqrt{q^2}}}F_0(q^2), \nn \\
 H_{V,\pm}(q^2) &= (m_B + m_{D^*})A_1(q^2)\mp \frac{\sqrt{\lambda_{D^*}}}{m_B + m_{D^*}}V(q^2), \nn \\
 H_{V,0}(q^2) &= \frac{(m_B + m_{D^*})}{2m_{D^*}\sqrt{q^2}} \nn \\
 & \Big[(m_B^2-m_{D^*}^2-q^2)A_1(q^2) \nn \\ 
 & + \frac{\lambda_{D^*}}{(m_B + m_{D^*})^2}A_2(q^2)\Big] \nn \\
 H_{V,t}(q^2) &=  \sqrt{\frac{\lambda_D(q^2)}{q^2}}A_0(q^2)\,.
\end{align}

The form-factors are defined as the matrix elements of various currents,  
\begin{align}
\nn \langle D(K)| \bar{c}\gamma_\mu b | \bar{B}(p)\rangle &= [(p+k)_\mu -\frac{m_B^2-m_D^2}{q^2}q_\mu] \\
 & F_1(q^2)+ q_\mu \frac{m_B^2-m_D^2}{q^2} F_0(q^2)\,,
 \label{matrixD}
\end{align}

and 
\begin{align} 
\nn \langle D^*(k,\varepsilon)|\bar{c}\gamma_\mu b &|\bar{B}(p)\rangle = i \epsilon_{\mu \nu\rho \sigma}
 \varepsilon^{\nu *}p^{\rho}k^{\sigma} \frac{2 V(q^2)}{m_B+m_{D^*}} \\ 
\langle D^*(k,\varepsilon)|\bar{c}\gamma_\mu \gamma_5 b &|\bar{B}(p)\rangle = \varepsilon_{\mu}^{*}(m_B + m_{D^*})A_1(q^2) \nn \\
 & -(p+k)_\mu (\varepsilon^* q)\frac{A_2(q^2)}{m_B+m_{D^*}} \nn \\
 & -q_\mu(\varepsilon^* q)\frac{2 m_{D^*}}{q^2}[A_3(q^2) - A_0(q^2]\,,
\label{matrixDst}
\end{align}
where 
\begin{equation}
 A_3(q^2) = \frac{m_B + m_{D^*}}{2 m_{D^*}} A_1(q^2) - \frac{m_B - m_{D^*}}{2 m_{D^*}} A_2(q^2)\,. 
\end{equation}
A direct comparison of the matrix elements in eq.(\ref{matrixD}) with those in heavy quark effective theory 
(HQET) gives us the relations
\bea
F_1(q^2) &=& \frac{1}{2\sqrt{m_B m_D}}\Big[(m_B+m_D)h_+(w(q^2)) \nn \\ 
& {}& -(m_B-m_D)h_-(w(q^2))\Big] \nn \\
F_0(q^2)&=& \frac{1}{2\sqrt{m_B m_D}}\Big[ \frac{(m_B + m_D)^2 - q^2}{m_B + m_D}h_+(w(q^2)) \nn \\
 &{}& - \frac{(m_B - m_D)^2 - q^2}{m_B - m_D}h_-(w(q^2))\Big]\,, 
\eea
where $h_{\pm}(w(q^2))$ are the HQET form-factors, with $w = v_B . v_{D^{(*)}} = \frac{m_B^2 + m_{D^{(*)}}^2 -q^2}{2 m_{D^{(*)}} m_B}$.
Following the parametrization given in \cite{CLN}, the HQET form-factors can be expressed as 
\bea
h_+(w) &=& \frac{1}{2(1+r_D^2-2 r_D w)}\Big[-(1+r_D)^2(w-1) V_1(w) \nn \\
&{}& + (1-r_D)^2(w+1) S_1(w)\Big] \nn \\
h_-(w) &=& \frac{(1-r_D^2)(w+1)}{2(1+r_D^2-2 r_D w)}[S_1(w) - V_1(w)] \,,
\eea
where $r_D = m_D / m_B$. The hadronic form-factors $V_1(w)$ and $S_1(w)$ coincide with the Isgur-Wise function $\xi(w)$ in the 
infinite mass limit of the heavy quark $m_Q$ ( = $m_b$ or $m_c$). This function is normalized 
to unity at zero recoil, i.e at $w=1$. In the Ref. \cite{CLN}, the $w$ dependence is parameterized as in eq.(\ref{ffV1}). 
The idea is to expand $V_1(w)$ around zero recoil point $w=1$. 
\begin{align}
 \nn V_1(w) = V_1(1) \times &\left[ 1 - 8 \rho_D^2 z(w) + (51 \rho_D^2 - 10) z(w)^2 \right.\\
 &\left. - (252 \rho_D^2 - 84) z(w)^3\right]
 \label{ffV1}
\end{align}
where  $z(w) = (\sqrt{w + 1} - \sqrt{2}) / (\sqrt{w + 1} + \sqrt{2})$. $V_1(1)$ includes
corrections at order $\alpha_s(m_Q)$ and $\Lambda_{QCD}/m_Q$ in HQET. Although $V_1(1)$ cancels in the ratio 
$R(D)$, it is better to note that lattice QCD can predict the value of $V_1(1) = 1.053 \pm 0.008$ \cite{Lattice:2015}. 
On the other hand, $\rho_D^2$ can be fitted directly from the data on $\Gamma(\bdnstell)$, where 
$\ell = e,~\mu$\footnote{From hereon, $\ell$ will mean light leptons, i.e. $e$ and $\mu$, unless specified otherwise.}. 
As of now, $\rho_D^2 = 1.186 \pm 0.054$, determined by the Heavy Flavor Averaging Group (HFAG) \cite{hfag}.

Following \cite{Tanaka:2010}, we parameterized the $w$ dependence of $S_1(w)$ as 
\begin{align}
 \nn S_1(w) = V_1(w) \times &\left\{1 + \Delta \left[ -0.019 + 0.041 \left(w - 1\right) \right.\right.\\
 &\left.\left. -0.015\left(w -1 \right)^2 \right]\right\}\,.
 \label{form-factors1}
\end{align}
Here, $\Delta$ parameterizes the unknown higher order corrections in HQET. In earlier analyses, for the prediction 
of the $R(D)$, $\Delta$ is assumed to have 100\% error. 
The decay rate $\Gamma(\bdnstell) $ is not useful to fit the parameters of $S_1(w)$, as it is not sensitive to the decay rates 
because of the negligible lepton masses. However, in our analysis, we fit $\Delta$ from the existing data on $R(D)$
%assuming that $R(D)$ is not affected by NP lets us fit $\Delta$ 
along with the other parameters defined earlier. 

As shown in eq. (\ref{matrixDst}), the $B \to D^* \tau \nu$ decays are described by four independent hadronic form-factors: 
$V$, $A_0$, $A_1$ and $A_2$, which are related to HQET form-factors by the following relations \cite{Fajfer:2012vx}:
\begin{align}
 \nn V(w) &= \frac{R_1(w)}{r_{D^*}} h_{A_1}(w)\,, \\
 \nn A_1(w) &= \frac{1}{2} r_{D^*}(w+1)h_{A_1}(w)\,, \\
 \nn A_2(w) &= \frac{R_2(w)}{r_{D^*}}h_{A_1}(w)\,, \\
 \nn A_0(w) &= \frac{R_0(w)}{r_{D^*}}h_{A_1}(w) \\
  \label{dstFFparam1}
\end{align}
where $r_{D^*} = 2\sqrt{m_B m_{D^*}} / (m_B + m_{D^*})$.
The $w$ dependencies of the HQET form-factors are parameterized following the 
ref. \cite{CLN}, 
\begin{align}
 \nn h_{A_1}(w) =& h_{A_1}(1) \left[ 1 - 8\rho_{D^*}^2 z(w) + (53\rho_{D^*}^2-15) z(w)^2 \right. \\
 \nn &\left.  - (231\rho_{D^*}^2-91) z(w)^3 \right] \,, \\
 \nn R_1(w) =& R_1(1) - 0.12(w-1) + 0.05(w-1)^2 \,, \\
 \nn R_2(w) =& R_2(1) + 0.11(w-1) - 0.06(w-1)^2 \,, \\
 R_0(w) =& R_0(1) - 0.11(w-1) + 0.01(w-1)^2 \,.
 \label{dstFFparam3}
\end{align}
Here, the current lattice prediction is $h_{A_1}(1) = 0.906 \pm 0.013$ \cite{bailey14}, the rest of the three parameters
like $\rho_{D^*}$, $R_1(1)$, $R_2(1)$ are fitted directly from the decay rate $\Gamma(\bdstell)$ \cite{hfag},
\begin{align}
 \nn \rho^2_{D^*} &= 1.207 \pm 0.026, ~~~ C\left(\rho^2_{D^*},~R_1(1)\right) = 0.568,\\
 \nn R_1(1) &= 1.406 \pm 0.033, ~~~ C\left(\rho^2_{D^*},~R_2(1)\right) = -0.809,\\
 R_2(1) &= 0.853 \pm 0.020, ~ C\left(R_1(1),~R_2(1)\right) = -0.758,
 \label{dstFFparam4}
\end{align} 
 where the second column lists the correlations between the parameters. As $B\to D^* \ell \nu$ decays are not sensitive to $R_0(w)$,
 there is only theoretical estimate available on $R_0(1) = 1.14 \pm 0.07$, based on HQET \cite{Fajfer:2012vx}. However, it can be 
 considered to be a free parameter in our analysis of $B\to D^* \tau \nu$ data.

\subsection{$\chi^2$ analysis}\label{latparfit}

\begin{figure*}\centering
\subfloat[$B \rightarrow D \tau \nu$(\Babar)]{
\includegraphics[scale=0.27]{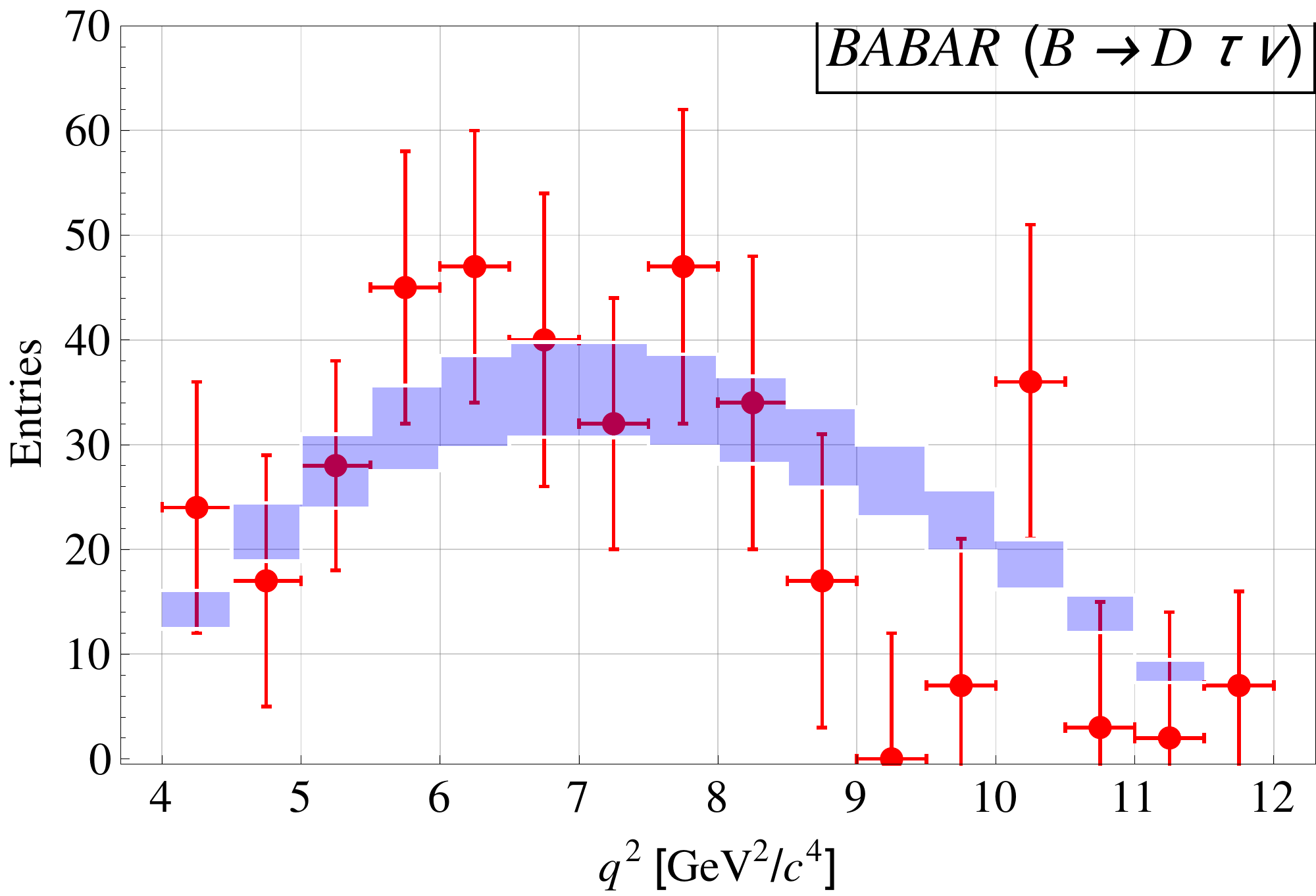}
 \label{fig:BABAR_dataD}}
\subfloat[$B \rightarrow D^* \tau \nu$(\Babar)]{
\includegraphics[scale=0.27]{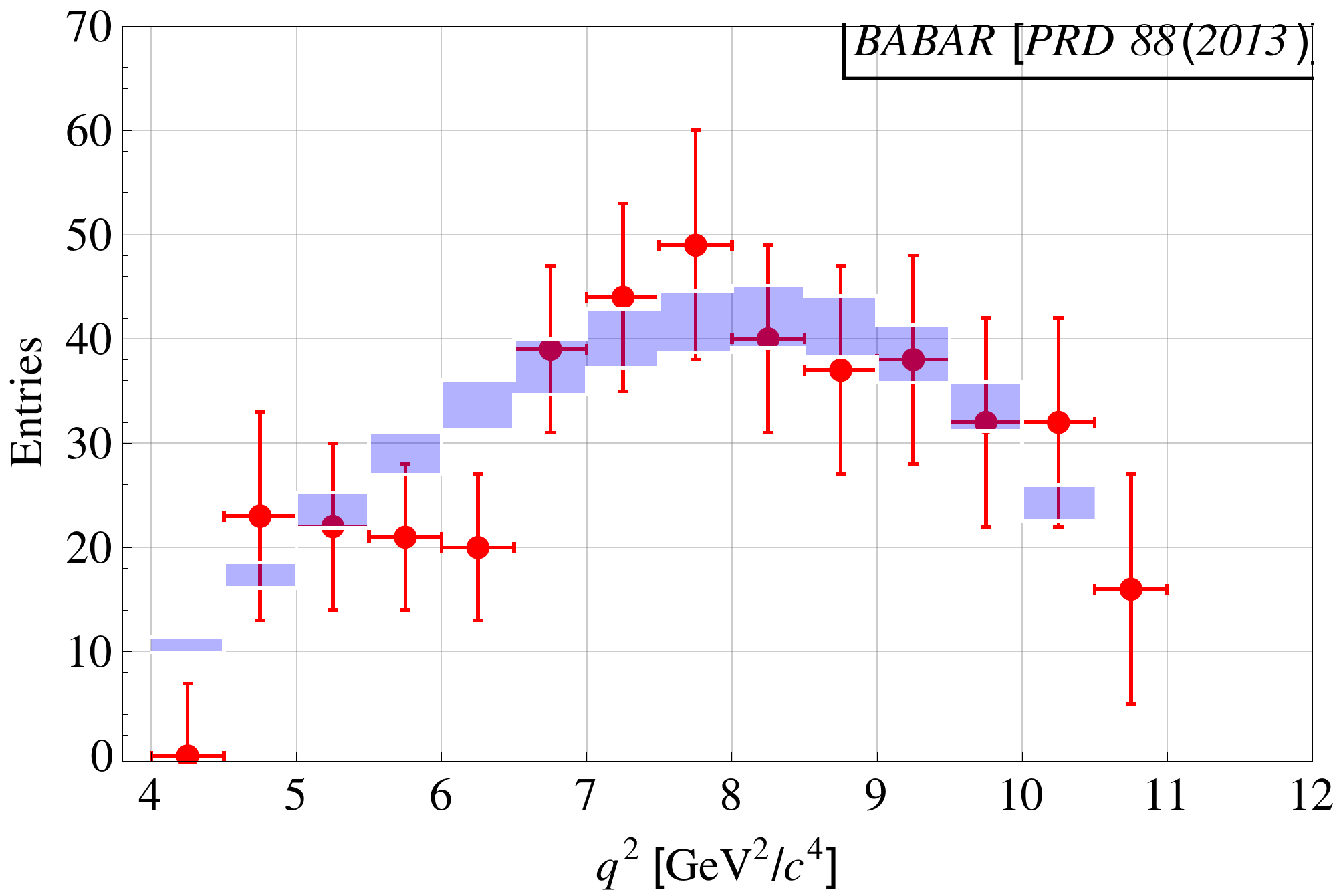}
 \label{fig:BABAR_dataDst}}
\subfloat[$B \rightarrow D^* \tau \nu$(Belle)]{
\includegraphics[scale=0.27]{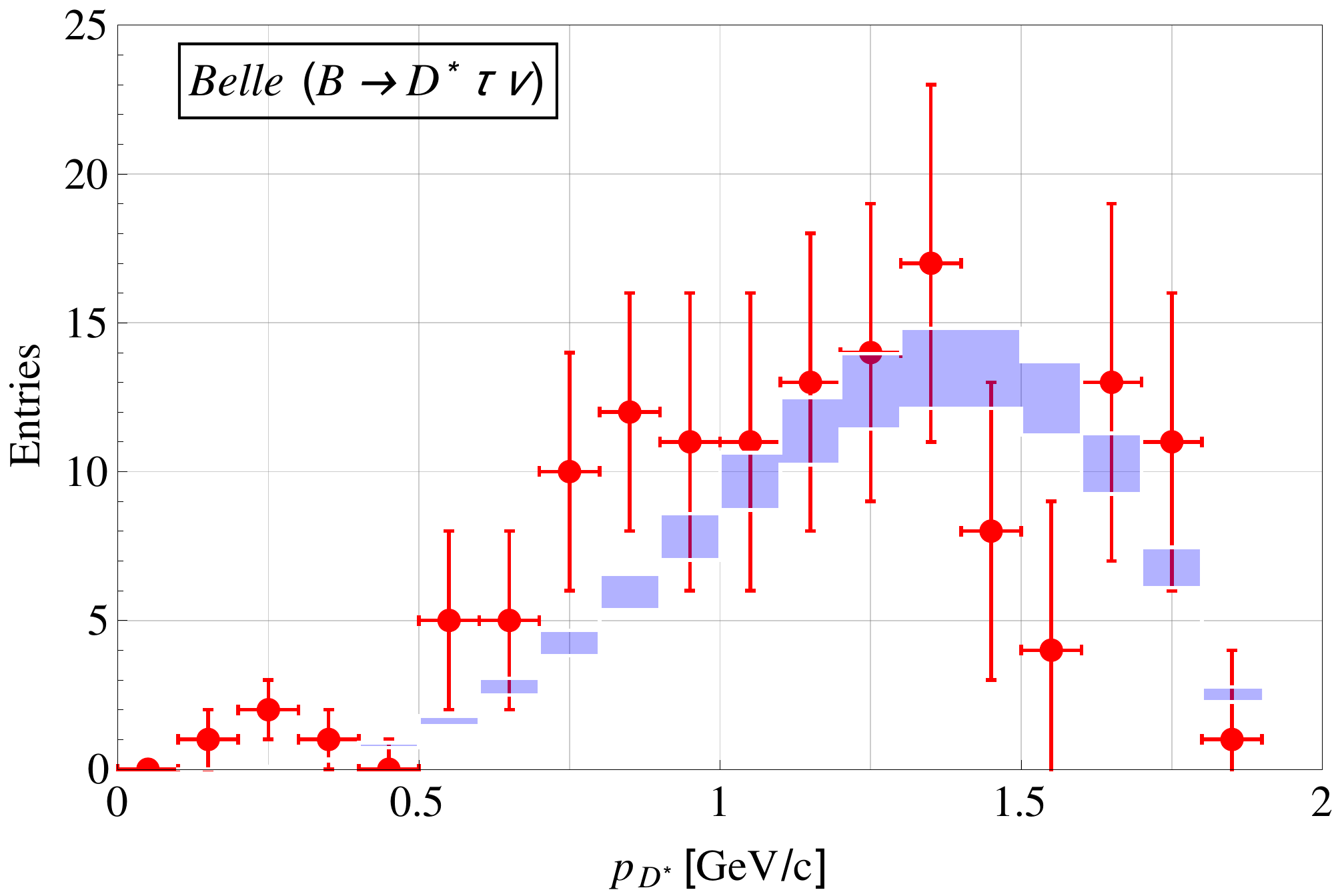}
 \label{fig:Belle_dataDst}} 
\caption{Fig.s \ref{fig:BABAR_dataD} and \ref{fig:BABAR_dataDst} are the measured background subtracted $q^2$-distributions 
for $\overline{B} \rightarrow D \tau \overline{\nu}_{\tau}$ and $\overline{B} \rightarrow D^* \tau \overline{\nu}_{\tau}$ 
events, extracted from the \Babar~ data \cite{babarexp}. Fig. \ref{fig:Belle_dataDst} is the background subtracted and 
normalized momentum distribution of $D^*$ extracted from the Belle data \cite{bellexp}}
\end{figure*}

Several parameters parameterizing the form-factors, otherwise not accessible in $\bar{B} \to D^{(*)} \ell^- \bar{\nu_{\ell}}$ decays, 
appear in $\bar{B} \to D^{(*)} \tau^- \bar{\nu_{\tau}}$ decays. By taking the binned data from the $q^2$-distribution of the decay rates 
in $\bar{B} \to D^{(*)} \tau^- \bar{\nu_{\tau}}$ normalized by $d\Gamma(\bdell)/dq^2$, we fit all the parameters given 
in section \ref{fftheory}. The only exceptions are $V_1(1)$ and $h_{A_1}(1)$ which will cancel in the ratios. 

Fig.s \ref{fig:BABAR_dataD} and \ref{fig:BABAR_dataDst} show efficiency-corrected $q^2$-distributions for 
$\overline{B} \rightarrow D \tau^- \overline{\nu}_{\tau}$ and $\overline{B} \rightarrow D^* \tau^- \overline{\nu}_{\tau}$ 
events with $m^2_{miss} > 1.5 ~\text{GeV}^2$, scaled to results of isospin-constrained fit extracted from the 
\Babar~\cite{babarexp} data. The $B^0$ and $B^+$ samples are combined and the normalization and background events are 
subtracted. The uncertainty on the data points includes the statistical uncertainties of data and simulation. 
%The distributions are normalized to the number of detected events. 
Fig. \ref{fig:Belle_dataDst} is the background subtracted and normalized momentum distribution of $D^*$ for 
$\overline{B} \rightarrow D^* \tau^- \overline{\nu}_{\tau}$ events extracted from the Belle \cite{bellexp} data. 
Here also, the $B^0$ and $B^+$ samples are combined and the normalization and background events are subtracted. 
The light blue histogram represent the SM prediction for the same in each individual bin. We note that both 
Belle and \Babar~binned data show deviations from SM predictions.  
\begin{table}
 \begin{center}
  \begin{tabular}{|c|c|c|c|}
    \hline
    Experiment & Channel & Input & Value\\
    \hline
     & $\overline{B} \rightarrow D \tau^- \overline{\nu}_{\tau}$ & $N_{sig}$ & $489 \pm 63$ \\
     \Babar~ & & $N_{norm}$ & $ 2981 \pm 65$ \\
     \cite{babarexp} & & $\epsilon_{sig} / \epsilon_{norm}$ & $ 0.372 \pm 0.010$ \\
      \cline{2-4}
     & $\overline{B} \rightarrow D^* \tau^- \overline{\nu}_{\tau}$ & $N^*_{sig}$ & $888 \pm 63$ \\
     & & $N^*_{norm}$ & $ 11953 \pm 122 $ \\
     & & $\epsilon^*_{sig} / \epsilon^*_{norm}$ & $ 0.224 \pm 0.004$ \\
     \hline
     & $\overline{B} \rightarrow D^* \tau^- \overline{\nu}_{\tau}$ & $N^*_{sig}$ & $231 \pm 23$ \\
     Belle(2016) & & $N^*_{norm}$ & $ 2800 \pm 57 $ \\
     \cite{bellexp} & & $\epsilon^*_{norm} / \epsilon^*_{sig}$ & $ 1.289 \pm 0.015$ \\
     \hline
     LHCb & $\overline{B} \rightarrow D^* \tau^- \overline{\nu}_{\tau}$ &
      $R(D^*)$ & $0.336\pm0.027$ \\
     \cite{lhcbexp} & & &  $\pm 0.030$ \\
     \hline
     & $\overline{B} \rightarrow D \tau^- \overline{\nu}_{\tau}$ &  $R(D)$ & $0.375\pm0.064$ \\
     Belle(2015) & & &  $\pm 0.026$ \\
     \cline{2-4}
     \cite{Huschle:2015rga} & $\overline{B} \rightarrow D^* \tau^- \overline{\nu}_{\tau}$ &  $R(D^*)$ & $0.293\pm0.038$ \\
     & & &  $\pm 0.015$ \\
     \hline
     Belle(Latest) & $\overline{B} \rightarrow D^* \tau^- \overline{\nu}_{\tau}$ &
      $R(D^*)$ & $0.276\pm0.034$ \\
     \cite{Abdesselam:2016xqt} & & & $^{+0.029}_{-0.026}$ \\
     \hline
  \end{tabular}
 \end{center}
 \caption{Experimental inputs for fits. Only statistical uncertainties are supplied for $N^{(*)}_{norm(sig)}$. Whenever two 
uncertainties are quoted, they are the statistical and systematic ones respectively.}
 \label{tab:expinput}
\end{table}

% \begin{table}[htbp]
%  \begin{center}
%   \begin{tabular}{|c|c|c|c|}
%     \hline
%     Experiment & Channel & Input & Value\\
%     \hline
%      & $\overline{B} \rightarrow D \tau^- \overline{\nu}_{\tau}$ & $N_{sig}$ & $489 \pm 63$ \\
%      \Babar~ & & $N_{norm}$ & $ 2981 \pm 65$ \\
%      & & $\epsilon_{sig} / \epsilon_{norm}$ & $ 0.372 \pm 0.010$ \\
%      & $\overline{B} \rightarrow D^* \tau^- \overline{\nu}_{\tau}$ & $N^*_{sig}$ & $888 \pm 63$ \\
%      & & $N^*_{norm}$ & $ 11953 \pm 122 $ \\
%      & & $\epsilon^*_{sig} / \epsilon^*_{norm}$ & $ 0.224 \pm 0.004$ \\
%      \hline
%      & $\overline{B} \rightarrow D^* \tau^- \overline{\nu}_{\tau}$ & $N^*_{sig}$ & $231 \pm 23$ \\
%      Belle & & $N^*_{norm}$ & $ 2800 \pm 57 $ \\
%      & & $\epsilon^*_{norm} / \epsilon^*_{sig}$ & $ 1.289 \pm 0.015$ \\
%      \hline
%   \end{tabular}
%  \end{center}
%  \caption{Experimental inputs for fits. Uncertainties of $N^{(*)}_{norm(sig)}$ are all statistical.}
%  \label{tab:expinput}
% \end{table}
%
To fit the parameters of the form-factors, we have performed a test of significance (goodness of fit) by 
defining a $\chi^2$ statistic, a function of the parameters parameterizing the form-factors, which is defined as
\begin{align}
\chi^2_{Lat} &= %\sum^{{\rm bins}}_i \frac{\left(R(D^{(*)})^{th}_{i} - R(D^{(*)})^{exp}_{i}\right)^2}{\sigma^2_{exp,~i}},  
\sum^{{\rm bins}}_{i,j = 1} \left(R(D^{(*)})^{exp}_i - R(D^{(*)})^{th}_i\right)\,. \nn \\
& V^{-1}_{i j}\,. \left(R(D^{(*)})^{exp}_j - R(D^{(*)})^{th}_j\right),
\label{chi2lat}
\end{align}
where 
\bea
 R(D^{(*)})^{th}_{bin} &= \frac{\int^{q^2_{{\rm max}}}_{q^2_{{\rm min}}} \left(d\Gamma\left(\overline{B} \rightarrow D^{(*)}
 \tau^- \overline{\nu}_{\tau}\right)/d q^2\right) d q^2}{\int_{{\rm full } ~q^2} \left(d\Gamma\left(\overline{B} \rightarrow
 D^{(*)} \ell \overline{\nu}_{\ell}\right) / d q^2 \right) d q^2},
 \label{Rth} 
 \eea
\begin{equation}
R(D^{(*)})^{exp}_{bin} = \begin{cases}  \frac{N^{(*)}_{bin}}{N^{(*)}_{norm}} \times \frac{\epsilon^{(*)}_{norm}}
                              {\epsilon^{(*)}_{sig}} \quad   \text{~\Babar} \\
                              \frac{1}{2 \mathcal{B}(\tau^- \to \ell^- \bar{\nu_{\ell}} \nu_{\tau})} 
                              ~\frac{N^{(*)}_{bin}}{N^{(*)}_{norm}} \times \frac{\epsilon^{(*)}_{norm}}{\epsilon^{(*)}_{sig}}
                              &  \text{Belle}.\\
 
\end{cases}
\label{RexpBaBe}
\end{equation}
$V_{i j}$ is the covariance matrix. It comprises of $\sigma^2_{exp,~bin}$, the experimental uncertainties obtained by 
propagating the uncertainties of individual parts in the r.h.s of eq.(\ref{RexpBaBe}).
As input, we consider the central values of number of events $N^{(*)}_{bin}$, along with their errors, for each $q^2$ or $p_{D^*}$ bin 
depending on whether we are analyzing the \Babar~or the Belle data. The 
total signal yield $N_{sig}^{(*)}$, along with the errors are given in table \ref{tab:expinput}.
For simplicity and due to lack of knowledge of $q^2$-distribution of the efficiencies, we have taken the 
ratio of efficiencies $\epsilon_{sig}^{(*)} / \epsilon_{norm}^{(*)}$ to be constant over all 
the $q^2$ regions and equal to the value shown in table \ref{tab:expinput}.  
In eqs. (\ref{Rth}) and (\ref{RexpBaBe}), $q^2_{max(min)}$ are end points of a particular bin. For the denominator 
in eq.(\ref{Rth}), we integrate over the whole allowed 
phase space (from $ q^2 = m^2_{\ell}$ to $q^2 = \left(m_B - m_{D^{(*)}}\right)^2$). 

In defining $V_{i j}$, we follow these procedures:
\begin{enumerate} 
\item Our $V$ comprises of two parts - the statistical covariance matrix $V^{stat}$ and the systematic one,
$V^{syst}$. So, $V^{exp} = V^{stat} + V^{syst}$. As there is no information available 
to us about the systematic uncertainties and their correlations on the binned data, we do two separate analyses.
\item The first analysis is done using only the data available to us, i.e. $V^{syst}$ is set to be zero and 
$V^{stat}_{i j} = \delta_{i j} ~\delta R^{exp}_{i} ~\delta R^{exp}_{j}$ (here $\delta_{i j}$ is the Kronecker delta). We will
call this ``Fit-1'' from hereon.
\item The second analysis is done assuming the systematic uncertainties to be the same as the statistical ones 
and $100\%$ systematic correlation, i.e. $V^{syst}_{i j} = \delta R^{exp}_{i} ~\delta R^{exp}_{j}$ and 
$V^{stat}_{i j}$ defined as earlier. We will call this ``Fit-2'' from hereon.
\end{enumerate}
The utility of considering the systematic uncertainties to be the same as statistical ones and considering 
$100\%$ systematic correlations in the second analysis are multi-pronged. First of all, as the statistical uncertainties
on the binned data are quite large, this makes the systematic errors similarly large and that in turn can conservatively account 
for the possible systematic errors coming from $a)$ the `model-dependence' of the `background-subtracted' binned data as mentioned in 
section \ref{sec:npmethod} and $b)$ the dependence of the shape of the $q^2$-distribution on the experimental cuts on the leptons 
and hadrons. Secondly, separately analyzing the data in both under-correlated and over-correlated ways and comparing them, gives 
us an idea of the dependence of the analysis on these unknown systematic bin-bin correlations.

The Belle results \cite{bellexp} used here is the first measurement of $R(D^*)$ 
using semileptonic tagging method for the ``other $B$'', referred to as $B_{tag}$ and instead of a $q^2$-distribution, 
the momentum distribution of $D^*$ and $\ell$ are given. For our analysis, we note that 
$p^2_{D^*} = \left(\frac{m^2_{B} + m^2_{D^*} - q^2}{2 m_{B}}\right)^2 -m^2_{D^*}$, and using this, eq.(\ref{Rth}) can 
be calculated for each bin in the $p_{D^*}$-distribution by converting the limits of integration appropriately. 
For $R(D^{*})^{exp}_{bin}$, we use eq.(\ref{RexpBaBe}). We do not use those bins for which central values of $ N^{(*)}_{bin} \leq 0$.

To utilize the fact that $V_1(1)$ and $h_{A_1}(1)$ get canceled respectively in $R(D)$ and $R(D^*)$, 
$R(D^{(*)})_{bin}$ is used instead of $N^{(*)}_{bin}$. So, the $\chi^2_{Lat}$ is a function of $\rho^2_D$ and 
$\Delta$ for $R(D)_{bin}$ and a function of $\rho^2_{D^*}$, $R_1(1)$, $R_2(1)$ and $R_0(1)$ for $R(D^{*})_{bin}$.

\subsection{Goodness of Fit}\label{sec:goodfit}

A true model with true parameter values will generate a $\chi^2 = d.o.f$ i.e. $\chi^2_{red} = 1$ as there is no fit involved.
But due to noise present in the data, this is not sufficient information to assess convergence or compare different models.
The obligatory step to assess the goodness-of-fit of an analysis after optimization is then to inspect the distribution of 
the residuals. For the true model, with a-priori known measurement errors, the distribution of normalized residuals 
(in our case, $\frac{R^{th}_{bin} - R^{exp}_{bin}}{\delta R_{bin}}$) is by definition a Gaussian with mean $\mu = 0$ and 
variance $\sigma^2 = 1$ \cite{dosdonts}. This fact is utilized to test the significance of the fit by objectively 
quantifying a significance test of fitting the distribution of residuals to this Gaussian. For this, 
we use Shapiro-Wilk's(S-W) test \cite{shapiro} for normality. The reasons for choosing 
S-W over other competing tests for normality are following: $a)$ Though we have used the algorithm $AS~R94$ by Royston \cite{royston}, 
which was developed for any sample size ($n$) $3-5000$, the original S-W test was specifically designed for $n<50$; this is precisely 
our case. $b)$ This is the first test which detected departures from normality using skewness and/or kurtosis and since then have 
been regularly corrected and developed. $c)$ It has repeatedly been shown \cite{comptest} that from low to medium sample sizes, where
degenerate values occur less, S-W is the `most powerful' parametric test for normality among other popular contenders like 
`Kolmogorov-Smirnov', `Anderson-Darling', `Cram\'{e}r-von Mises', `Jarque-Bera' etc.; as this identically applies to our case, 
we choose S-W test throughout this analysis. In all such tests, 
the validity of a hypothesis depends on whether the probability of the goodness of fit test is above or below 
the significance, which in our case is set at $5\%$. 
Across all the fitted models, the ones with the $p$-value of the residual-distribution above $5\%$ 
will be considered to fit the data well; all of the rest can be thrown out.
Therefore, if a particular model fitting analysis passes our normality test, we consider that model 
as the plausible explanation of the data.

\subsection{Fit Results}\label{ffresults}

\begin{figure*}[!htbp]
\centering
\subfloat[$R(D)_{bin}$]{
\includegraphics[scale=0.68]{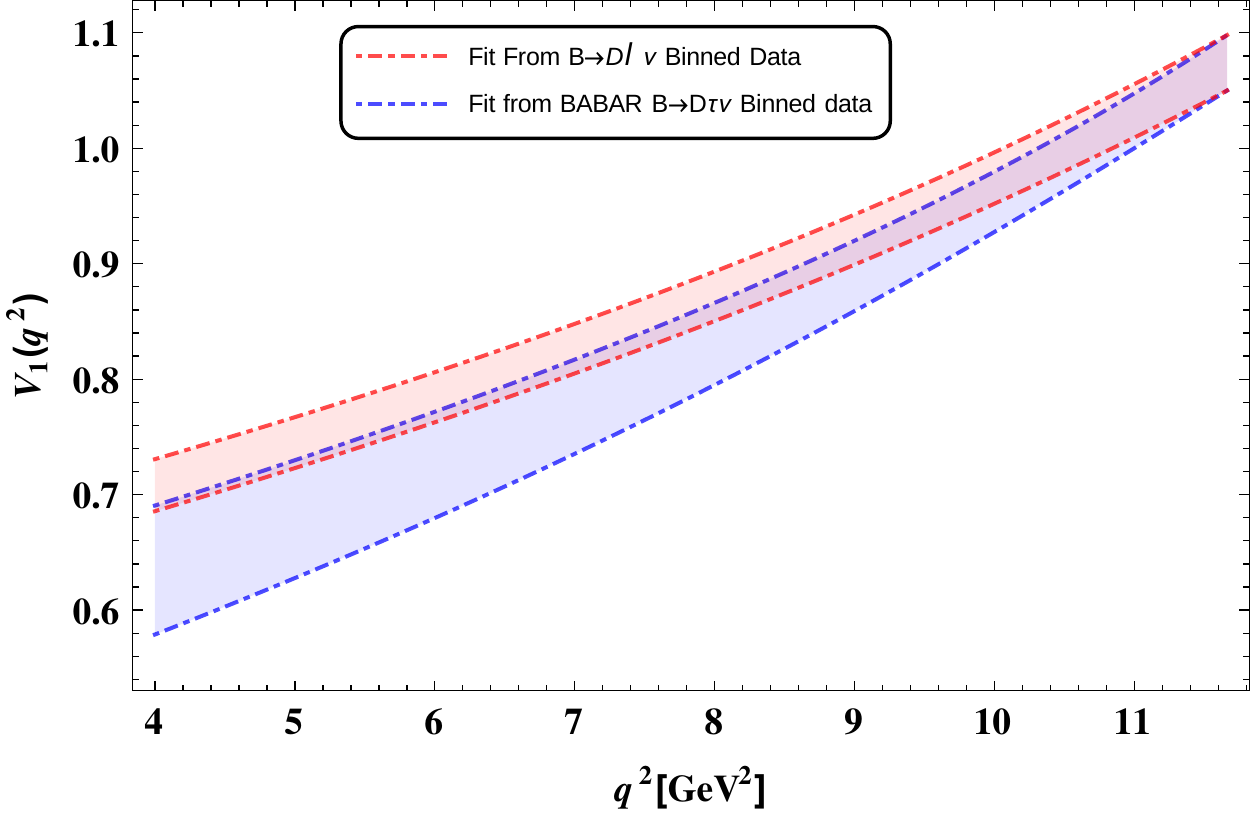}
 \label{fig:V1D}}
\subfloat[$R(D)_{bin}$]{
\includegraphics[scale=0.68]{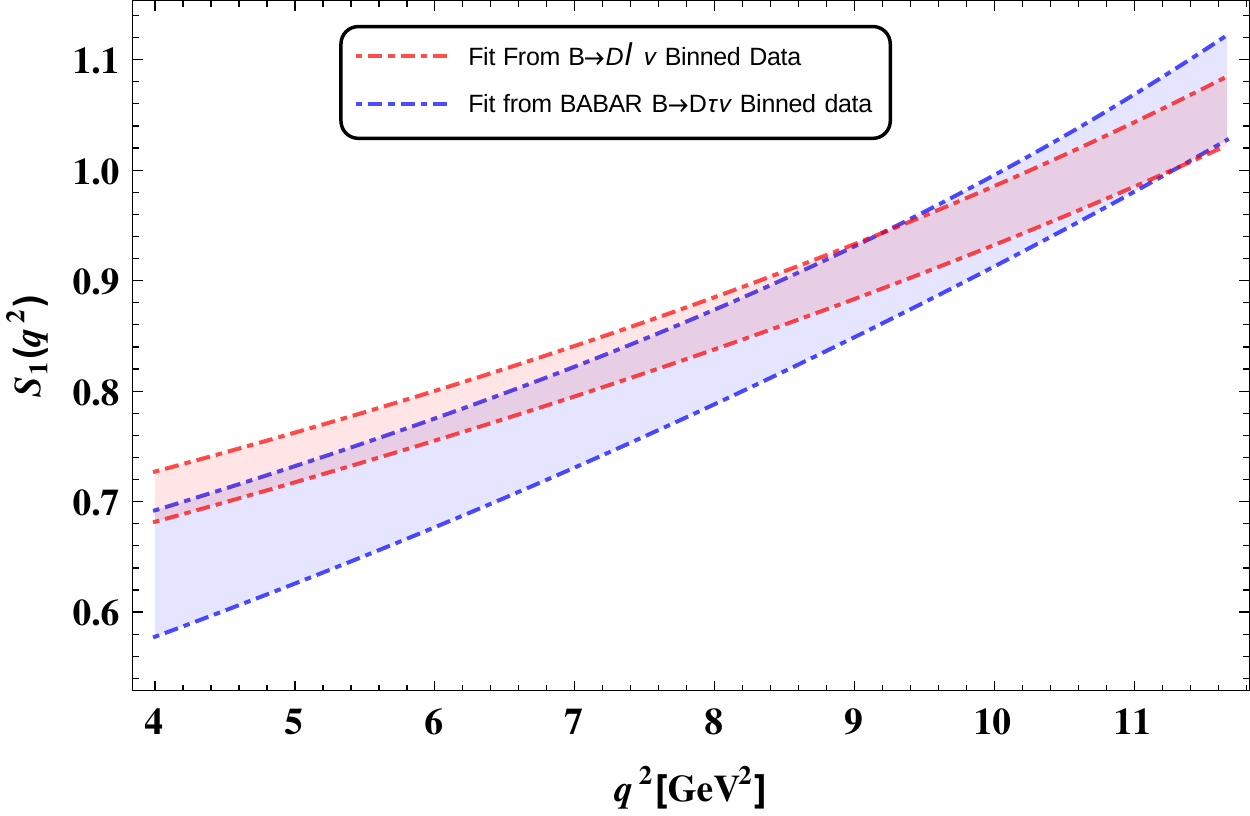}
 \label{fig:S1D}}\\
\subfloat[$R(D^{*})_{bin}$]{
\includegraphics[scale=0.68]{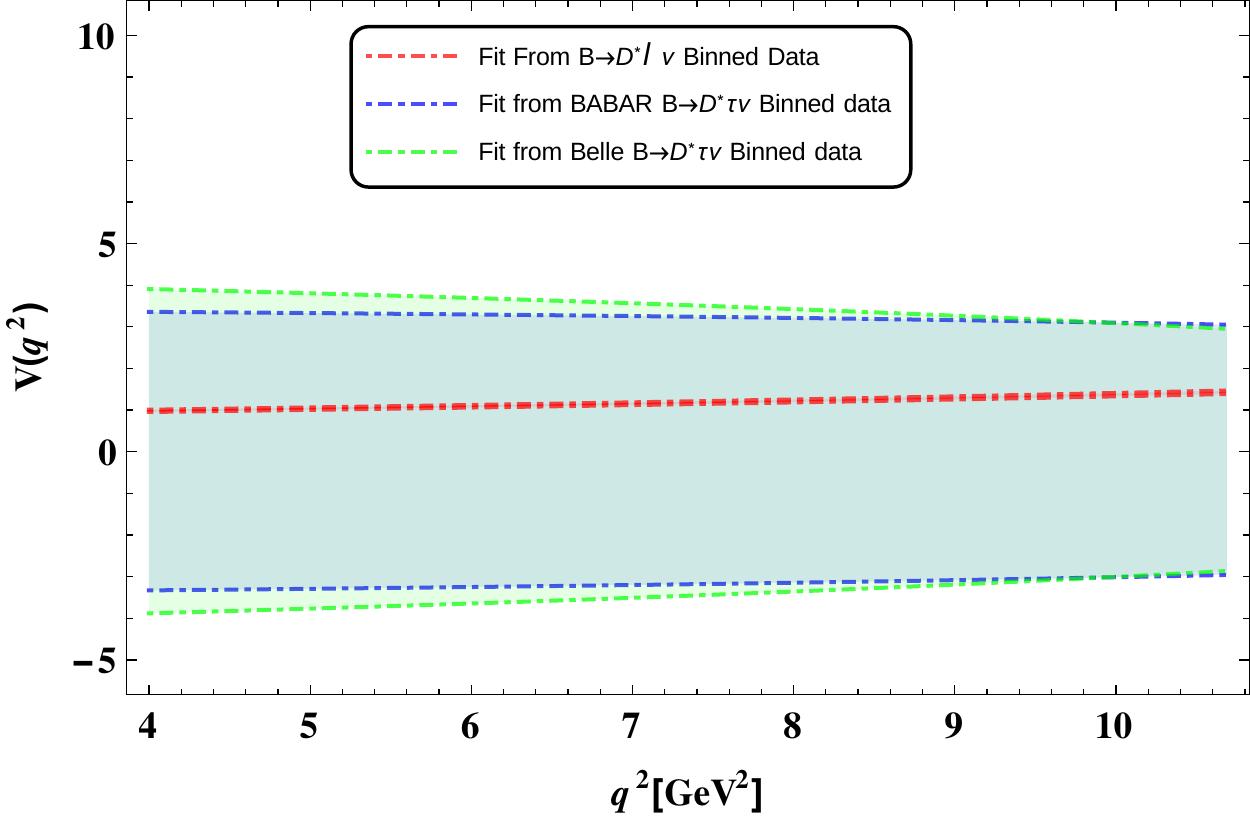}
 \label{fig:VDst}}
\subfloat[$R(D^{*})_{bin}$]{
\includegraphics[scale=0.68]{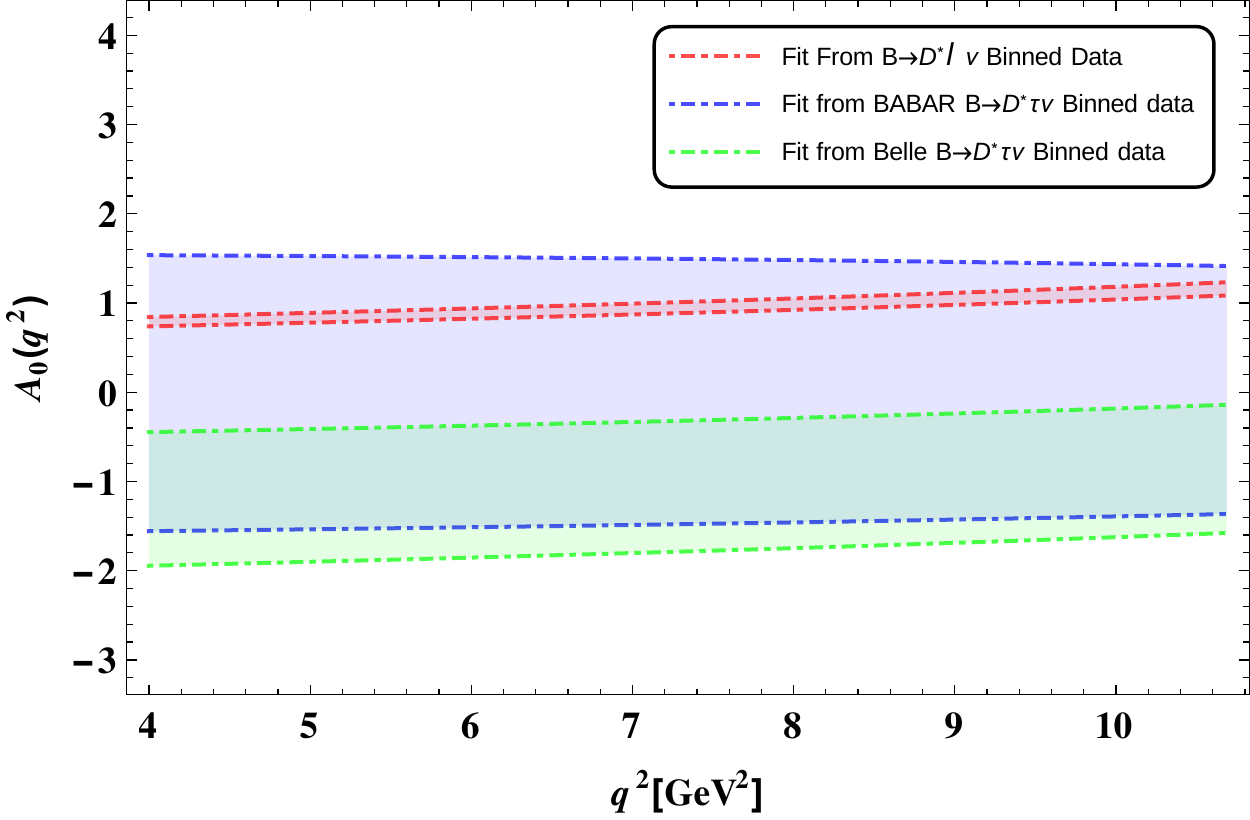}
 \label{fig:A0Dst}}\\
\subfloat[$R(D^{*})_{bin}$]{
\includegraphics[scale=0.68]{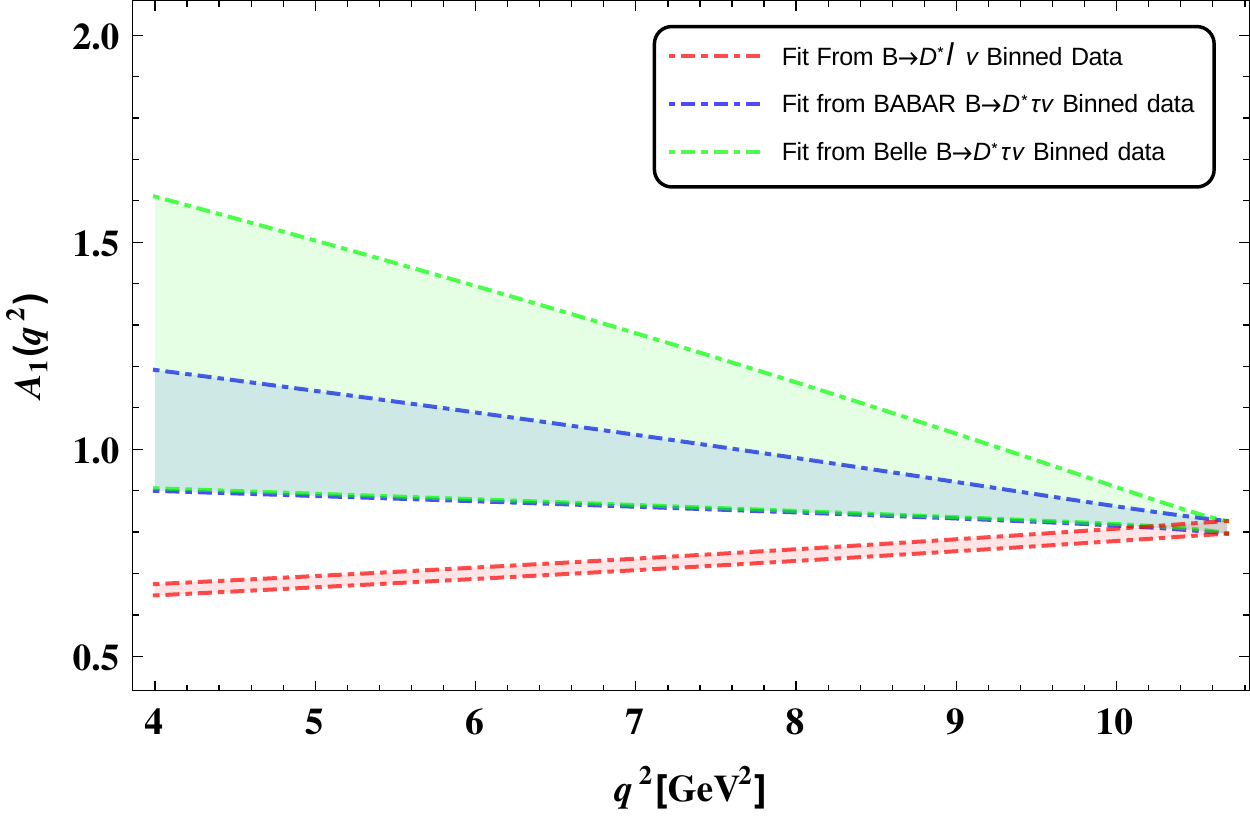}
 \label{fig:A1Dst}}
\subfloat[$R(D^{*})_{bin}$]{
\includegraphics[scale=0.68]{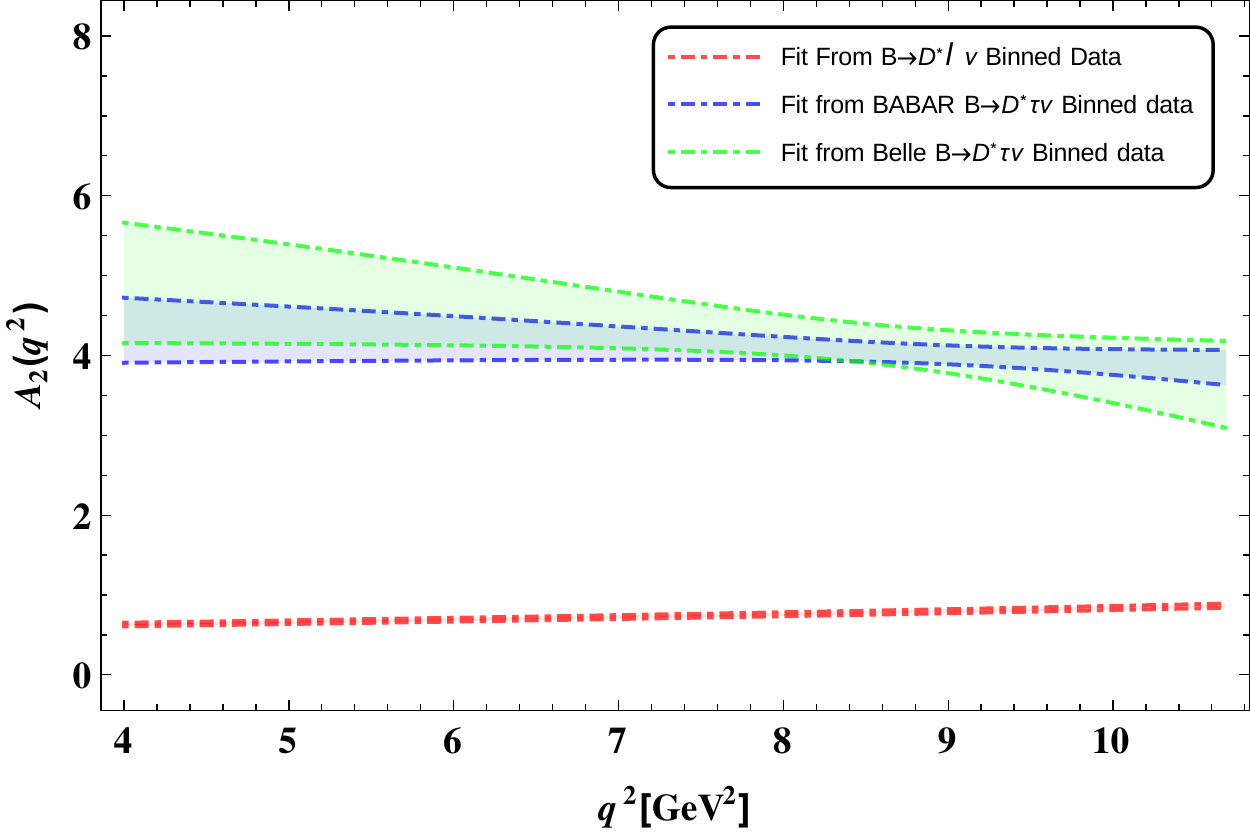}
 \label{fig:A2Dst}}
\caption{Results obtained from `Fit-1'. Fig.s \ref{fig:V1D} and \ref{fig:S1D} are the $q^2$ dependence of form-factors for semileptonic $b \to c$ transitions. Red (dotted) and blue (dot-dashed) lines enclose $\pm 1 \sigma$ regions for the form-factors with parameters fitted 
from $\overline{B} \rightarrow D \ell \overline{\nu}_{\ell}$ (world average) and 
$\overline{B} \rightarrow D \tau \overline{\nu}_{\tau}$ decays (\Babar) respectively. 
The rest of the figures are for form-factors for $\overline{B} \rightarrow D^* \ell \overline{\nu}_{\ell}$ and 
$\overline{B} \rightarrow D^* \tau \overline{\nu}_{\tau}$ decays. Here green (solid) lines enclose the region for 
$\overline{B} \rightarrow D^* \tau \overline{\nu}_{\tau}$ decays (Belle).}
\label{fig:form-factors}
\end{figure*}

\begin{table}[htbp]
 \begin{center}
  \begin{tabular}{|l|c|c|c|c|c|}
    \hline
     \, Obs. & Par.s & Value & $\chi^2_{min}$ & $d.o.f $ & Normality\\
%      \cline{6-8}
     & & & & & (S-W) \\
    \hline
    \Babar~ & & & & &  \\
     $R(D)_{bin}$ & $\Delta$ & $-0.04 \pm 2.00$ & $9.6 $ & $ 12$ & 0.20 \\
      & $\rho^2_D$ & $ 1.43 \pm 0.18 $ & & &\\
      \cline{2-6}
     $R(D^*)_{bin}$ & $\rho^2_{D^*}$ & $-0.55 \pm 0.66 $ & $5.4$ & $ 8$ & 0.25 \\
      & $R_1(1)$ & $ 0.04 \pm 2.96 $ & & &\\ 
      & $R_2(1)$ & $ 3.79 \pm 0.20 $ & & &\\
      & $R_0(1)$ & $ 0.02 \pm 1.37 $ & & & \\
     \hline
     Belle~ & & & & &  \\
     $R(D^*)_{bin}$ & $\rho^2_{D^*}$ & $ - 1.52 \pm 1.61 $ & $ 8.7 $ & $ 13 $ & 0.91  \\
      & $R_1(1)$ & $ 0.04 \pm 2.86 $ & & &  \\ 
      & $R_2(1)$ & $ 3.58 \pm 0.53 $ & & &  \\
      & $R_0(1)$ & $ -0.84 \pm 0.71 $ & & & \\
      \hline
  \end{tabular}
 \end{center}
 \caption{Fit-I Results of parameters parameterizing the form-factors in HQET. The last column lists the results of the 
 hypothesis test (Shapiro-Wilk) for assessment of goodness-of-fit.}
 \label{tab:latfitres}
\end{table}

\begin{table}[htbp]
 \begin{center}
  \begin{tabular}{|c|c|c|c|}
    \hline
    Channel & Correlation & \Babar & Belle (2016) \\
    \hline
   & $C\left(\rho^2_{D^*},~R_1(1)\right)$ & 0.057 & 0.023 \\
    \cline{2-4}
    $\overline{B} \rightarrow D^* \tau \overline{\nu}_{\tau}$ & $C\left(\rho^2_{D^*},~R_2(1)\right)$ & 0.907 & 0.928 \\
    \cline{2-4}
   & $C\left(\rho^2_{D^*},~R_0(1)\right)$  & -0.004 & -0.741 \\
    \cline{2-4}
   & $C\left(R_1(1),~R_2(1)\right)$  & 0.082 & 0.024 \\
    \cline{2-4}
   & $C\left(R_1(1),~R_0(1)\right)$  & 0.000 & -0.008 \\
    \cline{2-4}
   & $C\left(R_2(1),~R_0(1)\right)$  & 0.007 & -0.861 \\
   \hline
   $\overline{B} \rightarrow D \tau \overline{\nu}_{\tau}$ & $C\left(\Delta,~\rho^2_{D}\right)$ & 0.146 & - \\
   \hline
  \end{tabular}
\end{center}
\caption{Correlations between the fitted form-factor parameters from Fit-I.}
\label{tab:corrections}
\end{table}

\begin{table}[htbp]
 \begin{center}
  \begin{tabular}{|l|c|c|c|c|c|}
    \hline
     \, Obs. & Par.s & Value & $\chi^2_{min}$ & $d.o.f $ & Normality\\
%      \cline{6-8}
     & & & & & (S-W) \\
    \hline
    \Babar~ & & & & &  \\
     $R(D)_{bin}$ & $\Delta$ & $-0.03 \pm 2.25$ & $8.71 $ & $ 12$ & 0.14 \\
      & $\rho^2_D$ & $ 0.92 \pm 0.60 $ & & &\\
      \cline{2-6}
     $R(D^*)_{bin}$ & $\rho^2_{D^*}$ & $-0.54 \pm 0.73 $ & $5.13$ & $ 8$ & 0.55 \\
      & $R_1(1)$ & $ 0.04 \pm 1.99 $ & & &\\ 
      & $R_2(1)$ & $ 3.93 \pm 0.31 $ & & &\\
      & $R_0(1)$ & $ 0.03 \pm 0.76 $ & & & \\
     \hline
     Belle~ & & & & &  \\
     $R(D^*)_{bin}$ & $\rho^2_{D^*}$ & $ -3.03 \pm 2.24 $ & $ 6.62 $ & $ 13 $ & 0.68  \\
      & $R_1(1)$ & $ 0.04 \pm 2.31 $ & & &  \\ 
      & $R_2(1)$ & $ 3.78 \pm 0.45 $ & & &  \\
      & $R_0(1)$ & $ 0.03 \pm 0.93 $ & & & \\
      \hline
  \end{tabular}
 \end{center}
 \caption{Fit-II Results of parameters parameterizing the form-factors in HQET.}
 \label{tab:latfitres2}
\end{table}

\begin{table}[htbp]
 \begin{center}
  \begin{tabular}{|c|c|c|c|}
    \hline
    Channel & Correlation & \Babar & Belle (2016) \\
    \hline
   & $C\left(\rho^2_{D^*},~R_1(1)\right)$ & 0.031 & 0.015 \\
    \cline{2-4}
    $\overline{B} \rightarrow D^* \tau \overline{\nu}_{\tau}$ & $C\left(\rho^2_{D^*},~R_2(1)\right)$ & 0.698 & 0.563 \\
    \cline{2-4}
   & $C\left(\rho^2_{D^*},~R_0(1)\right)$  & 0.011 & 0.004 \\
    \cline{2-4}
   & $C\left(R_1(1),~R_2(1)\right)$  & 0.035 & 0.021 \\
    \cline{2-4}
   & $C\left(R_1(1),~R_0(1)\right)$  & 0.000 & 0.000 \\
    \cline{2-4}
   & $C\left(R_2(1),~R_0(1)\right)$  & 0.018 & 0.012 \\
   \hline
   $\overline{B} \rightarrow D \tau \overline{\nu}_{\tau}$ & $C\left(\Delta,~\rho^2_{D}\right)$ & 0.07 & - \\
   \hline
  \end{tabular}
\end{center}
\caption{Correlations between the fitted form-factor parameters from Fit-II.}
\label{tab:corrections2}
\end{table}

The fit results for the parameters of the form-factors are listed in tables \ref{tab:latfitres} and \ref{tab:latfitres2} 
for `Fit-1' and `Fit-2' respectively.
We find the distribution of the residuals for all those fits and check whether that 
distribution is accordant with a normal distribution with mean $0$ and variance $1$ (with the null hypothesis $H_0$ 
that this is true).  $p$-values obtained in our chosen normality test (S-W) quantify the 
probability of $H_0$ being true. 
% However, for comparison, in table  \ref{tab:latfitres}, we list the 
% $p$-values of two more representative tests, namely Kolmogorov-Smirnov and Anderson-Darling along with our chosen 
% test Shapiro-Wilk. 

After the minimization, we find the uncertainties of and correlations between the parameters around their best fit points. 
A general approach to find these is to construct the `Hessian Matrix' $H$, which is the matrix of second order 
partial-derivatives of the test-statistic with respect to the parameters; this describes the local curvature of 
a function of many variables, and find its inverse. This constitutes the `error matrix', square roots of whose 
diagonal elements give us the `standard error' of the parameters and the normalized matrix (w.r.t the errors) 
makes the `correlation matrix'. We list such errors in tables \ref{tab:latfitres} and \ref{tab:latfitres2} and relevant correlations in tables \ref{tab:corrections} and \ref{tab:corrections2}.

In the following we will discuss the outcome of our analysis, and compare our fit results with that determined by HFAG
\cite{hfag} (also given in eq. (\ref{dstFFparam4})):
\begin{itemize}
 \item We fit $\rho^2_D$ only using the \Babar~data, the obtained values are consistent 
  with that determined by the HFAG at 1$\sigma$. Our fitted values of $\Delta$ include  
 $\Delta = 1 \pm 1$, so far, which is used in the prediction of $R(D)$ by \Babar~ \cite{babarexp}.
 \item The analysis of the \Babar~bin data on $R(D^*)$ from both `Fit-1' and `Fit-2' shows that the fitted parameters like $\rho^2_{D^*}$ and $R_1(1)$
  are consistent within 2$\sigma$, with HFAG. However, $R_2(1)$ shows a 
  large deviation (more than 10$\sigma$ away). It is important to note that we can extract 
  $R_2(1)$ with relatively small error.
  \item After analyzing the data by Belle on $R(D^*)$ from `Fit-1', we obtain large errors on $\rho^2_{D^*}$ and $R_1(1)$, 
  and they are consistent with the fitted value by HFAG at 1$\sigma$. `Fit-2' increases both the best-fit value 
  and errors of $\rho^2_{D^*}$ even more.
  Also in this case, $R_2(1)$ fits with a small error, and shows a large deviation from that determined 
  by HFAG. 
  \item Whereas the analysis of $R(D^*)$ from `Fit-1' results obtained using \Babar~and Belle binned data (table 
  \ref{tab:latfitres}) are roughly consistent with each other, including the best-fit values of $R_0(1)$, the same analysis from `Fit-2' (table \ref{tab:latfitres2}) actually makes the results compatible. So much so, that the 
  $R_0(1)$ best-fit value becomes almost identical. This makes one inclined to think that Belle binned data is more correlated than 
  is assumed.
\end{itemize}

\begin{figure*}[!htbp]
\centering
\subfloat[$R(D)_{bin}$]{
\includegraphics[scale=0.68]{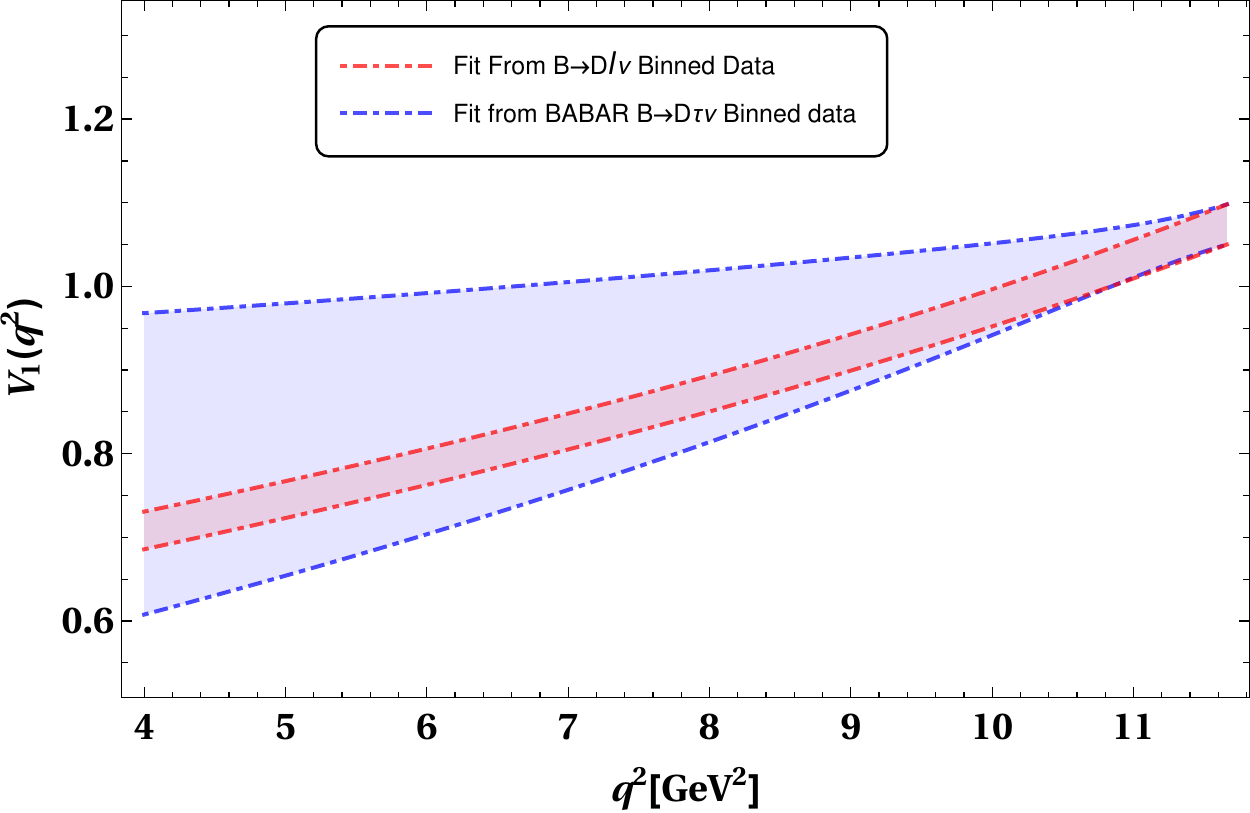}
 \label{fig:V1Dfit2}}
\subfloat[$R(D)_{bin}$]{
\includegraphics[scale=0.68]{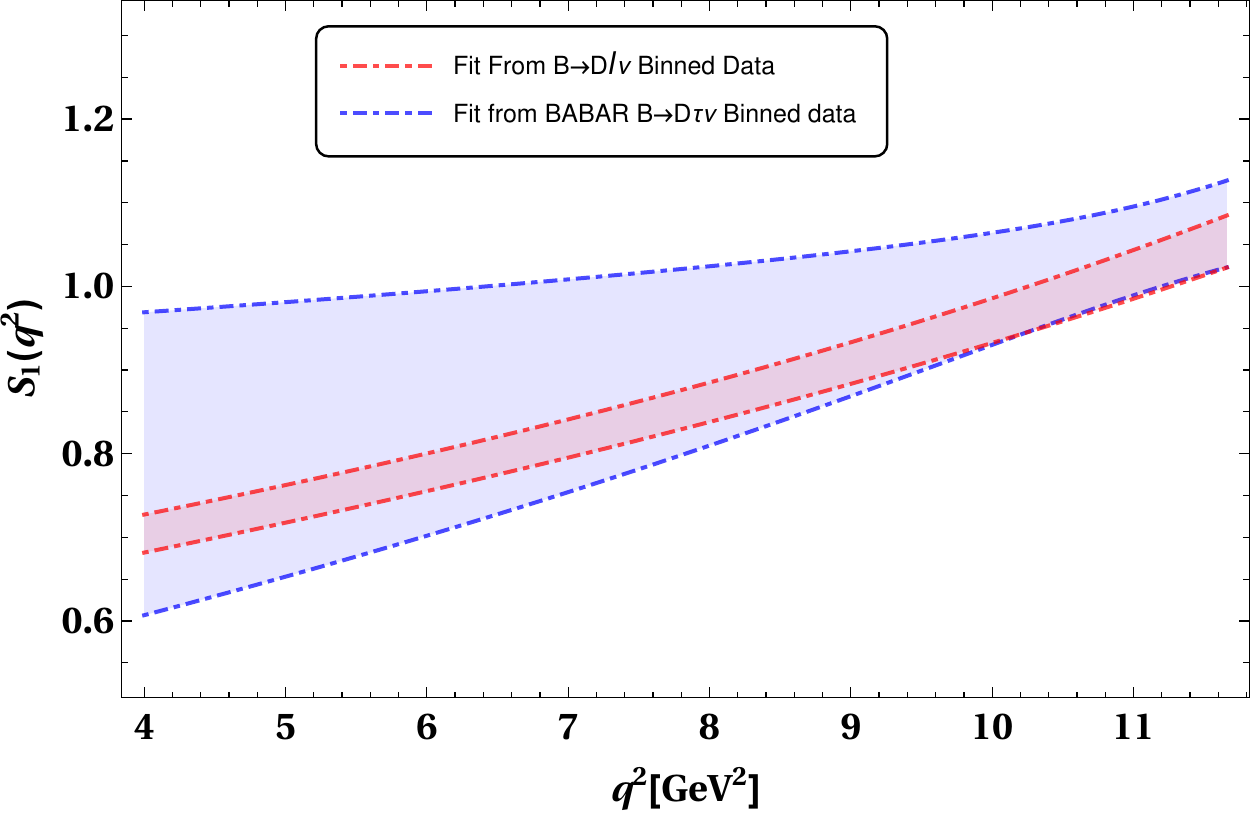}
 \label{fig:S1Dfit2}}\\
\subfloat[$R(D^{*})_{bin}$]{
\includegraphics[scale=0.68]{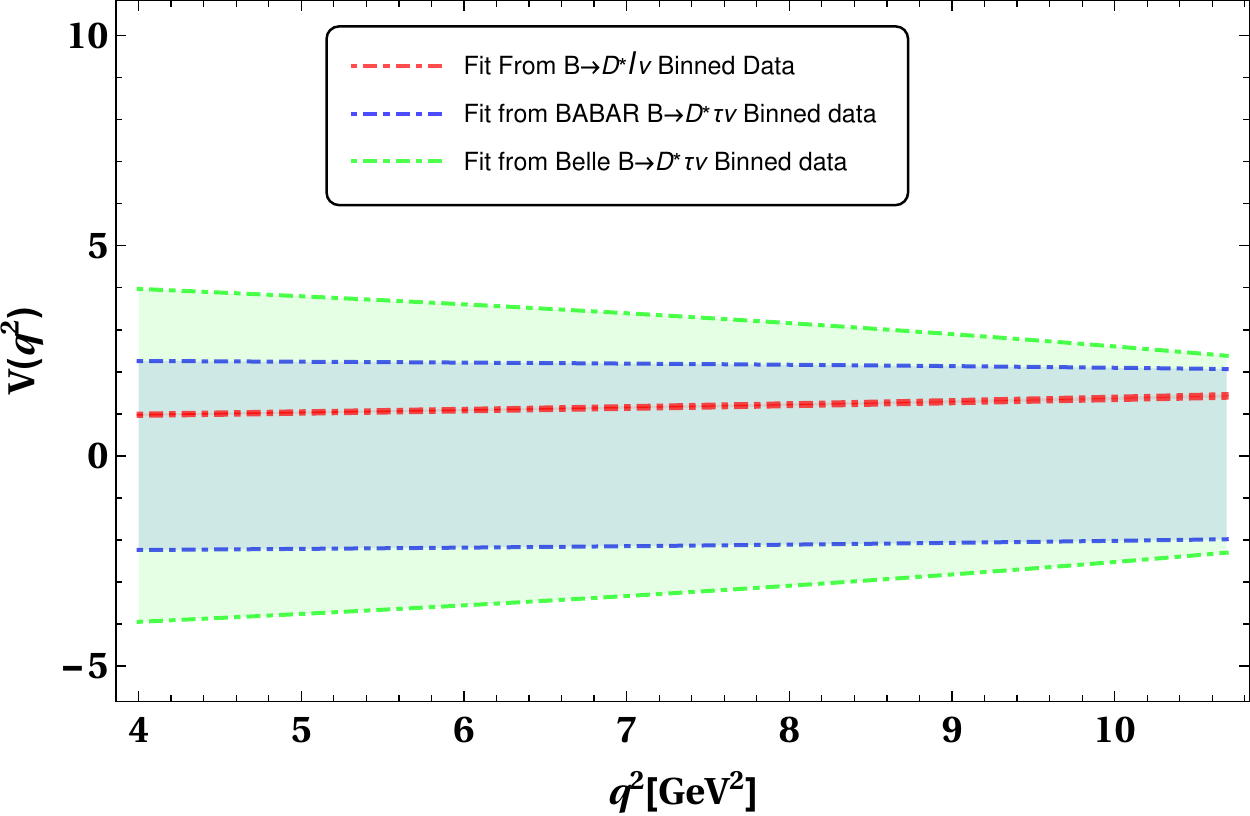}
 \label{fig:VDstfit2}}
\subfloat[$R(D^{*})_{bin}$]{
\includegraphics[scale=0.68]{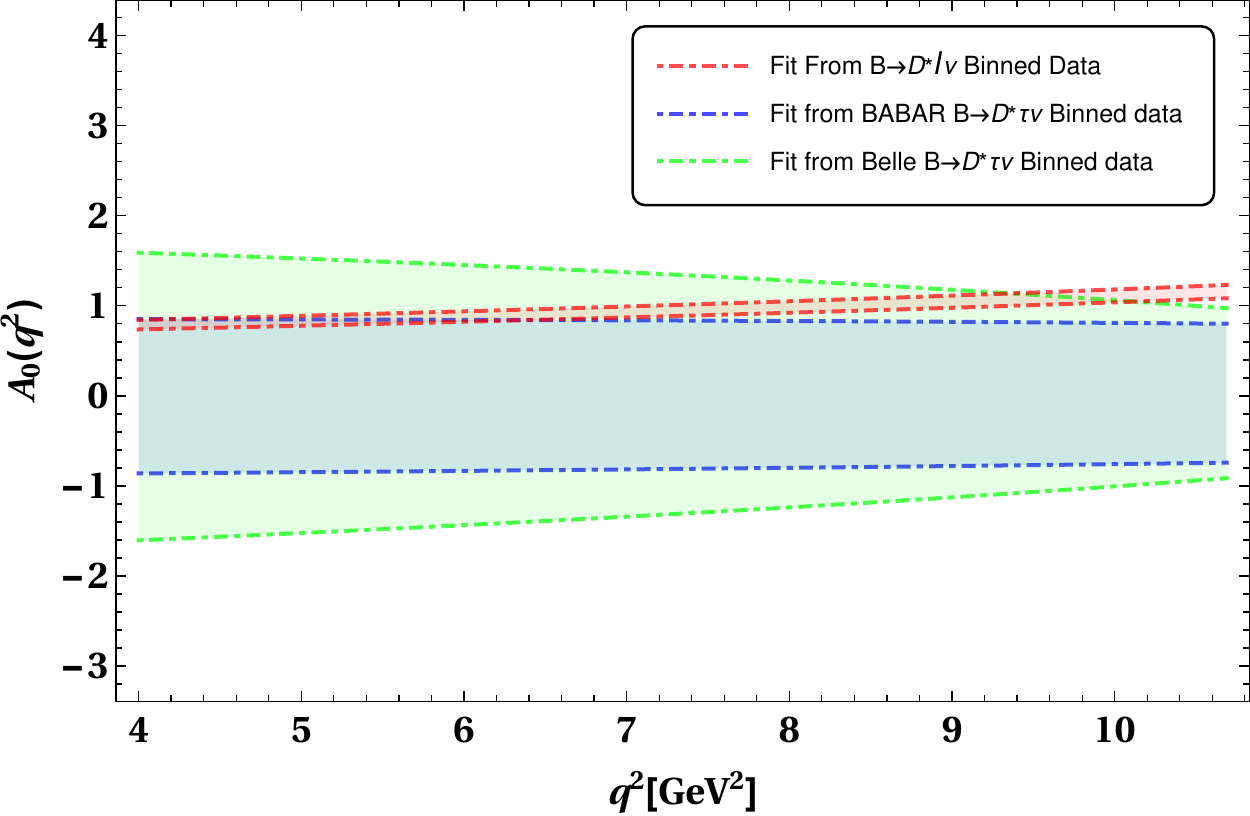}
 \label{fig:A0Dstfit2}}\\
\subfloat[$R(D^{*})_{bin}$]{
\includegraphics[scale=0.68]{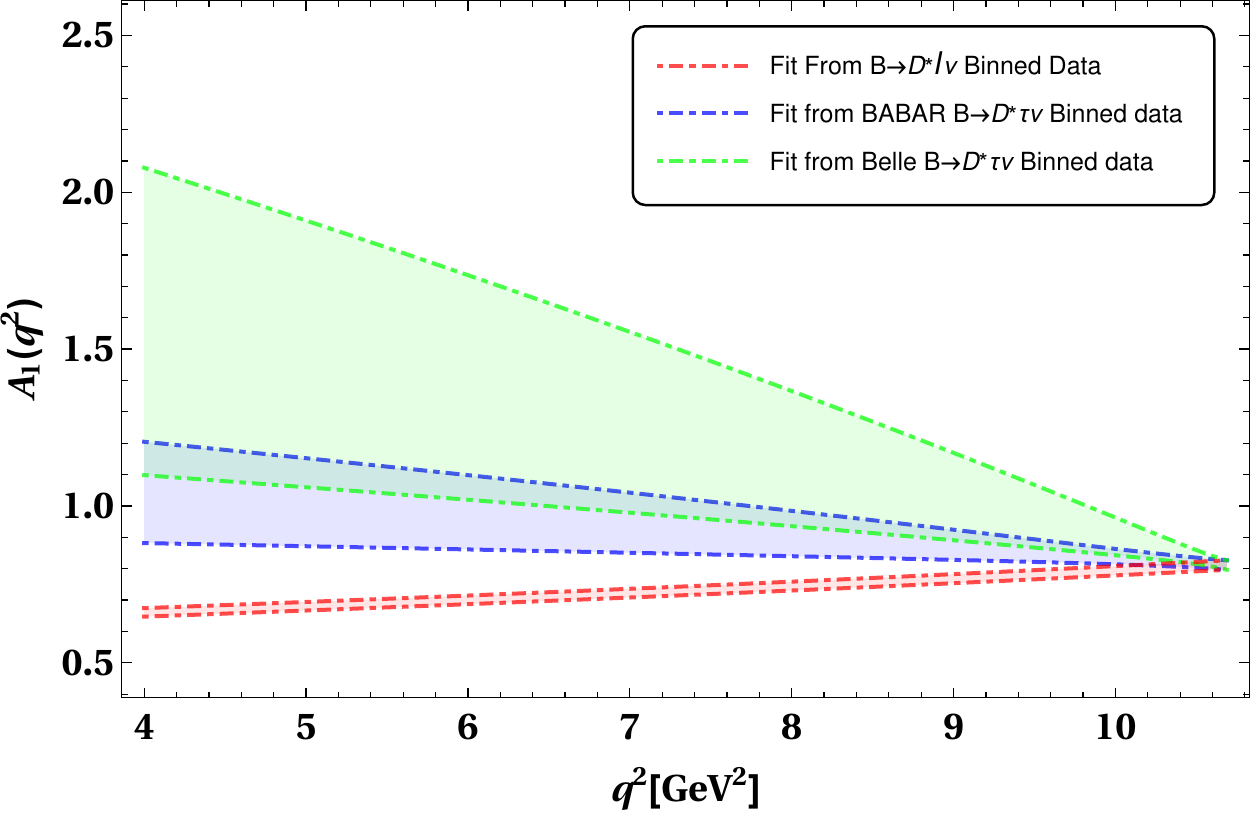}
 \label{fig:A1Dstfit2}}
\subfloat[$R(D^{*})_{bin}$]{
\includegraphics[scale=0.68]{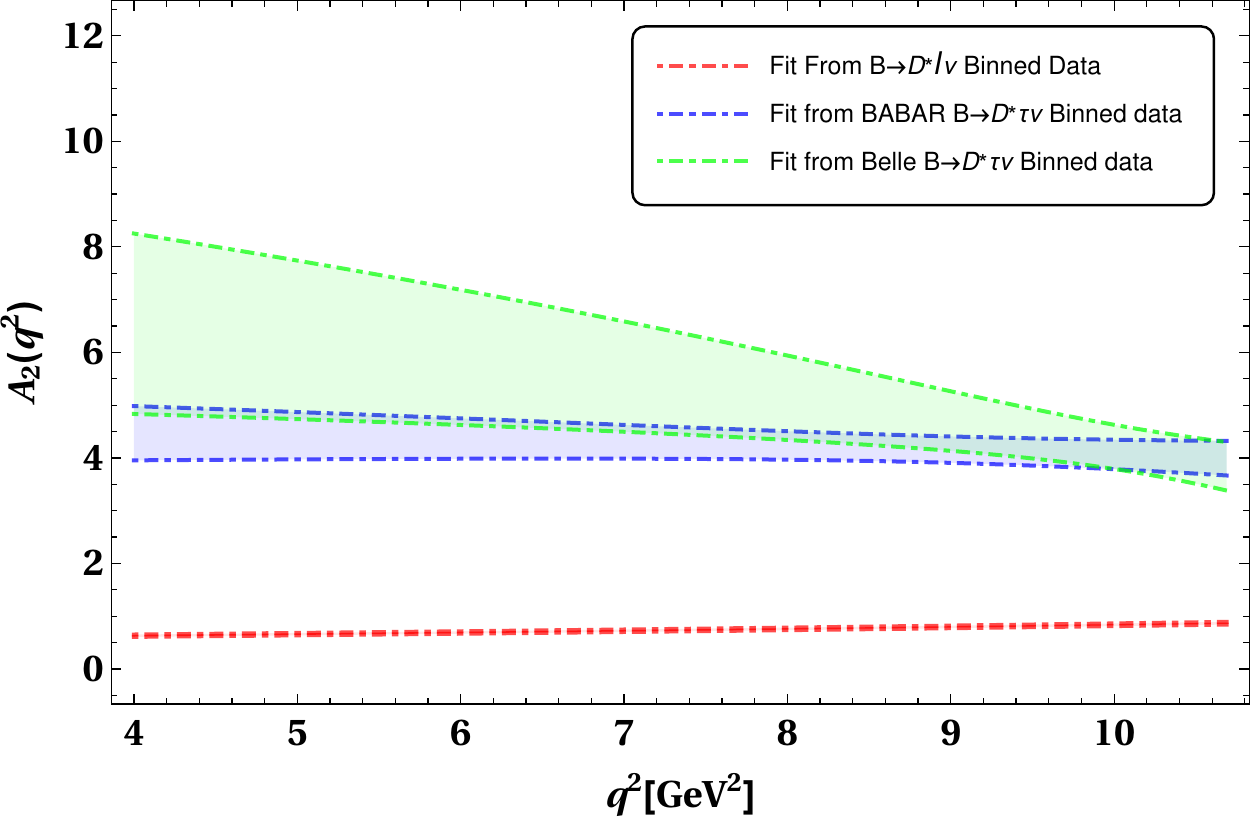}
 \label{fig:A2Dstfit2}}
\caption{Similar figures of $q^2$ dependence of form-factors as figure \ref{fig:form-factors}. These are obtained from `Fit-2'.}
\label{fig:form-factors2}
\end{figure*}

We note that across all the cases listed in tables \ref{tab:latfitres} and \ref{tab:latfitres2}, $R_2(1)$ can be fitted with 
a small error and has large deviations from the value obtained from the analysis of $\bdstell$ (eq. (\ref{dstFFparam4})). 
As the treatment of uncertainties in `Fit-1' and `Fit-2' are vastly different, we can conclude that this large deviation is not dependent on the fitting procedure, rather a consequence of the data-distribution.  
All other parameters are extracted with relatively larger errors and are consistent with the fit results obtained by HFAG 
within $68\%$ or $90\%$ confidence levels (C.L.).

The consequence of these results are reflected in the $q^2$ dependences of the various form-factors, as shown 
in figures \ref{fig:form-factors} and \ref{fig:form-factors2}. In these figures we have compared the $q^2$-distribution of the 
form-factors obtained from 
our fit results with those obtained using the values given in and around eq. (\ref{dstFFparam4}). 
As there is some agreement between the $\rho^2_D$ fitted from $\bd0tau$ and $\bdnstell$, 
the $q^2$-distributions of $V_1(q^2)$ and $S_1(q^2)$, shown 
in figs. \ref{fig:V1D} and \ref{fig:S1D} respectively, do not show any considerable deviation.  
In the analysis of $R(D^*)$, $V(q^2)$ depends on $R_1(1)$ and $\rho^2_{D^*}$, its $q^2$-distribution has large error 
and is consistent with those fitted from $\bdstell$. $A_1(q^2)$ depends on $\rho^2_{D^*}$ and its $q^2$-distribution does not show 
any considerable deviation from that obtained from $\bdstell$ fit. As $q^2$-distributions of both these form-factors obtained from our 
analysis have large errors, at the moment it is hard to conclude anything and we have to wait
for more precise data. On the other hand, among the form-factors associated with $\bdasttau$, $A_2(q^2)$ 
depend on $R_2(1)$ and hence it shows large deviation (in all the $q^2$ regions) from the analysis of $\bdstell$ decay. 
If we assume that the $\bdell$ decays are free from any kind of NP effects, 
which may be a natural assumption, then our results allow the possibility of a new contribution beyond the SM  
in $\bdasttau$ decay. In particular, it could be a beyond-the-SM (BSM) contribution from a pseudo-vector or 
 a pseudo-tensor \footnote{The pseudo-scalar and pseudo-tensor currents are related to the pseudo-vector currents following 
 the equation of motions, 
$i \partial_{\mu} (\bar{c}\gamma^{\mu} \gamma^5 b) = - (m_b + m_c) \bar{c}\gamma^{5} b$  and 
 $\partial_{\mu} (\bar{c}\sigma^{\mu\nu}\gamma^5 b) = (m_b - m_c) {\bar c}\gamma^{\nu}\gamma^{5} b - (i \partial^{\nu} {\bar c})\gamma^{5}b + 
 {\bar c}\gamma^{5}(i\partial^{\nu}b)$, respectively. Hence, the form-factors associated with the pseudo-scalar and 
 pseudo-tensor are related to $A_0(q^2)$ and/or $A_2(q^2)$. Therefore, a large 
 deviation in $A_2(q^2)$ can also be compensated by adding pseudo-tensor 
 current contributions, proportional to these form-factors, in the decay width.} current. On a similar note, 
we can comment that the SM contributions in $\bd0tau$ can explain the observed data.

\section{New Physics Analysis}\label{sec:np}

\subsection{Formalism: Theory}\label{sec:npform}

We follow a model independent approach in the search of the type of NP interactions that can best explain the present data 
on $\bdtau$. The most general effective Hamiltonian describing the 
$b\to c\ell \nu_{\ell}$ transitions (where $\ell = e$, $\mu$ or $\tau$) with all possible four-fermion operators 
in the lowest dimension is given by \cite{Bhattacharya:2015},

\begin{align}
\nn {\cal H}_{eff} &= \frac{4 G_F}{\sqrt{2}} V_{cb} \Big[( \delta_{\ell\tau} + C_{V_1}^{\ell}) {\cal O}_{V_1}^{\ell} + 
 C_{V_2}^{\ell} {\cal O}_{V_2}^{\ell} \\
 &+ C_{S_1}^{\ell} {\cal O}_{S_1}^{\ell} + C_{S_2}^{\ell} {\cal O}_{S_2}^{\ell}
 + C_{T}^{\ell} {\cal O}_{T}^{\ell}\Big],
 \label{eq1}
\end{align}
where the operator basis is defined as 
\bea
{\cal O}_{V_1}^{\ell} &=& ({\bar c}_L \gamma^\mu b_L)({\bar \tau}_L \gamma_\mu \nu_{\ell L}) \nn, \\
{\cal O}_{V_2}^{\ell} &=& ({\bar c}_R \gamma^\mu b_R)({\bar \tau}_L \gamma_\mu \nu_{\ell L}) \nn, \\
{\cal O}_{S_1}^{\ell} &=& ({\bar c}_L  b_R)({\bar \tau}_R \nu_{\ell L}) \nn, \\
{\cal O}_{S_2}^{\ell} &=& ({\bar c}_R b_L)({\bar \tau}_R \nu_{\ell L}) \nn, \\
{\cal O}_{T}^{\ell} &=& ({\bar c}_R \sigma^{\mu\nu} b_L)({\bar \tau}_R \sigma_{\mu\nu} \nu_{\ell L}),
\label{eq2}
\eea
and the corresponding Wilson coefficients are given by $C_W^{\ell}$ ( $W =V_1,V_2,S_1,S_2,T$ ). In this basis, 
neutrinos are assumed to be left handed. 
The complete expressions for the $q^2$-distributions of the differential decay rates $\diff$ in $\bdtau$ decays,
obtained using the effective Hamiltonian in eq.(\ref{eq1}), are given by \cite{Sakai:2013} 

\begin{widetext}
\begin{align}
 \nn &\frac{d\Gamma \left(\overline{B} \rightarrow D \tau \overline{\nu}_{\tau}\right)}{d q^2} = 
 \frac{G^2_F \left|V_{cb}\right|^2}{192 \pi^3 m^3_B} q^2 \sqrt{\lambda_D(q^2)} \left(1 - \frac{m^2_{\tau}}{q^2}\right)^2 
 \left\{ \left|1+ C_{V_1}+ C_{V_2}\right|^2 \left[ \left(1 + \frac{m^2_{\tau}}{2 q^2}\right) H^{s 2}_{V,0} + 
 \frac{3}{2} \frac{m^2_{\tau}}{q^2} H^{s 2}_{V,t}\right] \right.\\
 \nn &~~~\left. +\frac{3}{2} \left|C_{S_1} + C_{S_2}\right|^2 H^{s 2}_S + 8 \left|C_T \right|^2 \left(1 + 
 \frac{2 m^2_{\tau}}{q^2}\right) H^{s 2}_T  +3 \mathcal{R}e\left[\left(1+ C_{V_1}+ C_{V_2}\right) \left(C^*_{S_1} +
 C^*_{S_2}\right)\right] \frac{m_{\tau}}{\sqrt{q^2}} H^s_S H^s_{V,t} \right.\\
 &~~~\left. -12 \mathcal{R}e\left[\left(1+ C_{V_1}+ C_{V_2}\right) C^*_{T}\right] \frac{m_{\tau}}{\sqrt{q^2}} H^s_T H^s_{V,0}
 \right\}\,,
 \label{dgambd}
\end{align}

and

\begin{align}
 \nn &\frac{d\Gamma \left(\overline{B} \rightarrow D^* \tau \overline{\nu}_{\tau}\right)}{d q^2} = \frac{G^2_F 
 \left|V_{cb}\right|^2}{192 \pi^3 m^3_B} q^2 \sqrt{\lambda_D^*(q^2)} \left(1 - \frac{m^2_{\tau}}{q^2}\right)^2 
 \left\{ \left(\left|1 + C_{V_1}\right|^2 + \left|C_{V_2}\right|^2\right) \left[\left(1 + \frac{m^2_{\tau}}{2 q^2}\right)
 \left(H^2_{V,+} + H^2_{V,-} + H^2_{V,0}\right) \right.\right.\\
 \nn &~~~ \left.\left.+ \frac{3}{2} \frac{m^2_{\tau}}{q^2} H^{2}_{V,t}\right] - 2 \mathcal{R}e \left[\left(1+ C_{V_1}\right)
 C^*_{V_2}\right] \left[\left(1 + \frac{m^2_{\tau}}{2 q^2}\right) \left(H^2_{V,0} + 2 H_{V,+} H_{V,-} \right) + \frac{3}{2}
 \frac{m^2_{\tau}}{q^2} H^{2}_{V,t}\right] + \frac{3}{2} \left|C_{S_1} - C_{S_2}\right|^2 H^2_S \right. \\
 \nn &~~~ \left. + 8 \left|C_T\right|^2 \left(1 + \frac{2 m^2_{\tau}}{q^2}\right) \left(H^2_{T,+} + H^2_{T,-} + 
 H^2_{T,0}\right) + 3 \mathcal{R}e\left[ \left(1 + C_{V_1} - C_{V_2}\right) \left(C^*_{S_1} - C^*_{S_2}\right)\right] 
 \frac{m_{\tau}}{\sqrt{q^2}} H_S H_{V,t} \right.\\
 \nn & \left. -12 \mathcal{R}e\left[\left(1 + C_{V_1}\right) C^*_{T}\right] \frac{m_{\tau}}{\sqrt{q^2}}
 \left(H_{T,0} H_{V,0} + H_{T,+} H_{V,+}  - H_{T,-} H_{V,-}\right) \right.\\
 &~~~\left. + 12 \mathcal{R}e\left[C_{V_2} C^*_{T}\right] \frac{m_{\tau}}{\sqrt{q^2}} \left(H_{T,0} H_{V,0} + 
 H_{T,+} H_{V,-} - H_{T,-} H_{V,+}\right) \right\}\,.
 \label{dgambdst}
\end{align}

\end{widetext}

The $q^2$-distribution of the decay rate of the decays $\bdell$ are obtained from equations (\ref{dgambd}) and (\ref{dgambdst})
by setting $C_W = 0$ and $m_{\tau} = 0$. we define our observables as given in equations (\ref{Rth}) and  (\ref{RexpBaBe}).

\subsection{Methodology}\label{sec:npmethod}

We know that the yield in each bin depends on the probability density functions (PDFs) of different 
(56 in case of \Babar) signal and background sources. Considering any NP contribution changes these PDFs and they 
in turn change the two dimensional $m^2_{miss} - |\mathbf{p}^*_l|$ distributions. This change is reflected in 
the $q^2$-distribution as well, because of the following relation: $m^2_{miss} = (q - p_l)^2$. A complete and simultaneous 
fit to all PDFs can only be done for each specific NP model separately and the dependence of the shape and normalization 
of the PDFs on the NP parameters should be extracted rigorously using raw experimental data. Without the aid of 
simulation, we do not attempt to do such an analysis. Instead, we use the background subtracted and normalized 
binned data for $q^2$ and $p_{D^*}$-distributions as depicted in Fig.s \ref{fig:BABAR_dataD}, 
\ref{fig:BABAR_dataDst} and \ref{fig:Belle_dataDst} to perform a phenomenological analysis in a 
model independent way. Such an assumption can become a source of systematic errors in our analysis and the way we have 
dealt with that is discussed in section \ref{sec:npnumanalys}.

In addition to the binned data from \Babar~and Belle, we also have the total $R(D^{(*)})$ data from various 
experiments (see table \ref{tab:expinput}). Keeping in mind that the binned data is going to 
dominate the fit results, we take different combinations of these separate data points and do the whole 
analysis separately for them. 

At the beginning of our analysis, we have defined the most general scenario with contributions from all possible dimension 6 effective 
operators present simultaneously (with 10 parameters i.e. real and imaginary parts of all $C^l_W$s) as the global 
scenario. We have defined various sub-scenarios as different possible combinations of those operators. Including the global scenario, 
there are in total 31 such scenarios, which we are going to call ``cases'' from here onwards.

One of the main motivations of this paper is to do a multi-scenario analysis on the experimentally available binned 
data, to obtain a data-based selection of a `best' case and ranking and weighting of the remaining cases in 
the predefined set of 31. To that goal, we have made use of information-theoretic approaches, especially of 
AIC$_c$ in the analysis of empirical data. Such procedures lead to 
more robust inferences in simultaneous comparative analysis of multiple competing scenarios. 
Traditional statistical inference(e.g. confidence levels, errors on fit parameters, bias etc.) can then be obtained 
based on the selected best models.\footnote{One of the most powerful and most reliable methods for model comparison
(also computationally expensive) is 'cross-validation' \cite{dosdonts}. The most straightforward (and also most expensive) 
flavor of cross-validation is “leave-one-out cross-validation” (LOOCV). It simultaneously tests the predictive power of 
the model as well minimizes the bias and variance together. In LOOCV, one of the data points is left out and the rest of 
the sample (``training set'') is optimized. Then that result is used to find the predicted residual for the left out data point.
This process is repeated for all data points and a mean-squared-error (MSE) is obtained. For model selection, this 
MSE is minimized. It has been shown that this method is asymptotically equivalent to minimizing AIC \cite{shibata}}.

\subsubsection{A Short Introduction to AIC$_c$}

\begin{table} 
 \begin{center}
  \begin{tabular}{|c|c|}
    \hline
    $\Delta^{AIC}_i$ & Level of Empirical Support for Model $i$\\
    \hline
    $0 - 2$ & Substantial \\
    $4 - 7$ & Considerably Less \\
    $> 10$  & Essentially None \\
    \hline
  \end{tabular}
 \end{center}
 \caption{Rough rule-of-thumb values of $\Delta^{AIC}_i$ for analysis of nested models.}
 \label{tab:delAICrule}
\end{table}

\begin{table}
 \begin{center}
  \begin{tabular}{|c|c|}
    \hline
    Input & Value\\
    \hline
    $\Delta$ & $1 \pm 1$ \cite{Tanaka:2010}\\
    $\rho^2_D$ & $1.186 \pm 0.054$ \cite{hfag}\\
    $\rho^2_{D^*}$ & $1.207 \pm 0.026$ \cite{hfag}\\
    $R_1(1)$ & $1.406 \pm 0.033$ \cite{hfag}\\
    $R_2(1)$ & $0.853 \pm 0.020$ \cite{hfag}\\
    $R_0(1)$ & $1.14 \pm 0.07$ \cite{Fajfer:2012vx}\\
    $V_1(1)$ & $1.053 \pm 0.008$ \cite{Lattice:2015}\\
    $h_{A_1}(1)$ & $0.906 \pm 0.013$ \cite{bailey14}\\
    $m_{B_0}$ & $5.27958 \pm 0.00015 \pm 0.00028$ \cite{Aaij:2011}\\
    $m_{D_0}$ & $1.86484 \pm 0.00005$ \cite{Agashe:2014}\\
    $m_{b}$ & $4.18 \pm 0.03$ \cite{Agashe:2014}\\
    $m_{c}$ & $1.275 \pm 0.025$ \cite{Agashe:2014}\\
    $m_{\tau}$& $1.77682 \pm 0.00012$ \cite{Agashe:2014}\\
    \hline
  \end{tabular}
 \end{center}
 \caption{Inputs used in the fitting of new Wilson coefficients. All Masses are in GeV. Correlations between a few 
 form-factor parameters are listed in eq. (\ref{dstFFparam4}).}
 \label{tab:thinput}
\end{table}

\begin{figure*}[htbp]
\centering
%\subfloat[$R(D)_{bin}$]{
\includegraphics[scale=0.5]{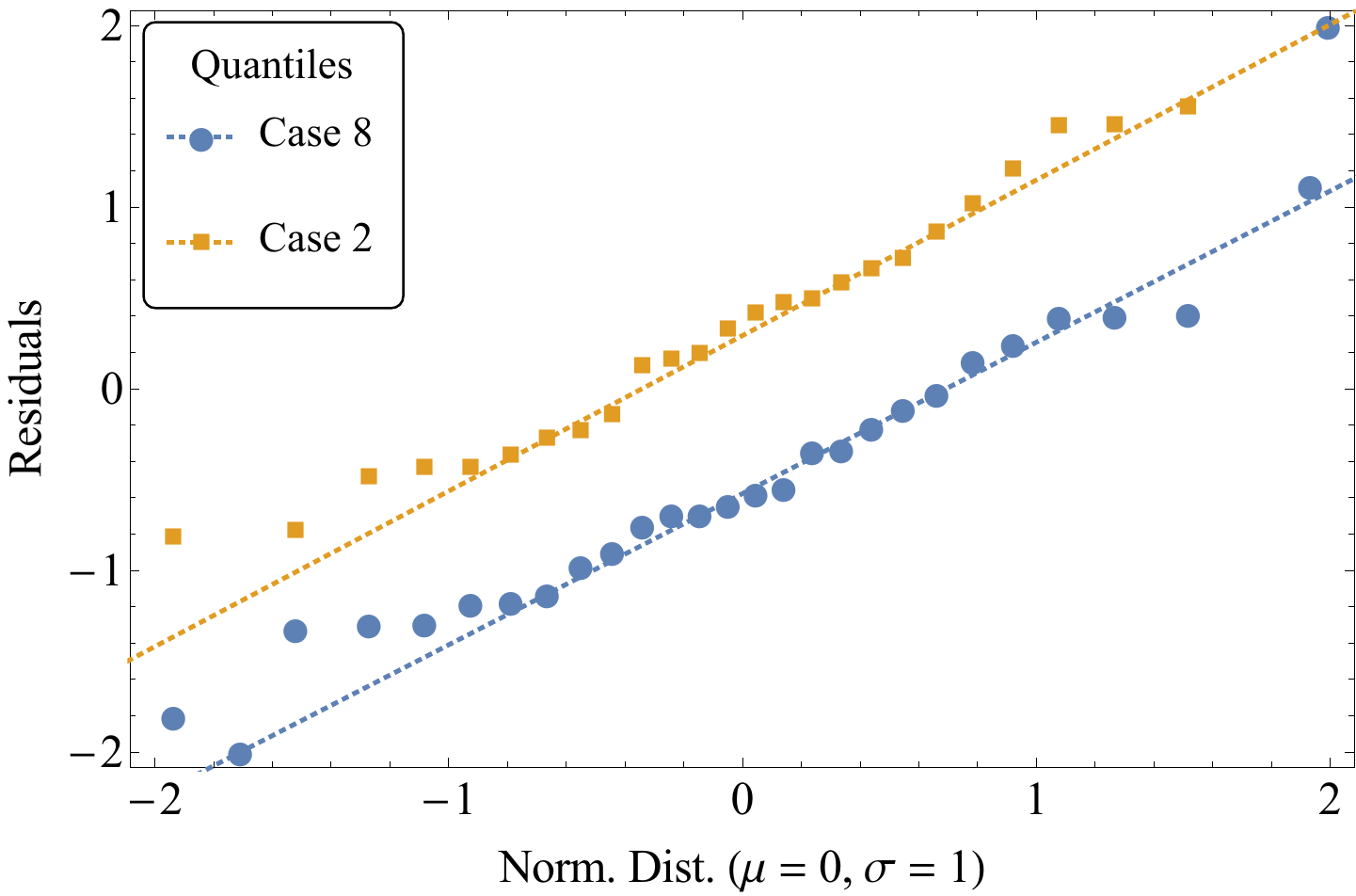}
% \label{fig:V1D}}
%\subfloat[$R(D)_{bin}$]{
\includegraphics[scale=0.5]{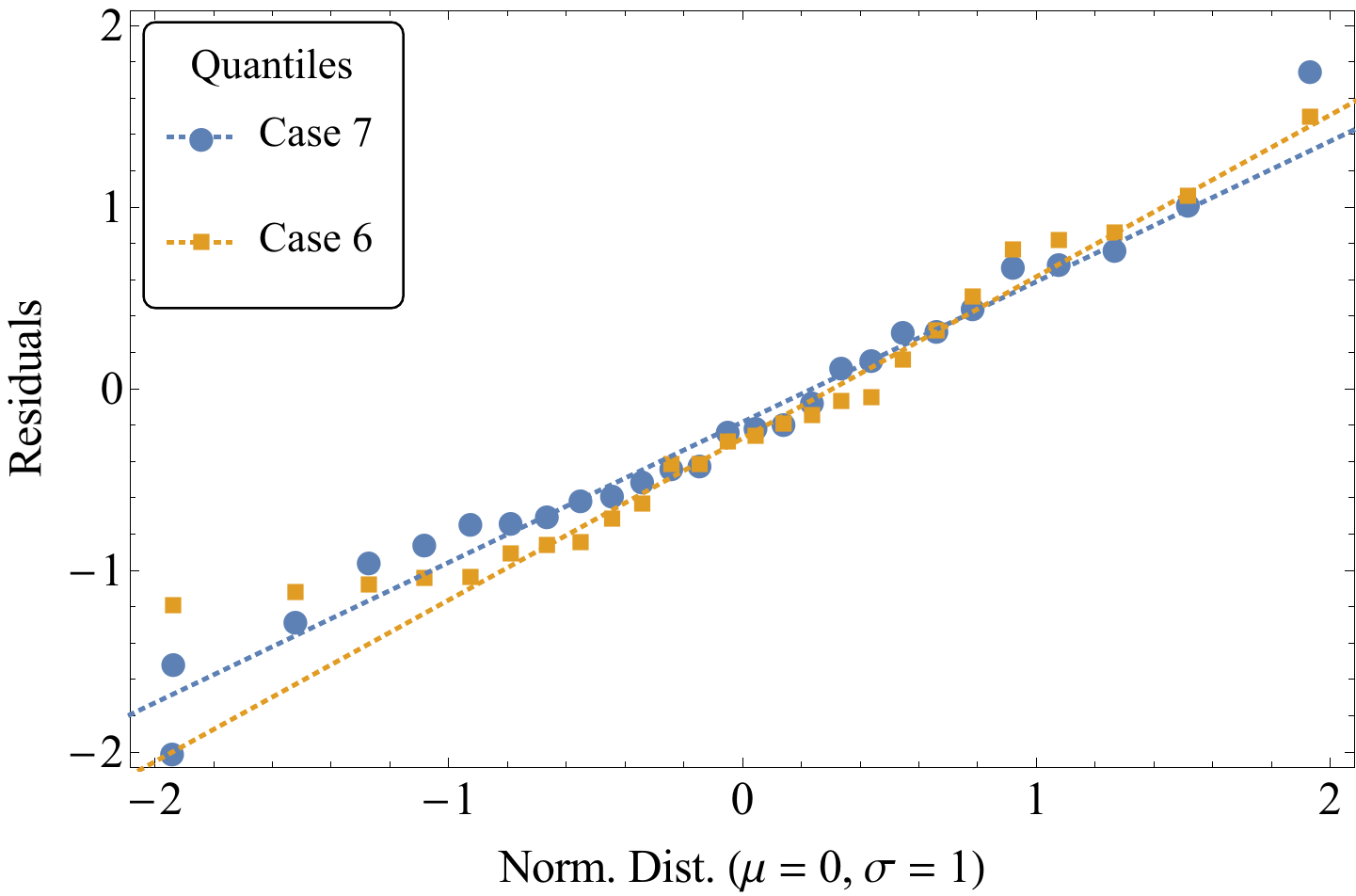}
% \label{fig:S1D}}
%\subfloat[$R(D^{*})_{bin}$]
% \includegraphics[scale=0.22]{hypoplt3.png}
 \caption{Q-Q Plot of the residuals of the best fits. Each plot compares the quantiles of the distribution of 
 the residuals with a Gaussian with $\mu = 0$ and $\sigma^2 = 1$. The closer the distribution o the points are to the corresponding
 dotted lines, the better they fit to the Gaussian. Here we show the best NP cases for data-set `3' 
 from table \ref{tab:Result1}}
\label{fig:hypotestnp}
\end{figure*}

\begin{table*}
 \begin{center}
  {\renewcommand{\arraystretch}{1.1}%
  \begin{tabular}{|c|c|c|c|c|c|c|c|c|c|}
    \hline
    ~Experiment~ & ~Dataset~ & ~Observables~ & ~Cases~ & $~\chi^2_{min}~$ & ~$d.o.f$~ & ~Parameters~ & ~Akaike Wgt.s~ & ~Normality~ & ~$\chi^2$ (SM)~\\
%     \cline{9-11}
     & Index. & & & & & & ($w_i$) & (S-W) &\\
    \hline
$\text{}$ & $\text{}$ & $\text{}$ & $5$ & $7.41$ & $12$ & $C_T$ & $0.26$ & $0.38$ & \\
$\text{}$ & $\text{}$ & $\text{}$ & $1$ & $7.79$ & $12$ & $C_{V_1}$ & $0.22$ & $0.29$ & \\
$\text{}$ & $\text{1}$ & $\text{$R(D)_{bin}$}$ & $2$ & $7.79$ & $12$ & $C_{V_2}$ & $0.22$ & $0.29$ & 10.31\\
$\text{}$ & $\text{}$ & $\text{}$ & $3$ & $9.17$ & $12$ & $C_{S_1}$ & $0.11$ & $0.18$ & \\
$\text{}$ & $\text{}$ & $\text{}$ & $4$ & $9.17$ & $12$ & $C_{S_2}$ & $0.11$ & $0.18$ & \\
\cline{2-10}
$\text{}$ & $\text{}$ & $\text{}$ & $1$ & $6.3$ & $10$ & $C_{V_1}$ & $0.56$ & $0.11$ & \\
$\text{\Babar}$ & $\text{2}$ & $\text{$R(D^*)_{bin}$}$ & $2$ & $7.18$ & $10$ & $C_{V_2}$ & $0.36$ & $0.12$ & 79.85\\
\cline{2-10}
$\text{}$ & $\text{}$ & $\text{}$ & $8$ & $13.01$ & $22$ & $C_{V_2}$, $C_{S_2}$ & $0.32$ & $0.86$ & \\
$\text{}$ & $\text{3}$ & $\text{Combined}$ & $2$ & $19.12$ & $24$ & $C_{V_2}$ & $0.22$ & $0.59$ & 90.16\\
$\text{}$ & $\text{}$ & $\text{}$ & $7$ & $14.23$ & $22$ & $C_{V_2}$, $C_{S_1}$ & $0.17$ & $0.79$ & \\
$\text{}$ & $\text{}$ & $\text{}$ & $6$ & $14.61$ & $22$ & $C_{V_1}$, $C_{V_2}$ & $0.14$ & $0.19$ & \\
\hline
$\text{}$ & $\text{}$ & $\text{}$ & $2$ & $9.07$ & $15$  & $C_{V_2}$ & $0.47$ & $0.95$ & \\
$\text{Belle(2016)}$ & $\text{4}$ & $\text{$R(D^*)_{bin}$}$ & $1$ & $9.43$ & $15$ & $C_{V_1}$ & $0.39$ & $1.00$ & 26.20\\
\hline
$\text{}$ & $\text{}$ & $\text{}$ & $8$ & $22.59$ & $39$ & $C_{V_2}$, $C_{S_2}$ & $0.31$ & $0.95$ & \\
$\text{\Babar+}$ & $\text{}$ & $\text{}$ & $2$ & $28.33$ & $41$ & $C_{V_2}$ & $0.19$ & $0.94$ & \\
$\text{Belle(2016)}$ & $\text{5}$ & $\text{Combined}$ & $7$ & $23.69$ & $39$ & $C_{V_2}$, $C_{S_1}$ & $0.18$ & $0.89$ & 116.36\\
$\text{}$ & $\text{}$ & $\text{}$ & $6$ & $24.16$ & $39$ & $C_{V_1}$, $C_{V_2}$ & $0.14$ & $0.69$ & \\
\hline
$\text{\Babar+}$ & $\text{}$ & $\text{}$ & $2$ & $48.54$ & $28$ & $C_{V_2}$ & $0.52$ & $0.02$ & \\
$\text{Belle(2015)+}$ & $\text{}$ & $\text{}$ & $8$ & $45.71$ & $26$ & $C_{V_2}$, $C_{S_2}$ & $0.16$ & $0.01$ & \\
$\text{LHCb+}$ & $\text{6}$ & $\text{Combined}$ & $7$ & $46.87$ & $26$ & $C_{V_2}$, $C_{S_1}$ & $0.09$ & $0.02$ & 96.68\\
$\text{Belle(Latest)}$ & $\text{}$ & $\text{}$ & $6$ & $47.24$ & $26$ & $C_{V_1}$, $C_{V_2}$ & $0.08$ & $0.04$ & \\
\hline
$\text{Belle(2016)+}$ & $\text{}$ & $\text{}$ & $2$ & $28.81$ & $19$ & $C_{V_2}$ & $0.34$ & $0.64$ & \\
$\text{Belle(2015)+}$ & $\text{}$ & $\text{}$ & $1$ & $30.81$ & $19$ & $C_{V_1}$ & $0.13$ & $0.77$ & \\
$\text{LHCb+}$ & $\text{7}$ & $\text{Combined}$ & $4$ & $31.29$ & $19$ & $C_{S_2}$ & $0.1$ & $0.83$ & 32.72\\
$\text{Belle New}$ & $\text{}$ & $\text{}$ & $3$ & $31.48$ & $19$ & $C_{S_1}$ & $0.09$ & $0.91$ & \\
$\text{}$ & $\text{}$ & $\text{}$ & $5$ & $31.52$ & $19$ & $C_{T}$ & $0.09$ & $0.82$ & \\
\hline
  \end{tabular}}
 \end{center}
 \caption{The best selected scearios for ``Fit-1'' (section \ref{sec:resultfit1}). 
 The cases listed in order in the fourth column  for each dataset have passed through the selection criteria 
 $0 \le\Delta^{AIC}_i \le 4$, where $\Delta^{AIC}_1 = 0$ in each dataset. 
 Note that the case-index values represent a specific set of parameters and each parameter listed here is considered to be complex, so 
 the number of parameters is actually double. $w_i$ in the eighth column is defined in eq.(\ref{omegai}). 
 The next column lists the results of the S-W normality test 
 for the assessment of goodness-of-fit. The last column lists the $\chi^2$ value corresponding to the SM for each dataset. Note that AIC$_c$ value for SM is same as the $\chi^2$ as no. of fit parameters $K=0$ for SM.}
 \label{tab:Result1}
\end{table*}
% \vspace{5mm}

The `concept of parsimony' \cite{boxjenkins} dictates that a model representing the truth should be obtained with 
``... the smallest possible number of parameters for adequate representation of the data.'' %Statisticians view the 
%principle of parsimony as a bias versus variance trade-off. 
In general, bias decreases and variance increases as the 
dimension of the model increases. Often, the number of parameters in a model is used as a measure of the degree of 
structure inferred from the data. The fit of any model can be improved by increasing the number of parameters.
 Parsimonious models achieve a proper trade-off between bias and variance. All model selection methods are based to 
 some extent on the principle of parsimony \cite{breiman}.

In information theory, the Kullback-Leibler (K-L) Information or measure $I(f,g)$ denotes the information lost when 
$g$ is used to approximate $f$. Here $f$ is a notation for full reality or truth and $g$ denotes an approximating 
model in terms of probability distribution. $I(f,g)$ can also be defined between the `best' approximating model 
and a competing one. Akaike, in his seminal paper \cite{akaike73} proposed the use of the K-L information as a fundamental 
basis for model selection. However, K-L distance cannot be computed without full knowledge of both $f$ (full reality) 
and the parameters ($\Theta$) in each of the candidate models $g_i(x|\Theta)$ (a model $g_i$ with parameter-set $\Theta$ 
explaining data $x$). Akaike found a rigorous way to estimate 
K-L information, based on the empirical log-likelihood function at its maximum point.

`Akaike's information criterion'(AIC) with respect to our analysis can be defined as,
\be
 {\rm AIC} = \chi^2_{min} + 2 K\,
\label{aic}
\ee
where $K$ is the number of estimable parameters. In application, one computes AIC for each of the candidate models and 
selects the model with the smallest value of AIC. It is this model that is estimated to be ``closest'' to the unknown 
reality that generated the data, from among the candidate models considered.

While Akaike derived an estimator of K-L information, AIC may perform poorly if there are too many parameters in relation 
to the size of the sample. Sugiura\cite{sugiura78} derived a second-order variant of AIC,
\be
 {\rm AIC}_c = \chi^2_{min} + 2 K + \frac{2 K (K+1)}{n - K -1}\,
\label{aicc}
 \ee
where $n$ is the sample size. As a rule of thumb, Use of AIC$_c$ is preferred in literature when $n/K < 40$. There are 
various other such information criteria defined later on, e.g. QAIC, QAIC$_c$, TIC etc. In this analysis, we consistently 
use AIC$_c$.

Whereas AIC$_c$ are all on a relative (or interval) scale and are strongly dependent on sample size, simple differences of 
AIC$_c$ values ($\Delta^{AIC}_i = {\rm AIC}^i_c - {\rm AIC}^{min}_c$) allow estimates of the relative expected K-L 
differences between $f$ and $g_i(x|\Theta)$. This allows a quick comparison and ranking of candidate models. 
The model estimated to be best has $\Delta^{AIC}_i \equiv \Delta^{AIC}_{min} = 0$. The larger $\Delta^{AIC}_i$ is, 
the less plausible it is that the fitted model $g_i(x|\Theta)$ is the
K-L best model, given the data $x$. Table \ref{tab:delAICrule} lists rough rule-of-thumb values of $\Delta^{AIC}_i$ 
for analysis of nested models.

While the $\Delta^{AIC}_i$ are useful in ranking the models, it is possible to quantify the plausibility of each model as 
being the actual K-L best model. This can be done by extending the concept of the likelihood of the parameters given both 
the data and model, i.e. $\mathcal{L}(\Theta|x, g_i)$, to the concept of the likelihood of the model given the data, 
hence $\mathcal{L}(g_i|x)$;
\be
 \mathcal{L}(g_i|x) \propto e^{(-\Delta^{AIC}_i / 2)}\,.
\ee
Such likelihoods represent the relative strength of evidence for each model \cite{akaike83a}. 

To better interpret the relative likelihood of a model, given the data and the set of $R$ models, we normalize the 
$\mathcal{L}(g_i|x)$ to be a set of positive “Akaike weights,” $w_i$ , adding up to $1$:
\be\label{omegai}
 w_i = \frac{e^{(-\Delta^{AIC}_i / 2)}}{\sum_{r = 1}^R e^{(-\Delta^{AIC}_r / 2)}}
\ee
A given $w_i$ is considered as the weight of evidence in favor of model $i$ being the actual K-L best model for the 
situation at hand, given that one of the $R$ models must be the K-L best model of that set. 
The $w_i$ depend on the entire set; therefore, if a model is added or dropped during a post hoc analysis, 
the $w_i$ must be recomputed for all the models in the newly defined set.

\subsubsection{Numerical Multi-parameter Optimization}\label{sec:npnumanalys}

To compare the latest \Babar~and Belle binned data with a specific model, we devise a $\chi^2$ defined as:
\be
 \chi^2_{NP} = \sum^{n_b}_{i,j = 1} \left(R^{exp}_i - R^{th}_i\right) \left(V^{exp} + V^{th}\right)^{-1}_{i j} 
 \left(R^{exp}_j - R^{th}_j\right)\,,
\ee
where $R^{th}_{bin}$ and $R^{exp}_{bin}$ are defined in eq.s (\ref{Rth}) and (\ref{RexpBaBe}). $i$ and 
$j$ vary over the number of bins ($n_b$) taken into account in the analysis. For the calculation of $R^{th}_{bin}$, 
central values of HQET hadronic form-factors and the quark masses are used (listed in table \ref{tab:thinput}).
The standard bin-width for the \Babar~analysis is $ 0.5 (\text{GeV}^2 / c^4)$ and due to this the last bin exceeds 
the allowed phase space($q^2_{max} = \left(m_B - m_{D^{(*)}}\right)^2$) in both channels. Instead of changing the 
bin width for those last bins, we drop these bins from our analysis. We follow this same philosophy for Belle bins too.
$V^{th}_{i j}$ and $V^{exp}_{i j}$ are the theoretical and experimental covariance matrices respectively. 
For the analysis of any specific NP model, the uncertainties of the HQET hadronic form-factors and the 
quark masses(table \ref{tab:thinput}) are taken into account in the calculation of $V^{th}_{i j}$.

To calculate the errors $\delta R^{exp}_{bin}$, we use eq.(\ref{RexpBaBe}) according to the 
case and propagate the errors listed in table \ref{tab:expinput}. Following the reasoning stated in section \ref{latparfit}, 
we break the NP analysis in two parts: `Fit-1' and `Fit-2'. In addition to $V^{th}_{ij}$, here we treat the $V^{exp}_{ij}$ exactly 
as the $V_{ij}$ in section \ref{latparfit}. 

We define the $\chi^2$ statistic for each of the 31 cases, a function of the NP Wilson coefficients.
The definition and usage of the observables closely follow the fitting process in section \ref{latparfit}. 
Here, we take the existing world-averages of the parameters of the form-factors \cite{hfag}.
If we include all the NP interactions, we have total 10 unknown NP parameters and 26 observables 
for \Babar~(14 bins for $B \rightarrow D \tau \nu$ and 12 bins for $B \rightarrow D^* \tau \nu$) and 17 observables for 
Belle. We then minimize the $\chi^2$ for different cases and different set of observables. 
Though we have varied the process for various global optimization methods to optimize the minimization, 
due to the presence of large uncertainties, this is not important for the present analysis.
To glean any information of goodness-of-fit from $\chi^2_{min}$, we need to know the degrees of freedom 
($d.o.f = N_{Obs} - N_{Params}$). A reduced statistic $\chi^2_{red} = \chi^2_{min} / d.o.f$ can thus be defined.

In many cases in our optimization problem, the minimum is not an isolated single point, rather a contour 
in the parametric dimensions. For these cases, Hessian in not positive definite and the errors thus obtained 
are meaningless. In those cases, the $1 \sigma$ uncertainties have to found from the contours in the parameter 
space and we have done that for all cases with $2-3$ parameters. As contours are impossible to draw when number of 
parameters $> 3$, we have devised a numerical method to obtain the range of a parameter. 
In this method, we sequentially minimize or maximize each parameter by scanning along the enclosing 
$1 \sigma$ $\chi^2_{NP}$  hyper-contour-surface (the method can be extended to any number of $n \sigma$ contours). 
These values give us the range of each parameter while taking their correlation into account all along. 
These errors, for obvious reasons, are asymmetric. We have also systematically found these uncertainties 
for all cases. We will in general quote them in our results.

In our present analysis, after optimizing the $\chi^2_{NP}$ for all $31$ cases, we make use of $\Delta^{AIC}_i$ and 
$w_i$ to find the `best' set of cases, which are more favorable compared to others, and do further analysis on them.
After selecting a class of models describing the data with optimum bias and variance 
with AIC$_c$, we check the significance of them to find most suited model to describe the data.

\subsubsection{Note on Model Selection Criteria}\label{sec:result3}

Unlike the AIC$_c$ or the Schwarz-Bayesian Criterion (BIC) \cite{bic}, which incorporate the concept of 
parsimony and can be applied to nested as well as non-nested models, Likelihood-Ratio test - more commonly known as 
$\Delta \chi^2$ test, can only be applied to nested models. When the model with the fewer free parameters (null, 
in this case) is true, and when certain conditions are satisfied, Wilks' Theorem \cite{Wilks:1938dza} says that this 
difference ($\Delta\chi^2$) should have a 
$\chi^2$ distribution with the number of degrees of freedom equal to the difference in the number of free parameters in the 
two models. This lets one compute a $p$-value and then compare it to a critical value to decide whether to reject the null 
model in favor of the alternative model.

\begin{table}[!hbt] 
 \begin{center}
  \begin{tabular}{|c|c|c|c|}
    \hline
    Case Index. & Params & $\chi^2_{min}$ & $d.o.f$\\
    \hline
    $A$ & $C_{V_2}$ & $27.35$ & $41$ \\
    $B$ & $C_{V_2}$, $C_{S_2}$ & $19.11$ & $39$ \\
    $C$ & $C_{V_2}$, $C_{S_1}$, $C_{S_2}$ & $18.68$ & $37$ \\
    $D$ & $C_{V_2}$, $C_{S_1}$, $C_{S_2}$, $C_{T}$ & $18.39$ & $35$\\
    \hline
  \end{tabular}
 \end{center}
 \caption{The cases with lowest $\chi^2_{min}$ values for different sets according to number of parameters for dataset-5 of
 `Fit-2' (e.g. scenario $A$ is the case with lowest $\chi^2$ among all the 2 parameter cases analyzed in `Fit-2', 
 $B$ is the best among all 4 parameter cases and so on.)}
 \label{tab:modselnote1}
\end{table}

For a demonstration of this method, as an example, we have taken dataset-5 from `Fit-2' (table \ref{tab:Result1}) as our 
experimental input and we have separated all the cases in different sets according to their number of parameters. This means 
that in this method, all the cases in such a set have same number of parameters and the best among them has the lowest 
$\chi^2$ at its best-fit point. Only the best cases with their $\chi^2$ values and $d.o.f$s are listed in table 
\ref{tab:modselnote1}. Whereas the AIC$c$ analysis picked up a group of best possible scenarios, here 
we have used all the cases for comparison. Then in table \ref{tab:modselnote2}, we have compared different combinations of 
these best cases from table \ref{tab:modselnote2} to do a $\Delta\chi^2$ test. From table \ref{tab:modselnote2} it can be seen 
that case $A$ (i.e. with only $C_{V_2}$) is disfavored in comparison to $B$ and $C$ (though it cannot be be discarded at 
a significance of $5\%$ in comparison with $D$, the $p$-value obtained is pretty small), whereas $B$ is favored with very 
high $p$-values when compared to cases with larger number of parameters. This analysis thus picks out the 
case with both $C_{V_2}$ and $C_{S_2}$ as the winning model among others.

\begin{table}[!hbt] 
 \begin{center}
  \begin{tabular}{|c|c|c|c|}
    \hline
    Cases Compared & $\Delta\chi^2$ & $\Delta d.o.f$ & $p$-value\\
    \hline
    $A$, $B$ & 8.24 & 2 & 0.02 \\
    $A$, $C$ & 8.67 & 4 & 0.07 \\
    $A$, $D$ & 8.96 & 6 & 0.18 \\
    $B$, $C$ & 0.42 & 2 & 0.81 \\
    $B$, $D$ & 0.72 & 4 & 0.95 \\
    $C$, $D$ & 0.30 & 2 & 0.86 \\
    \hline
  \end{tabular}
 \end{center}
 \caption{$\Delta\chi^2$ analysis of the best models obtained from table \ref{tab:modselnote1}.}
 \label{tab:modselnote2}
\end{table}

Though the system of all competing models are nested in our analysis, merely being able to reject one of the models compared to another is clearly not enough.

On the other hand, BIC, (also defined with the help of the likelihood function) can be defined as:
\begin{align}
{\rm BIC} =  \chi^2 + (\log{n})~p
\label{bic}
\end{align}
where $n$ is the sample size and $p$ is the number of parameters. We can then define $\Delta$BIC in a similar manner as 
$\Delta$AIC. In  \cite{kass}, the authors have shown that $0 < \Delta\text{BIC} < 2$ selects the best models.

AIC$c$ and BIC were originally derived under different assumptions and are useful under different settings. 
AIC$c$ was derived under the assumption that the true model requires an infinite number of parameters and attempts to 
minimize the information lost by using a given finite dimensional model to approximate it. BIC was derived as a 
large-sample approximation to Bayesian selection among a fixed set of finite dimensional models. The only difference 
between the two criteria extended to take number of samples into account. 

As can be seen from eqs. \ref{aic}, \ref{aicc} and \ref{bic}, the two criteria may produce quite different 
results for large $n$.

The reasons we prefer AIC$c$ over BIC are as follows:
\begin{enumerate}
 \item BIC applies a much larger penalty for complex models, and hence may lead to a simpler model than AIC$c$. 
 In general, BIC penalizes models with more parameters than AIC$c$ does and thus leads to choosing more parsimonious models than AIC$c$.
 \item While AIC compares the cases as approximations of some true model, BIC tries to assign the best model as the true model. 
 This is one of the prevalent arguments against BIC. 
 \item For realistic sample sizes, BIC selected models may underfit the data.
\end{enumerate} 
For a comparative study, we have included table \ref{tab:Result4}, which lists the best scenarios obtained from 
``Fit-2'' using both AIC$c$ and BIC. To make BIC selection at par with AIC$_c$, i.e. more lenient, we have chosen a 
range $0 - 4$. This is same as for $\Delta$AIC. We note that, in our case, the same sets of scenarios/models are 
selected in both the selection criteria.

Both AIC, BIC and such criteria fail when the models being compared have same number of independent parameters and 
comparable likelihood. In such cases, something like `parametric bootstrap' \cite{paraboot} can be used, but such 
analysis is out of scope for the present work.

\begin{table}
 \begin{center}
  \begin{tabular}{|c|c|c|c|c|}
    \hline
    DataSet Ind. & Cases & Param.s & B.F. Val & $\pm$ Err. \\
    \hline
$\text{}$ & $5$ & $\text{$Re(C_T)$}$ & $0.27$ & $0.10$ \\
$\text{}$ & & $\text{$Im(C_T)$}$ & $0.00$ & $1.06$ \\
\cline{2-5}
$\text{1}$ & $3$ & $\text{$Re(C_{S_1})$}$ & $0.09$ & $0.06$ \\
$\text{}$ &  & $\text{$Im(C_{S_1})$}$ & $0.00$ & $0.30$ \\
\cline{2-5}
$\text{}$ & $4$ & $\text{$Re(C_{S_2})$}$ & $0.09$ & $0.06$ \\
$\text{}$ &  & $\text{$Im(C_{S_2})$}$ & $0.00$ & $0.30$ \\
\hline
$\text{}$ &  & $\text{$Re(C_{V_2})$}$ & $0.27$ & $0.03$ \\
$\text{2}$ & $2$ & $\text{$Im(C_{V_2})$}$ & $0.00$ & $0.40$ \\
\hline
$\text{}$ &  & $\text{$Re(C_{V_2})$}$ & $0.34$ & $0.12$ \\
$\text{}$ &  & $\text{$Im(C_{V_2})$}$ & $-0.33$ & $0.20$ \\
$\text{}$ & $8$ & $\text{$Re(C_{S_2})$}$ & $-0.52$ & $0.43$ \\
$\text{}$ &  & $\text{$Im(C_{S_2})$}$ & $-0.16$ & $0.21$ \\
\cline{2-5}
$\text{3}$ & $2$ & $\text{$Re(C_{V_2})$}$ & $0.23$ & $0.02$ \\
$\text{}$ &  & $\text{$Im(C_{V_2})$}$ & $0.00$ & $0.09$ \\
\cline{2-5}
$\text{}$ &  & $\text{$Re(C_{V_2})$}$ & $0.29$ & $0.03$ \\
$\text{}$ &  & $\text{$Im(C_{V_2})$}$ & $0.00$ & $0.54$ \\
$\text{}$ & $7$ & $\text{$Re(C_{S_1})$}$ & $-0.23$ & $0.09$ \\
$\text{}$ &  & $\text{$Im(C_{S_1})$}$ & $0.00$ & $0.48$ \\
\hline
$\text{}$ &  & $\text{$Re(C_{V_2})$}$ & $0.58$ & $0.69$ \\
$\text{4}$ & $2$ & $\text{$Im(C_{V_2})$}$ & $-0.59$ & $0.37$ \\
\hline
$\text{}$ &  & $\text{$Re(C_{V_2})$}$ & $0.34$ & $0.12$ \\
$\text{}$ &  & $\text{$Im(C_{V_2})$}$ & $-0.35$ & $0.21$ \\
$\text{}$ & $8$ & $\text{$Re(C_{S_2})$}$ & $-0.51$ & $0.46$ \\
$\text{}$ &  & $\text{$Im(C_{S_2})$}$ & $-0.14$ & $0.22$ \\
\cline{2-5}
$\text{5}$ &  & $\text{$Re(C_{V_2})$}$ & $0.23$ & $0.02$ \\
$\text{}$ & $2$ & $\text{$Im(C_{V_2})$}$ & $0.00$ & $0.09$ \\
\cline{2-5}
$\text{}$ &  & $\text{$Re(C_{V_2})$}$ & $0.28$ & $0.03$ \\
$\text{}$ &  & $\text{$Im(C_{V_2})$}$ & $0.00$ & $0.70$ \\
$\text{}$ & $7$ & $\text{$Re(C_{S_1})$}$ & $-0.22$ & $0.08$ \\
$\text{}$ &  & $\text{$Im(C_{S_1})$}$ & $0.00$ & $0.55$ \\
\hline
 & $2$ & $Re(C_{V_2})$ & $0.10$ & $0.05$ \\
 & & $Im(C_{V_2})$ & $-0.23$ & $0.17$ \\
\cline{2-5}
 & $4$ & $Re(C_{S_2})$ & $-0.93$ & $0.73$ \\
 & & $Im(C_{S_2})$ & $-0.69$ & $0.32$ \\
\cline{2-5}
7 & $3$ & $Re(C_{S_1})$ & $0.14$ & $0.08$ \\
 & & $Im(C_{S_1})$ & $0.00$ & $0.43$ \\
\cline{2-5}
 & $5$ & $Re(C_{T})$ & $0.04$ & $0.03$ \\
 & & $Im(C_{T})$ & $0.00$ & $0.03$ \\
\hline
\end{tabular}
 \end{center}
 \caption{Best-fit values and Gaussian errors of all parameters for the selected `best' cases for `Fit-1', listed in table 
 \ref{tab:Result1}. Some cases are omitted due to the reason explained in section \ref{sec:goodfit} and corresponding plots are 
 tabulated in fig. \ref{fig:contours}.}
 \label{tab:Result2}
\end{table}

\begin{table}
 \begin{center}
  {\renewcommand{\arraystretch}{1.1}%
  \begin{tabular}{|c|c|c|}
    \hline
    ~Dataset~ & ~Cases with~ & ~Cases with~ \\
    Index. & $0 < \Delta\text{AIC}_c < 4$ & $0 < \Delta\text{BIC} < 4$\\
  \hline
 & $C_{T}$ & $C_{T}$ \\
 & $C_{V_1}$ & $C_{V_1}$ \\
1 & $C_{V_2}$ & $C_{V_2}$ \\
 & $C_{S_1}$ & $C_{S_1}$ \\
 & $C_{S_2}$ & $C_{S_2}$ \\
\hline
2 & $C_{V_1}$ & $C_{V_1}$ \\
 & $C_{V_2}$ & $C_{V_2}$ \\
\hline
 & $C_{V_2}$ & $C_{V_2}$ \\
3 & $C_{V_2}$, $C_{S_2}$ & $C_{V_2}$, $C_{S_2}$ \\
 & $C_{V_1}$ & $C_{V_1}$ \\
 & $C_{V_2}$, $C_{S_1}$ & $-$ \\
\hline
4 & $C_{V_2}$ & $C_{V_2}$ \\
 & $C_{V_1}$ & $C_{V_1}$ \\
\hline
 & $C_{V_2}$, $C_{S_2}$ & $C_{V_2}$, $C_{S_2}$ \\
 & $C_{V_2}$, $C_{S_1}$ & $C_{V_2}$ \\
5 & $C_{V_1}$, $C_{V_2}$ & $C_{V_1}$ \\
 & $C_{V_2}$ & $C_{V_2}$, $C_{S_1}$ \\
 & $C_{V_1}$ & $C_{V_1}$, $C_{V_2}$ \\
\hline
 & $C_{V_2}$ & $C_{V_2}$ \\
6 & $C_{V_2}$, $C_{S_2}$ & $C_{V_2}$, $C_{S_2}$ \\
 & $C_{V_2}$, $C_{S_1}$ & $-$ \\
\hline
 & $C_{S_1}$ & $C_{S_1}$ \\
 & $C_{T}$ & $C_{T}$ \\
7 & $C_{S_2}$ & $C_{S_2}$ \\
 & $C_{V_2}$ & $C_{V_2}$ \\
 & $C_{V_1}$ & $C_{V_1}$ \\
\hline
  \end{tabular}}
 \end{center}
 \caption{The best selected scenarios for ``Fit-2'' (section \ref{sec:resultfit2}). Here we compare the performance of AIC$_c$ 
 with BIC in model selection.}
 \label{tab:Result4}
\end{table}

\begin{table*}
 \begin{center}
  {\renewcommand{\arraystretch}{1.1}%
  \begin{tabular}{|c|c|c|c|c|c|c|c|c|c|}
    \hline
    ~Experiment~ & ~Dataset~ & ~Observables~ & ~Cases~ & $~\chi^2_{min}~$ & ~$d.o.f$~ & ~Parameters~ & ~Akaike Wgt.s~ & ~Normality~ & ~$\chi^2$ (SM)~ \\
%     \cline{9-11}
     & Index. & & & & & & ($w_i$) & (S-W) & \\
  \hline
$\text{}$ & $\text{}$ & $\text{}$ & $5$ & $7.28$ & $12$ & $C_T$ & $0.25$ & $0.54$ & \\
$\text{}$ & $\text{}$ & $\text{}$ & $1$ & $7.65$ & $12$ & $C_{V_1}$ & $0.20$ & $0.38$ & \\
$\text{}$ & $\text{1}$ & $\text{$R(D)_{bin}$}$ & $2$ & $7.65$ & $12$ & $C_{V_2}$ & $0.20$ & $0.38$ & 8.63\\
$\text{}$ & $\text{}$ & $\text{}$ & $3$ & $8.56$ & $12$ & $C_{S_1}$ & $0.13$ & $0.15$ & \\
$\text{}$ & $\text{}$ & $\text{}$ & $4$ & $8.56$ & $12$ & $C_{S_2}$ & $0.13$ & $0.15$ & \\
\cline{2-10}
$\text{\Babar}$ & $\text{2}$ & $\text{$R(D^*)_{bin}$}$ & $1$ & $5.71$ & $10$ & $C_{V_1}$ & $0.57$ & $0.17$ & 20.20\\
$\text{}$ & $\text{}$ & $\text{}$ & $2$ & $6.75$ & $10$ & $C_{V_2}$ & $0.34$ & $0.09$ & \\
\cline{2-10}
$\text{}$ & $\text{}$ & $\text{}$ & $2$ & $15.68$ & $24$ & $C_{V_2}$ & $0.48$ & $0.53$ & \\
$\text{}$ & $\text{3}$ & $\text{Combined}$ & $8$ & $12.32$ & $22$ & $C_{V_2}$, $C_{S_2}$ & $0.18$ & $0.73$ & 70.44\\
$\text{}$ & $\text{}$ & $\text{}$ & $1$ & $19.03$ & $24$ & $C_{V_1}$ & $0.09$ & $0.3$ & \\
$\text{}$ & $\text{}$ & $\text{}$ & $7$ & $14.13$ & $22$ & $C_{V_2}$, $C_{S_1}$ & $0.07$ & $0.81$ & \\
\hline
$\text{Belle(2016)}$ & $\text{4}$ & $\text{$R(D^*)_{bin}$}$ & $2$ & $6.44$ & $15$ & $C_{V_2}$ & $0.47$ & $0.74$ & 17.76\\
$\text{}$ & $\text{}$ & $\text{}$ & $1$ & $6.92$ & $15$ & $C_{V_1}$ & $0.37$ & $0.86$ & \\
\hline
$\text{}$ & $\text{}$ & $\text{}$ & $8$ & $19.11$ & $39$ & $C_{V_2}$, $C_{S_2}$ & $0.41$ & $0.72$ & \\
$\text{\Babar+}$ & $\text{}$ & $\text{}$ & $7$ & $21.54$ & $39$ & $C_{V_2}$, $C_{S_1}$ & $0.12$ & $0.72$ & \\
$\text{Belle(2016)}$ & $\text{5}$ & $\text{Combined}$ & $6$ & $21.75$ & $39$ & $C_{V_1}$, $C_{V_2}$ & $0.11$ & $0.12$ & 87.91\\
$\text{}$ & $\text{}$ & $\text{}$ & $2$ & $27.35$ & $41$ & $C_{V_2}$ & $0.07$ & $0.96$ & \\
$\text{}$ & $\text{}$ & $\text{}$ & $1$ & $27.42$ & $41$ & $C_{V_1}$ & $0.07$ & $0.65$ & \\
\hline
$\text{\Babar+}$ & $\text{}$ & $\text{}$ & $2$ & $47.88$ & $28$ & $C_{V_2}$ & $0.49$ & $0.03$ & \\
$\text{Belle(2015)+}$ & $\text{6}$ & $\text{Combined}$ & $8$ & $44.97$ & $26$ & $C_{V_2}$, $C_{S_2}$ & $0.16$ & $0.01$ & 85.47\\
$\text{LHCb+}$ & $\text{}$ & $\text{}$ & $7$ & $46.47$ & $26$ & $C_{V_2}$, $C_{S_1}$ & $0.08$ & $0.01$ & \\
Belle(Latest) & & & & & & & & & \\
\hline
$\text{Belle(2016)+}$ & $\text{}$ & $\text{}$ & $3$ & $27.68$ & $19$ & $C_{S_1}$ & $0.21$ & $0.56$ & \\
$\text{Belle(2015)+}$ & $\text{}$ & $\text{}$ & $5$ & $27.83$ & $19$ & $C_{T}$ & $0.19$ & $0.44$ & \\
$\text{LHCb+}$ & $\text{7}$ & $\text{Combined}$ & $4$ & $27.93$ & $19$ & $C_{S_2}$ & $0.18$ & $0.84$ & 29.85\\
$\text{Belle(Latest)}$ & $\text{}$ & $\text{}$ & $2$ & $28.00$ & $19$ & $C_{V_2}$ & $0.18$ & $0.86$ & \\
$\text{}$ & $\text{}$ & $\text{}$ & $1$ & $29.82$ & $19$ & $C_{V_1}$ & $0.07$ & $0.75$ & \\
\hline
  \end{tabular}}
 \end{center}
 \caption{The best selected scearios for ``Fit-2'' (section \ref{sec:resultfit2}). For clarification of columns, please see 
 the caption of table \ref{tab:Result1}}
 \label{tab:Result3}
\end{table*}

\begin{table}
 \begin{center}
  \begin{tabular}{|c|c|c|c|c|}
    \hline
    DataSet Ind. & Cases & Param.s & B.F. Val & $\pm$ Err. \\
    \hline
 & $5$ & $Re(C_T)$ & $0.36$ & $0.19$ \\
 & & $Im(C_T)$ & $0.00$ & $1.14$ \\
\cline{2-5}
1 & $3$ & $Re(C_{S_1})$ & $-0.05$ & $0.16$ \\
 &  & $Im(C_{S_1})$ & $0.00$ & $0.28$ \\
\cline{2-5}
 & $4$ & $Re(C_{S_2})$ & $-0.05$ & $0.16$ \\
 &  & $Im(C_{S_2})$ & $0.00$ & $0.28$ \\
\hline
% $\text{BABAR$\_$BDst}$ & $1$ & $Re(C_{V_1})$ & $-0.573$ & $6235.68$ \\
% $\text{BABAR$\_$BDst}$ & $1$ & $Im(C_{V_1})$ & $0.564$ & $4725.69$ \\
% \cline{2-5}
2 & $2$ & $Re(C_{V_2})$ & $0.34$ & $0.08$ \\
 &  & $Im(C_{V_2})$ & $0.00$ & $0.38$ \\
\hline
 & $2$ & $Re(C_{V_2})$ & $0.22$ & $0.02$ \\
 &  & $Im(C_{V_2})$ & $0.00$ & $0.20$ \\
\cline{2-5}
 & & $Re(C_{V_2})$ & $0.31$ & $0.11$ \\
 & & $Im(C_{V_2})$ & $0.04$ & $2.16$ \\
3 & $8$ & $Re(C_{S_2})$ & $-0.52$ & $0.37$ \\
 & & $Im(C_{S_2})$ & $0.004$ & $0.46$ \\
\cline{2-5}
% $\text{BABAR$\_$Tot}$ & $1$ & $Re(C_{V_1})$ & $-0.554$ & $1325.23$ \\
% $\text{BABAR$\_$Tot}$ & $1$ & $Im(C_{V_1})$ & $0.48$ & $1230.32$ \\
% \cline{2-5}
 & & $Re(C_{V_2})$ & $0.31$ & $0.06$ \\
 & $7$ & $Im(C_{V_2})$ & $0.00$ & $0.36$ \\
 & & $Re(C_{S_1})$ & $-0.31$ & $0.21$ \\
 & & $Im(C_{S_1})$ & $0.00$ & $0.40$ \\
\hline
4 & $2$ & $Re(C_{V_2})$ & $0.67$ & $0.69$ \\
 & & $Im(C_{V_2})$ & $-0.36$ & $0.47$ \\
% \cline{2-5}
% $\text{Bell$\_$BDst}$ & $1$ & $Re(C_{V_1})$ & $-0.579$ & $\text{Not Pos.Def!}$ \\
% $\text{Bell$\_$BDst}$ & $1$ & $Im(C_{V_1})$ & $0.446$ & $\text{Not Pos.Def!}$ \\
\hline
 & & $Re(C_{V_2})$ & $0.34$ & $0.04$ \\
 & & $Im(C_{V_2})$ & $0.00$ & $0.62$ \\
 & $8$ & $Re(C_{S_2})$ & $-0.61$ & $0.22$ \\
 & & $Im(C_{S_2})$ & $0.00$ & $0.34$ \\
\cline{2-5}
 & & $Re(C_{V_2})$ & $0.37$ & $0.06$ \\
5 & & $Im(C_{V_2})$ & $0.00$ & $0.23$ \\
 & $7$ & $Re(C_{S_1})$ & $-0.51$ & $0.20$ \\
 & & $Im(C_{S_1})$ & $0.00$ & $0.30$ \\
\cline{2-5}
% $\text{BaBe}$ & $6$ & $Re(C_{V_1})$ & $-1.236$ & $\text{Not Pos.Def!}$ \\
% $\text{BaBe}$ & $6$ & $Im(C_{V_1})$ & $-0.187$ & $\text{Not Pos.Def!}$ \\
% $\text{BaBe}$ & $6$ & $Re(C_{V_2})$ & $-0.741$ & $\text{Not Pos.Def!}$ \\
% $\text{BaBe}$ & $6$ & $Im(C_{V_2})$ & $0.282$ & $\text{Not Pos.Def!}$ \\
% \cline{2-5}
 & $2$ & $Re(C_{V_2})$ & $0.22$ & $0.02$ \\
 & & $Im(C_{V_2})$ & $0.00$ & $0.11$ \\
% \cline{2-5}
% $\text{BaBe}$ & $1$ & $Re(C_{V_1})$ & $-1.65$ & $\text{Not Pos.Def!}$ \\
% $\text{BaBe}$ & $1$ & $Im(C_{V_1})$ & $-0.113$ & $\text{Not Pos.Def!}$ \\
\hline
% $\text{BaBe15LBeNew}$ & $2$ & $Re(C_{V_2})$ & $0.161$ & $0.02$ \\
% $\text{BaBe15LBeNew}$ & $2$ & $Im(C_{V_2})$ & $0.$ & $0.189$ \\
% \cline{2-5}
% $\text{BaBe15LBeNew}$ & $11$ & $Re(C_{V_2})$ & $0.277$ & $0.102$ \\
% $\text{BaBe15LBeNew}$ & $11$ & $Im(C_{V_2})$ & $-0.29$ & $0.26$ \\
% $\text{BaBe15LBeNew}$ & $11$ & $Re(C_{S_2})$ & $-0.485$ & $0.365$ \\
% $\text{BaBe15LBeNew}$ & $11$ & $Im(C_{S_2})$ & $-0.066$ & $0.27$ \\
% \cline{2-5}
% $\text{BaBe15LBeNew}$ & $10$ & $Re(C_{V_2})$ & $0.234$ & $0.051$ \\
% $\text{BaBe15LBeNew}$ & $10$ & $Im(C_{V_2})$ & $0.$ & $0.618$ \\
% $\text{BaBe15LBeNew}$ & $10$ & $Re(C_{S_1})$ & $-0.235$ & $0.162$ \\
% $\text{BaBe15LBeNew}$ & $10$ & $Im(C_{S_1})$ & $0.$ & $0.459$ \\
% \hline
 & $3$ & $Re(C_{S_1})$ & $-0.45$ & $0.47$ \\
 & & $Im(C_{S_1})$ & $0.80$ & $0.15$ \\
\cline{2-5}
 & $5$ & $Re(C_T)$ & $0.09$ & $0.03$ \\
 & & $Im(C_T)$ & $0.00$ & $0.06$ \\
\cline{2-5}
7 & $4$ & $Re(C_{S_2})$ & $0.18$ & $0.08$ \\
 & & $Im(C_{S_2})$ & $0.00$ & $0.87$ \\
\cline{2-5}
 & $2$ & $Re(C_{V_2})$ & $0.09$ & $0.05$ \\
 & & $Im(C_{V_2})$ & $0.39$ & $0.14$ \\
% \cline{2-5}
% $\text{BeBe15LBeNew}$ & $1$ & $Re(C_{V_1})$ & $-1.993$ & $\text{Not Pos.Def!}$ \\
% $\text{BeBe15LBeNew}$ & $1$ & $Im(C_{V_1})$ & $-0.195$ & $\text{Not Pos.Def!}$ \\
\hline
  \end{tabular}
 \end{center}
 \caption{Best-fit values and Gaussian errors of all parameters for the selected `best' cases for `Fit-2', listed in table 
 \ref{tab:Result3}. Some cases are omitted due to the reason explained in section \ref{sec:goodfit}% and corresponding plots are 
 %tabulated in fig. \ref{fig:contours}.
 }
 \label{tab:Result3a}
\end{table}

\begin{figure*}[htbp]
\centering
\subfloat[Dataset $1$, Case $1$]{
\includegraphics[scale=0.25]{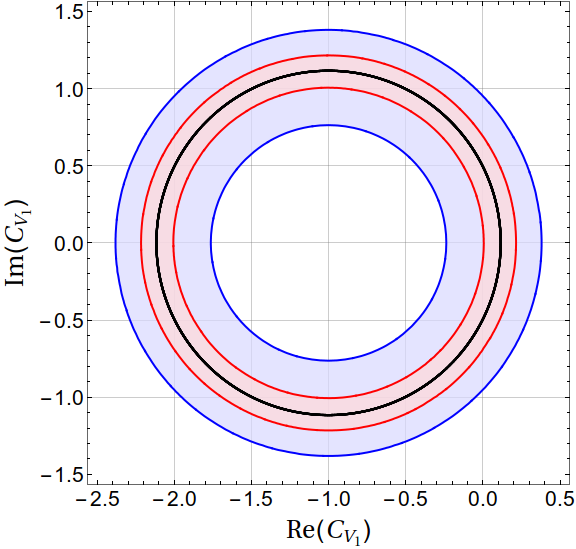}
 \label{fig:BaBD1}}
\subfloat[Dataset $1$, Case $2$]{
\includegraphics[scale=0.25]{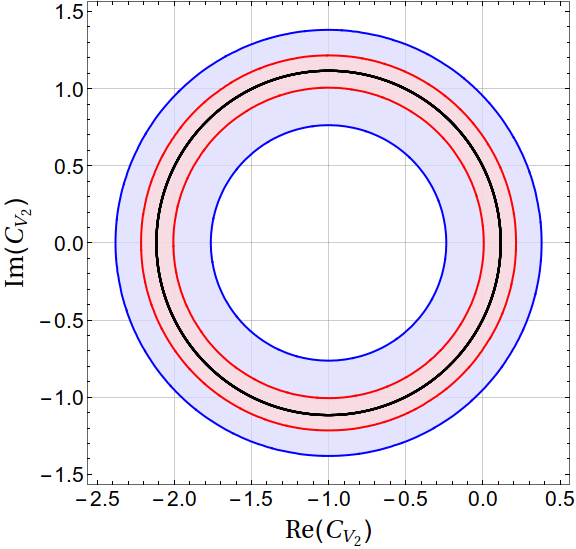}
 \label{fig:BaBD2}}
\subfloat[Dataset $2$, Case $1$]{
\includegraphics[scale=0.25]{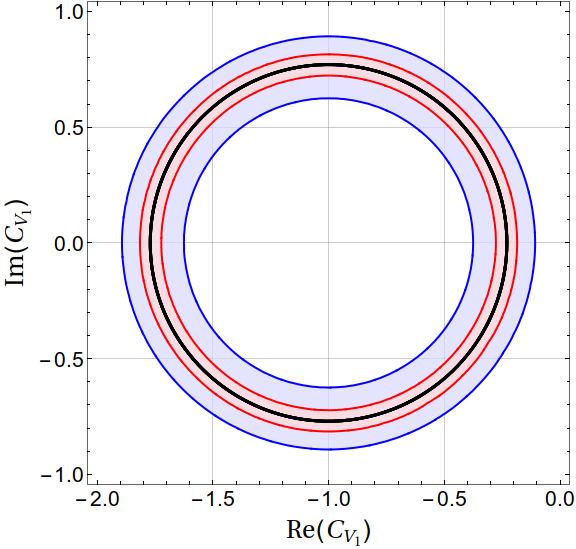}
 \label{fig:BaBDst1}}\\
\subfloat[Dataset $3$, Case $6$]{
\includegraphics[scale=0.25]{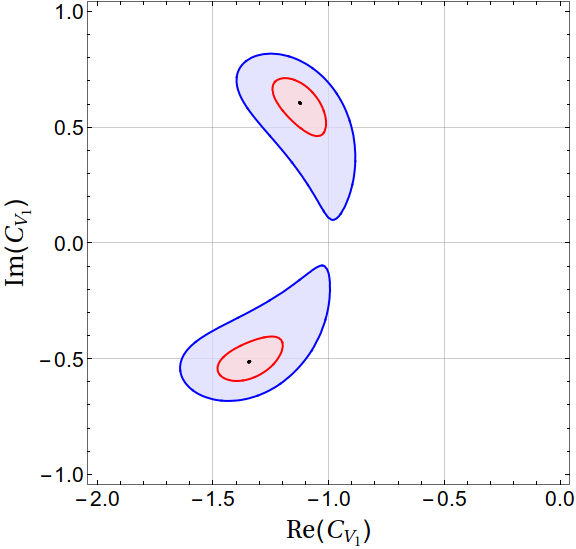}
 \label{fig:BaTot6a}}
\subfloat[Dataset $3$, Case $6$]{
\includegraphics[scale=0.25]{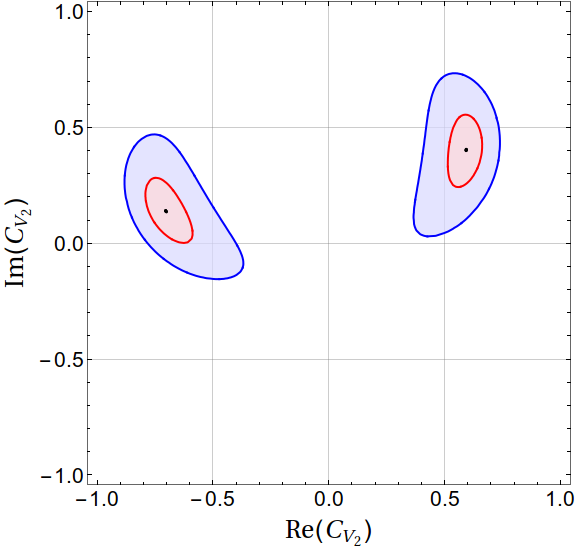}
 \label{fig:BaTot6b}}
\subfloat[Dataset $4$, Case $1$]{
\includegraphics[scale=0.25]{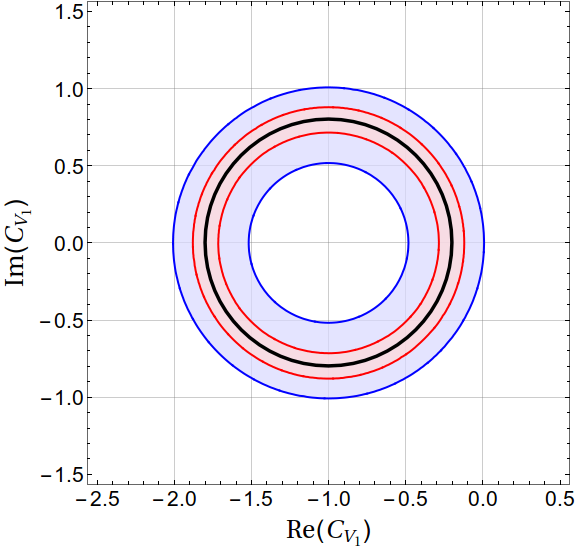}
 \label{fig:BeBDst1}}\\
\subfloat[Dataset $5$, Case $6$]{
\includegraphics[scale=0.25]{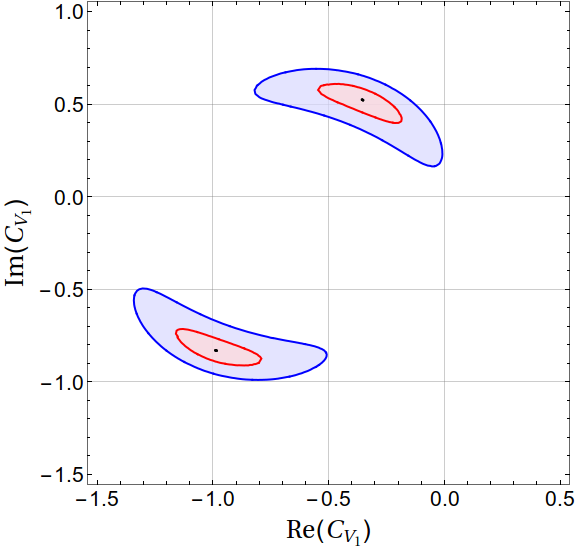}
 \label{fig:BaBem6a}}
\subfloat[Dataset $5$, Case $6$]{
\includegraphics[scale=0.25]{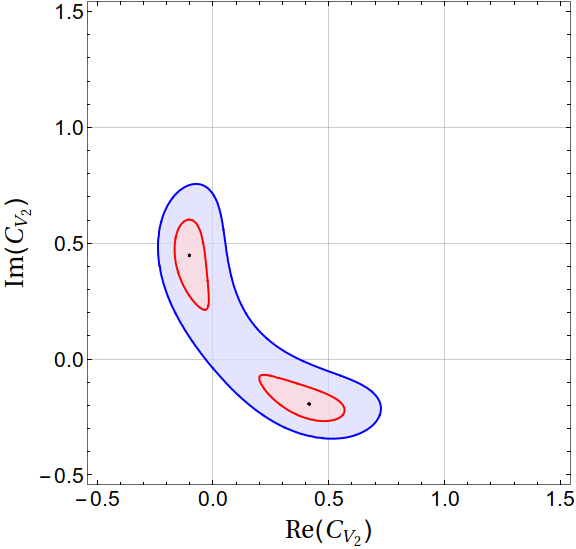}
 \label{fig:BaBem6b}}
% \subfloat[Dataset $10$, Case $1$]{
% \includegraphics[scale=0.4]{BeBe15LBeNewm1.pdf}
%  \label{fig:BeBe15LBeNewm1}}\\
\subfloat[Dataset $7$, Case $1$]{
\includegraphics[scale=0.25]{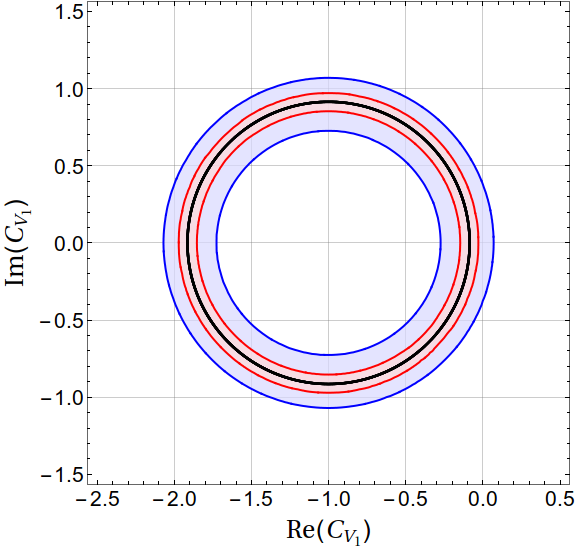}
 \label{fig:BeLBeNewm1}}
\caption{The `cases' for different datasets listed in table \ref{tab:Result1}, which pass the goodness-of-fit hypothesis tests but 
could not be listed in table \ref{tab:Result2} as for these cases, the minimum, instead of being an isolated point, is actually 
a contour in the parameter-space. Though this is true for all plots listed here, some cases have four parameters and we are only 
able to show the two-parameter cross-sections of these.(e.g. plots \ref{fig:BaTot6a} and \ref{fig:BaTot6b} are actually cross-sections of 
a single four-dimensional plot. Same is true for \ref{fig:BaBem6a} and \ref{fig:BaBem6b}).}
\label{fig:contours}
\end{figure*}

\begin{figure*}[htbp] 
\centering
\subfloat[Dataset $1$, Case $5$]{
\includegraphics[scale=0.3]{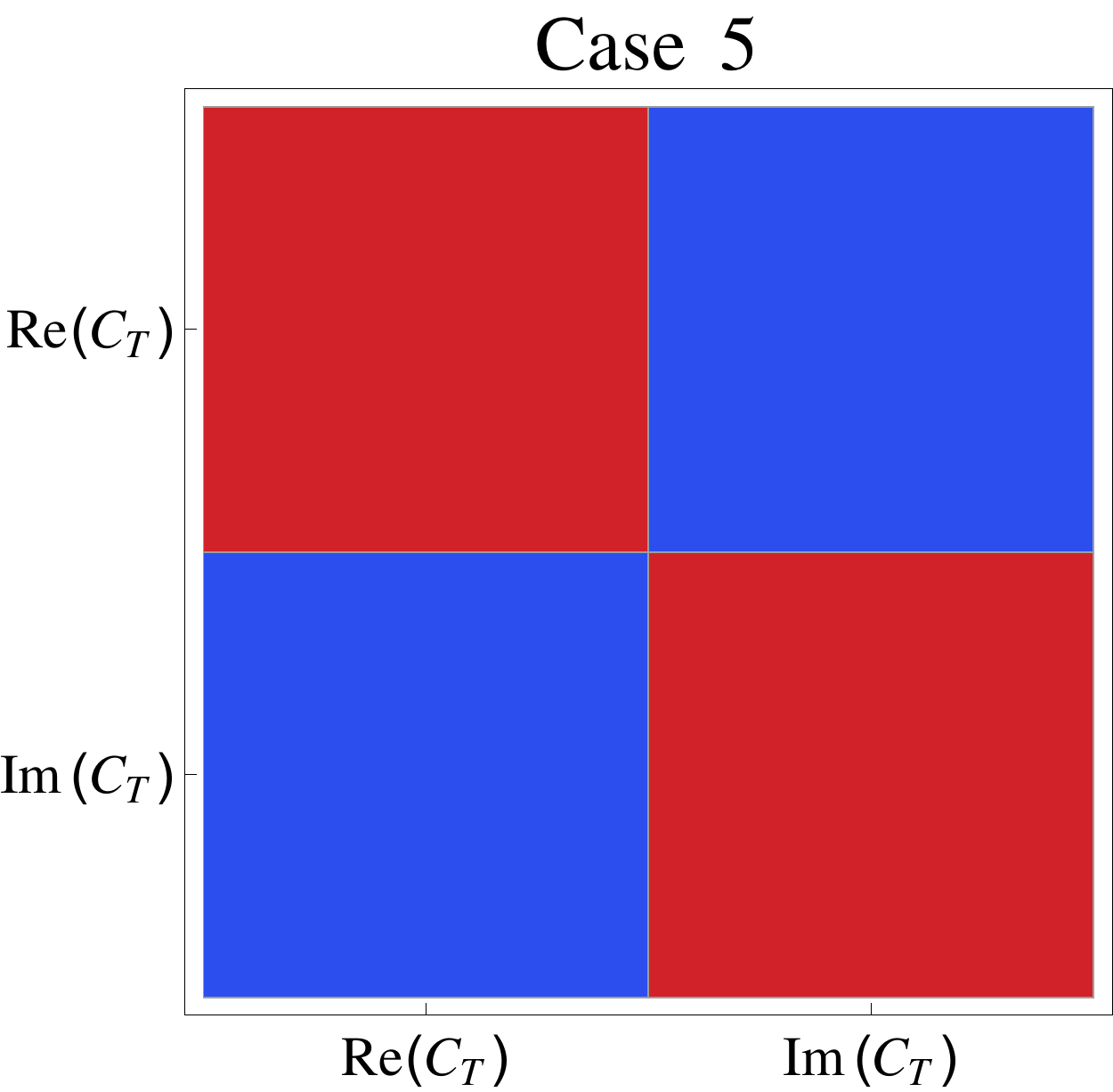}
 \label{fig:D1C5}}
\subfloat[Dataset $1$, Case $3$]{
\includegraphics[scale=0.3]{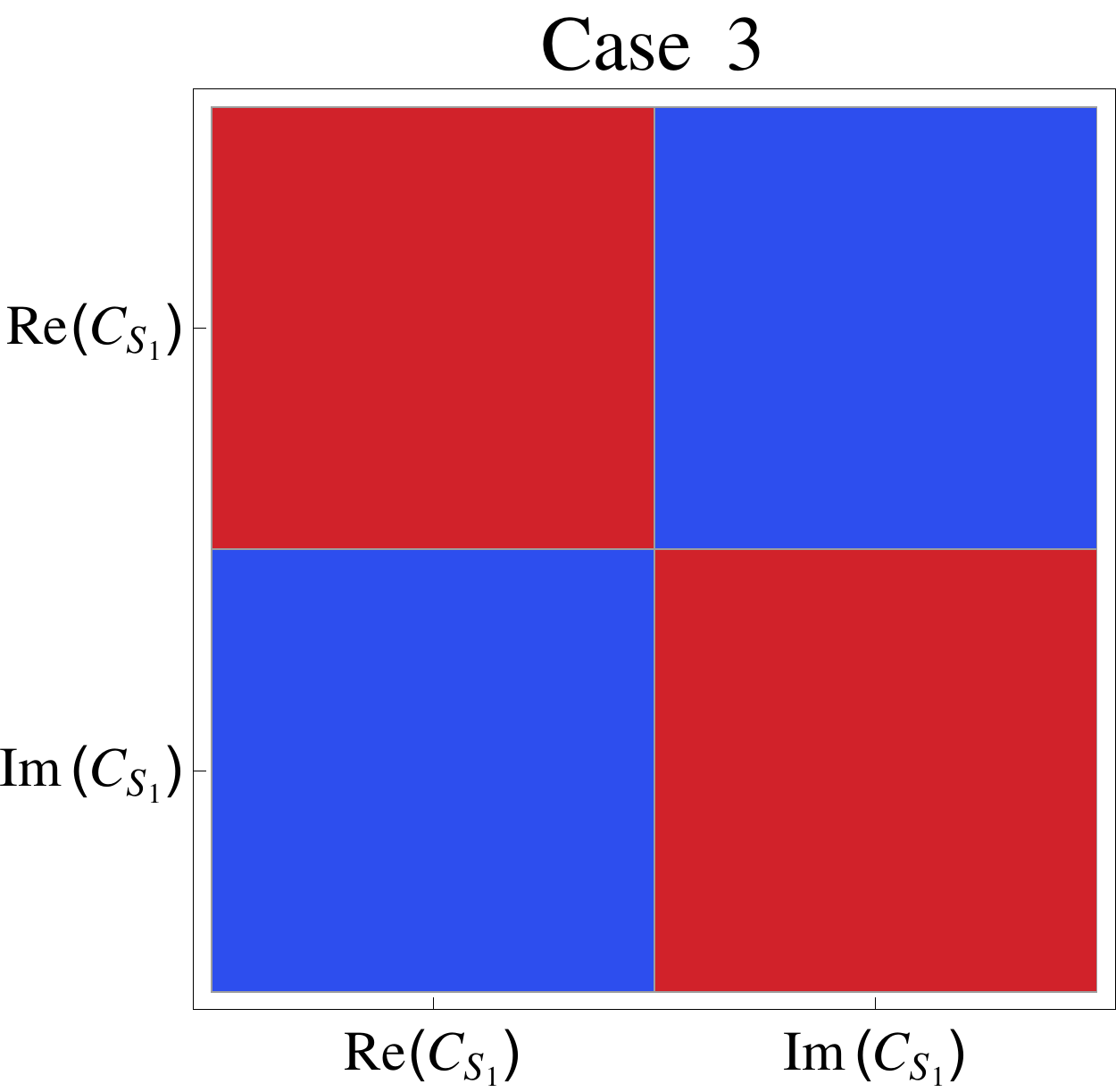}
 \label{fig:D1C3}}
\subfloat[Dataset $1$, Case $4$]{
\includegraphics[scale=0.3]{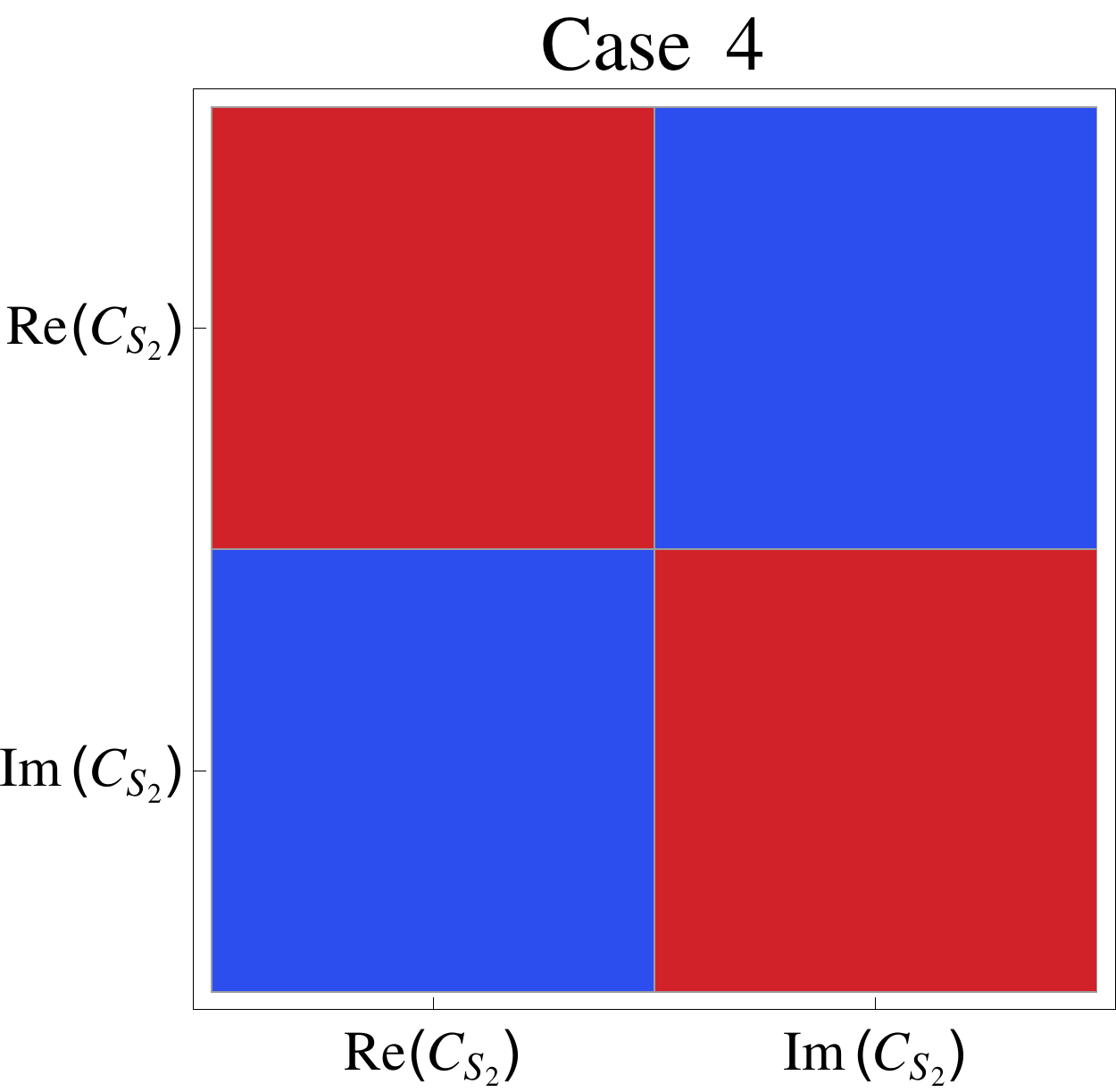}
 \label{fig:D1C4}}
\subfloat[Dataset $2$, Case $2$]{
\includegraphics[scale=0.3]{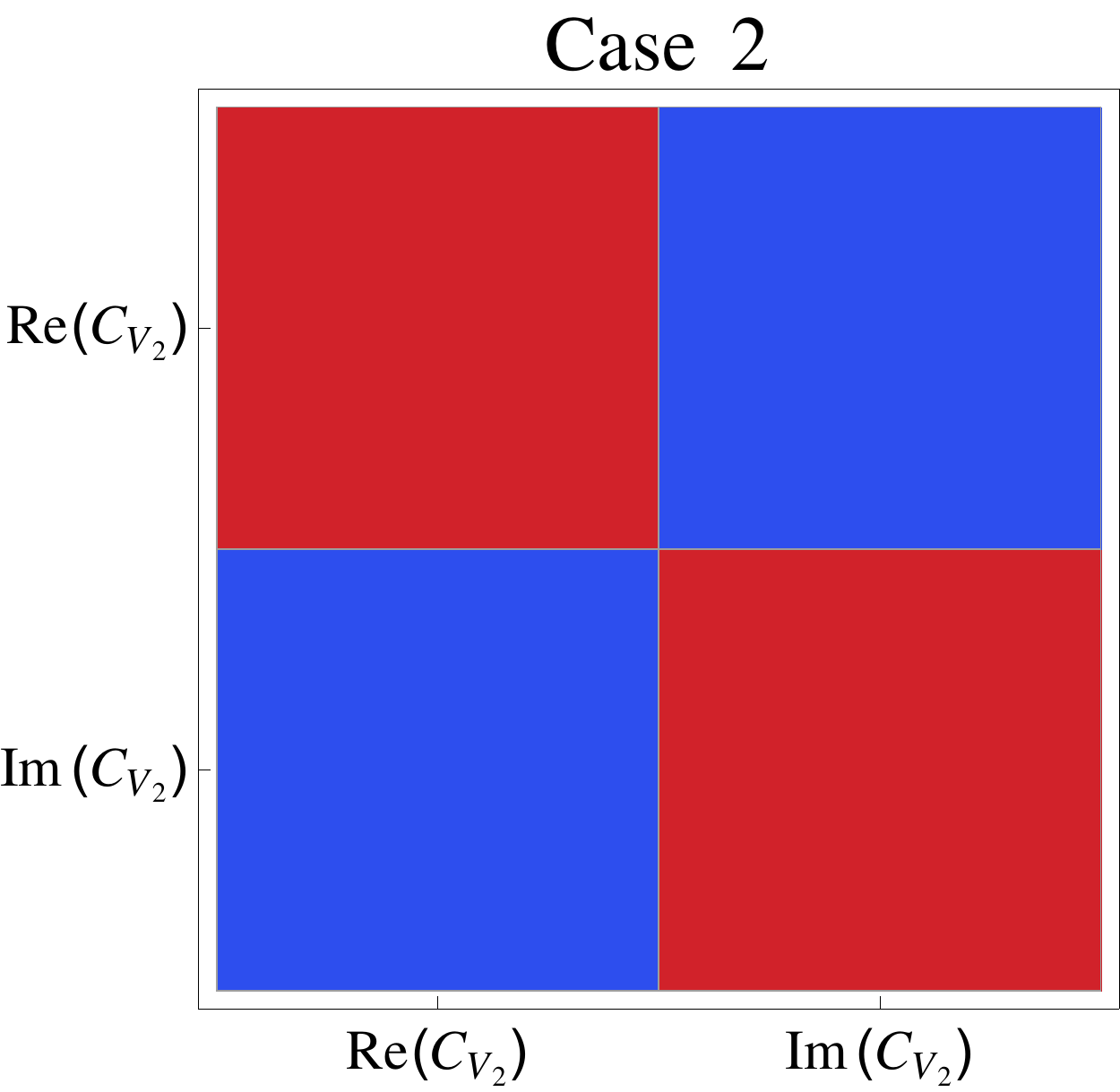}
 \label{fig:D2C2}}\\
\subfloat[Dataset $3$, Case $8$]{
\includegraphics[scale=0.3]{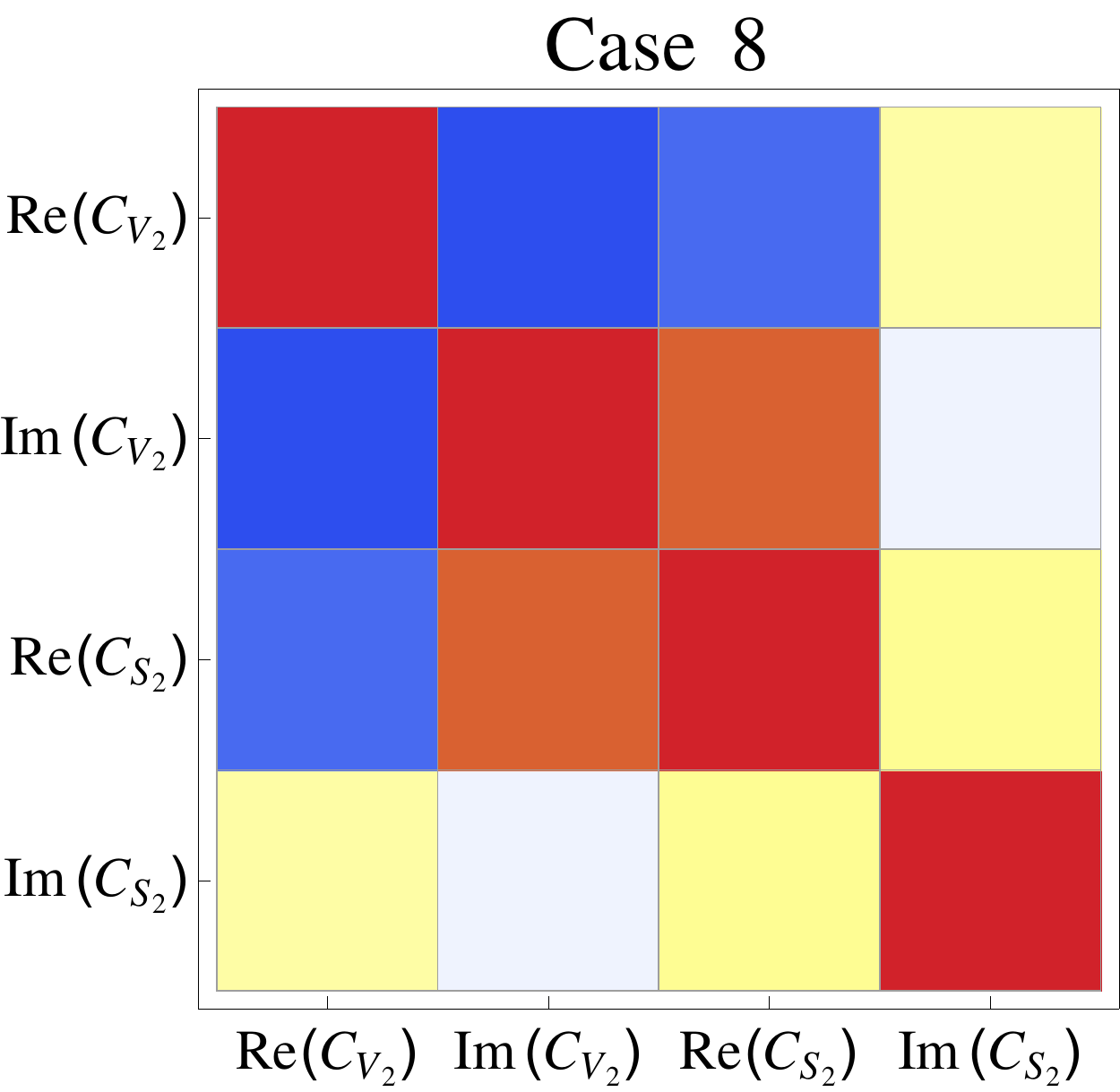}
 \label{fig:D3C8}}
\subfloat[Dataset $3$, Case $2$]{
\includegraphics[scale=0.3]{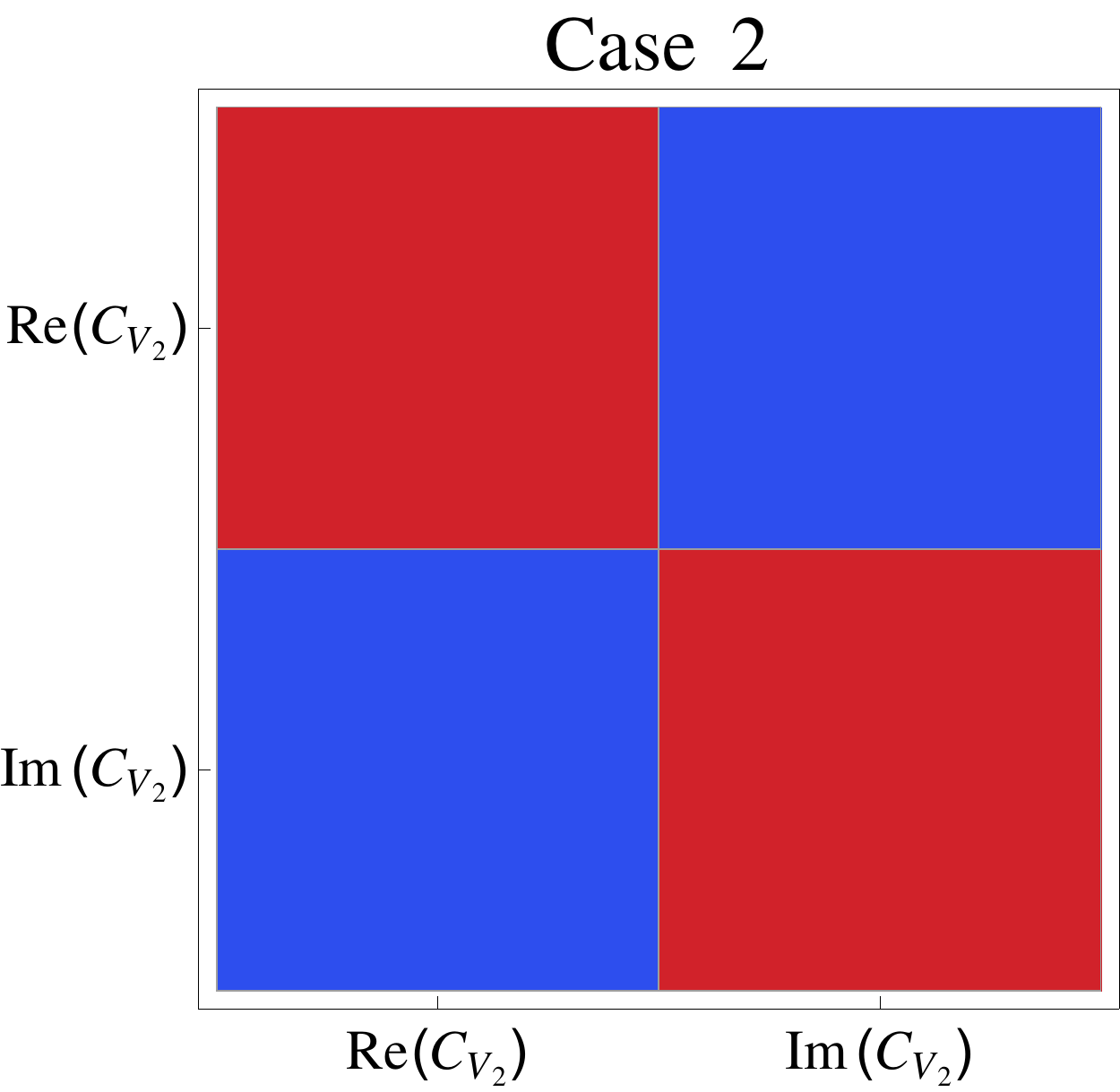}
 \label{fig:D3C2}}
\subfloat[Dataset $3$, Case $7$]{
\includegraphics[scale=0.3]{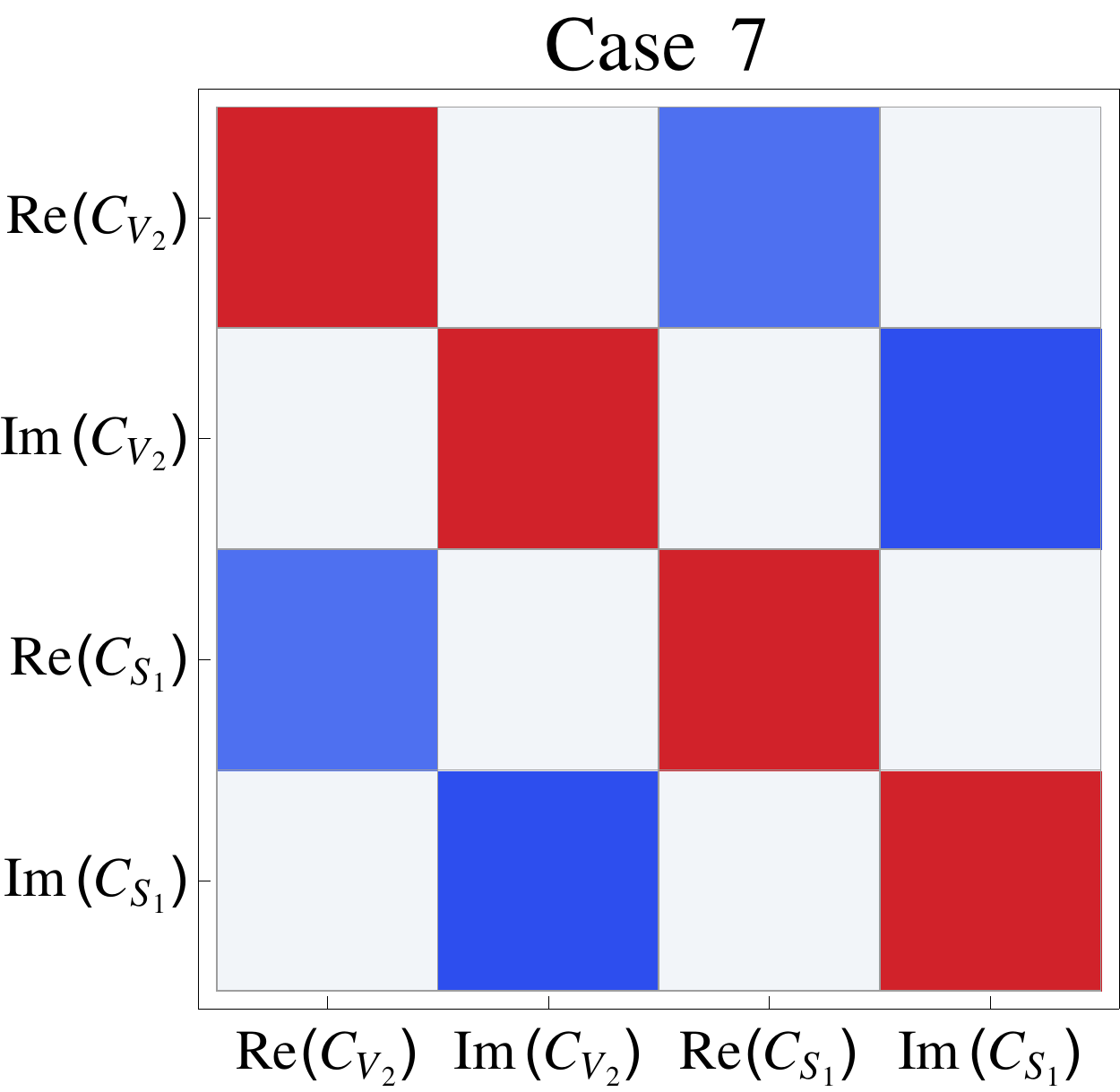}
 \label{fig:D3C7}}
\subfloat[Dataset $4$, Case $2$]{
\includegraphics[scale=0.3]{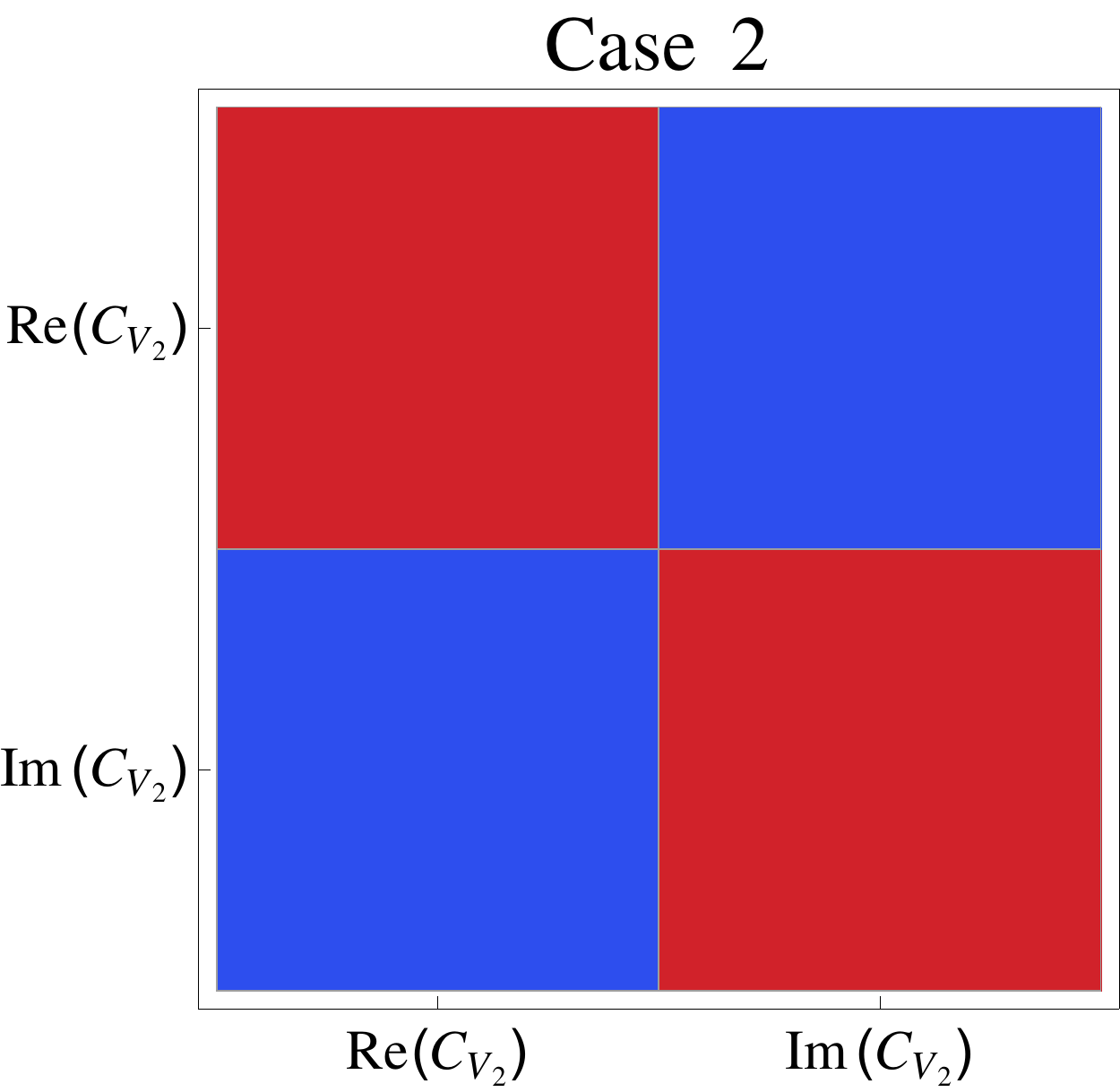}
 \label{fig:D4C2}}\\
\subfloat[Dataset $5$, Case $8$]{
\includegraphics[scale=0.3]{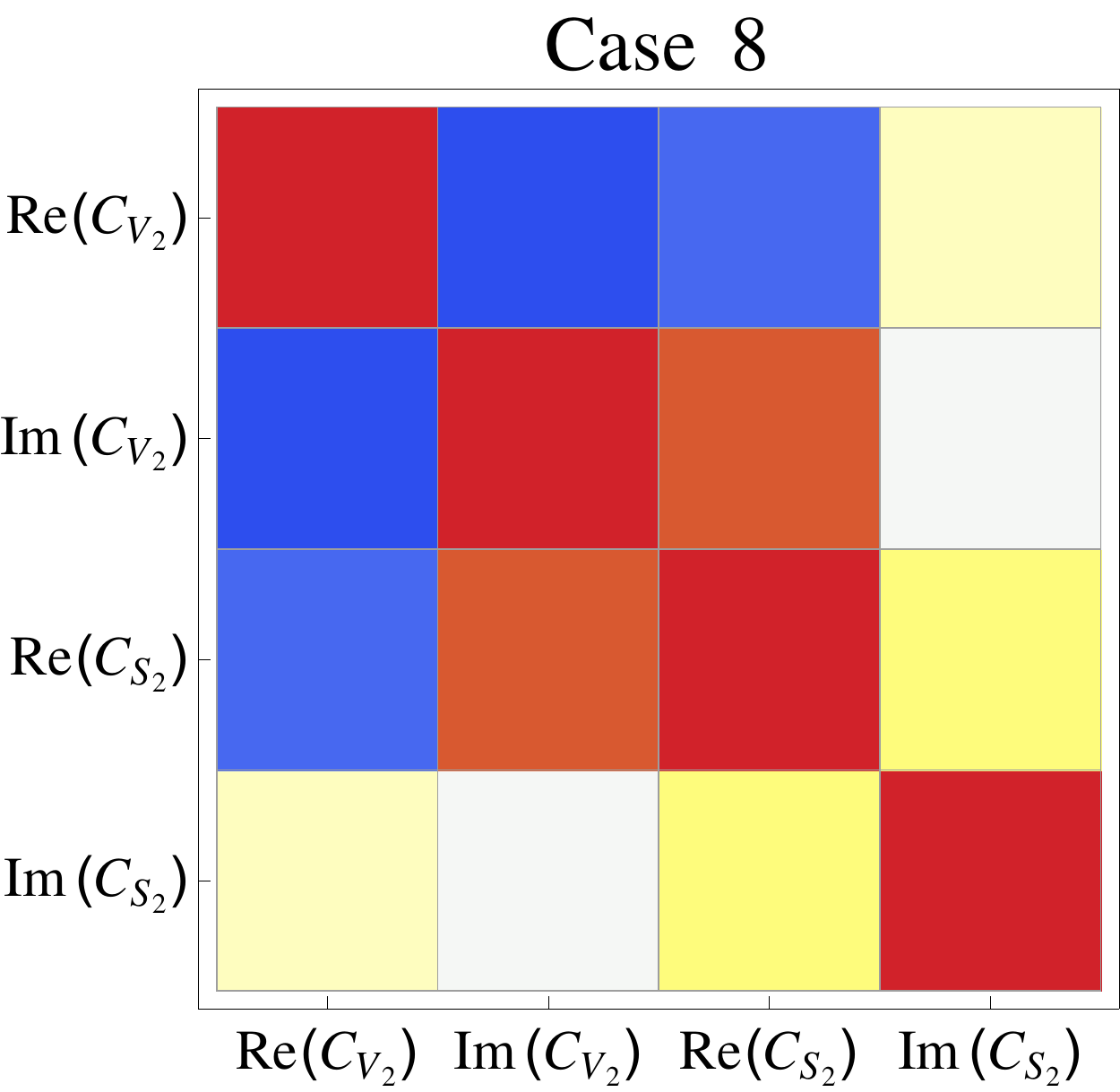}
 \label{fig:D5C8}}
\subfloat[Dataset $5$, Case $2$]{
\includegraphics[scale=0.3]{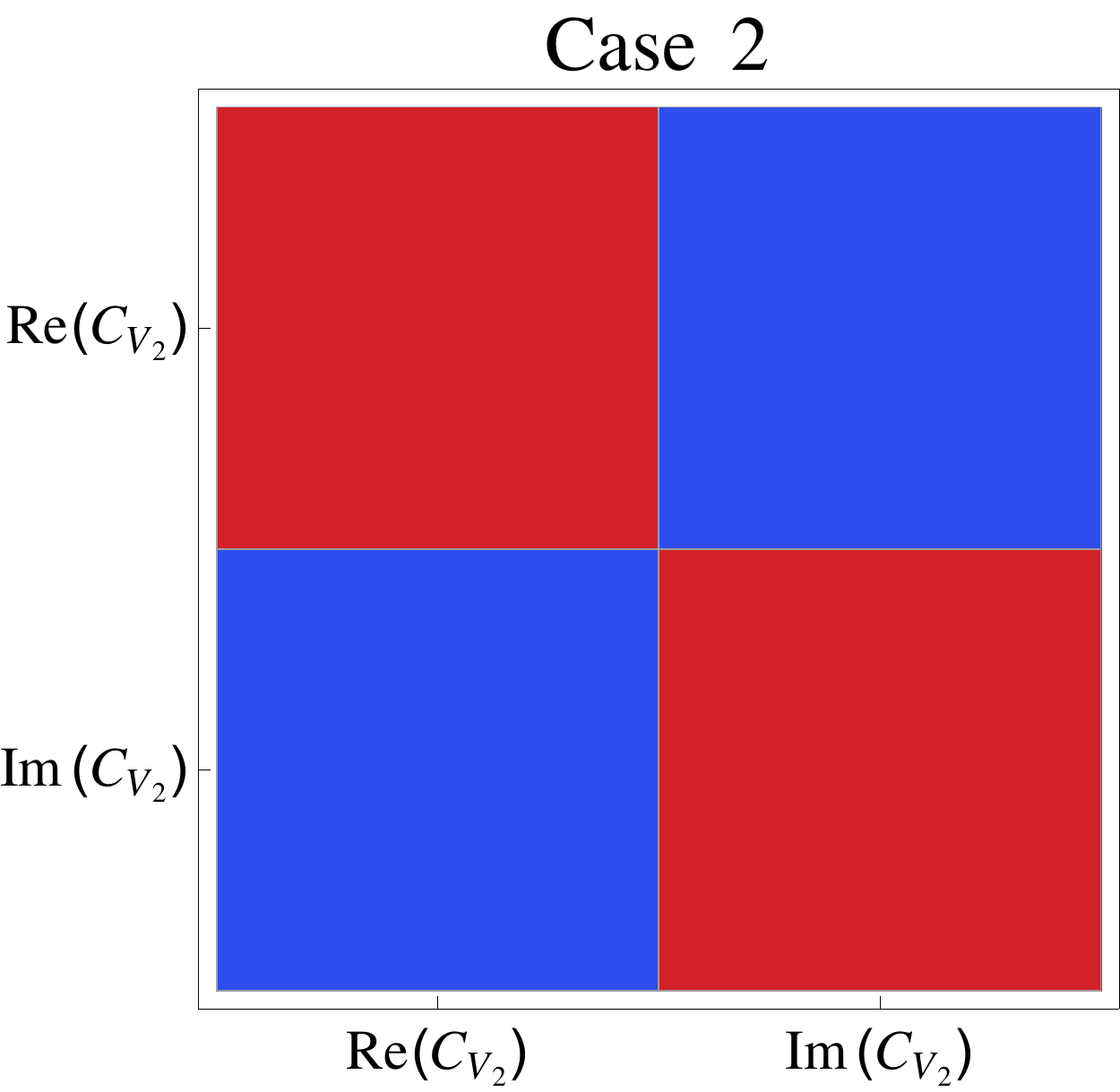}
 \label{fig:D5C2}}
\subfloat[Dataset $5$, Case $7$]{
\includegraphics[scale=0.3]{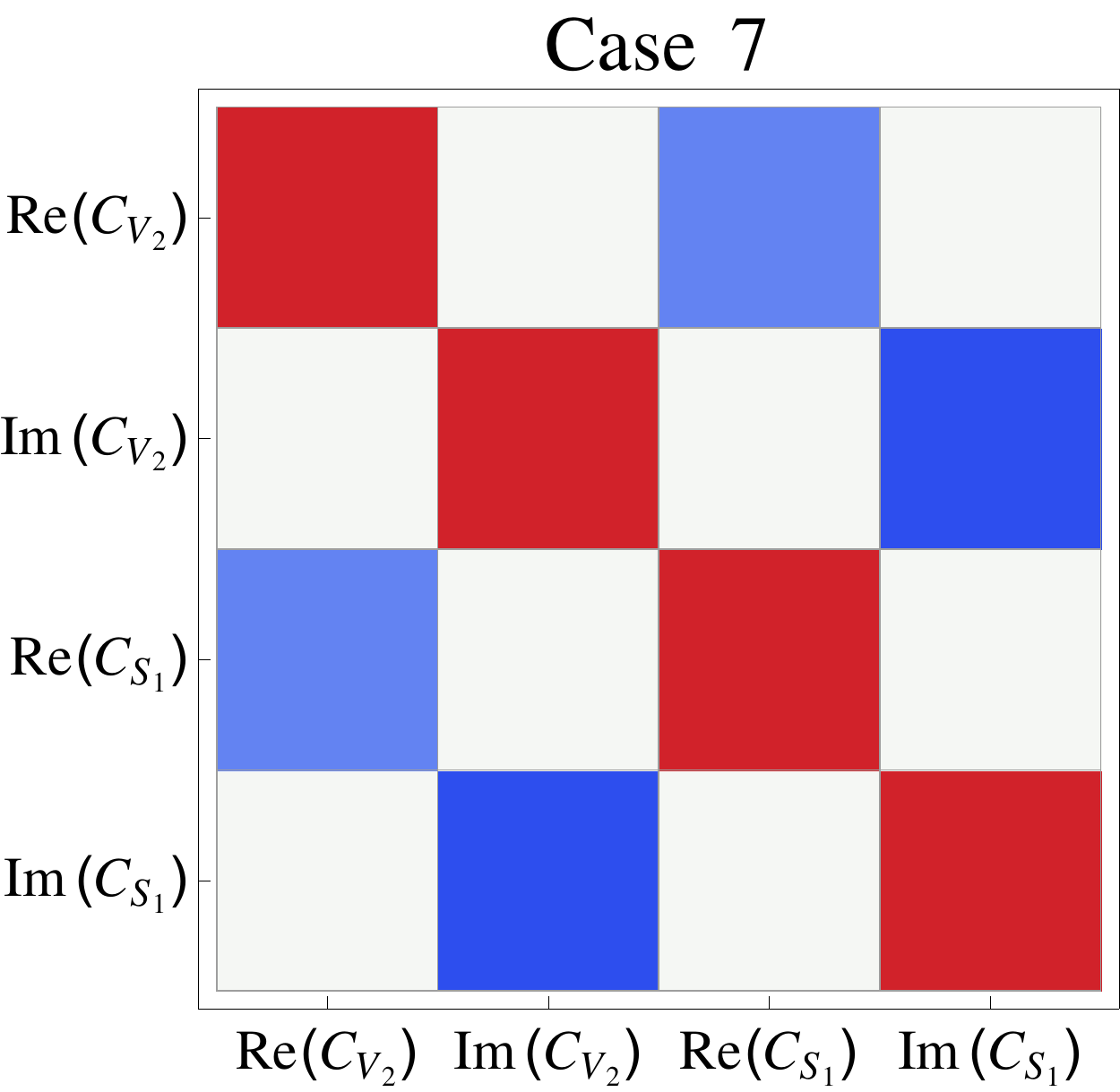}
 \label{fig:D5C7}}
\subfloat[Dataset $11$, Case $2$]{
\includegraphics[scale=0.3]{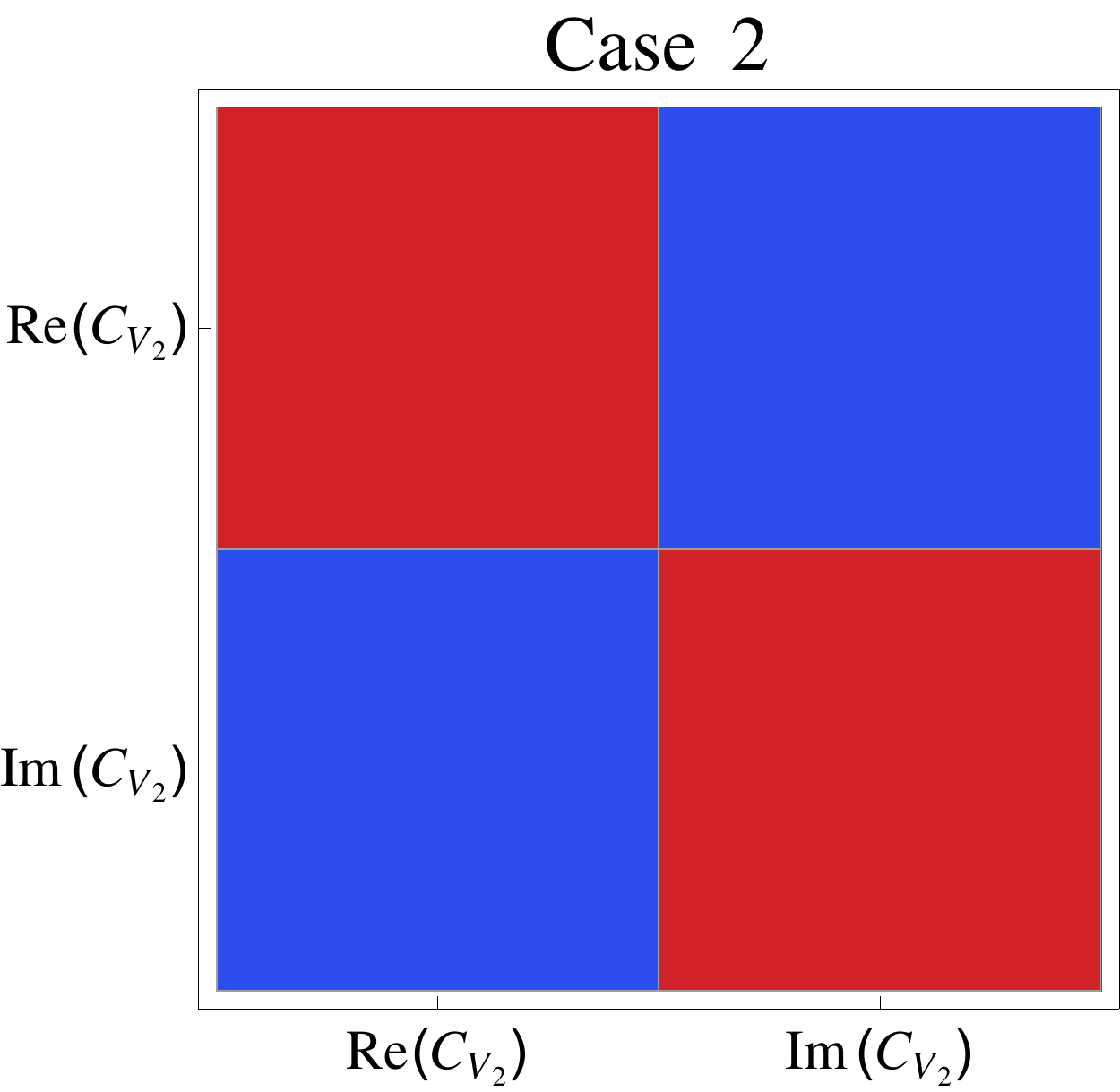}
 \label{fig:D11C2}}\\
\includegraphics[scale=0.8]{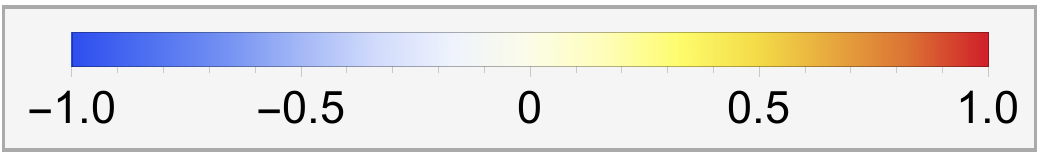}
\caption{Array-plots showcasing the correlations between the fitted parameters of separate `cases' for different datasets 
listed in table \ref{tab:Result2}. The color-coding is explained in the horizontal legend. As can be seen, for the cases with only 
two independent parameters, the parameters are strongly (negatively) correlated, compared to other cases, as expected.}
\label{fig:corrplt}
\end{figure*}

\subsection{Results}\label{sec:resultsnp}

\subsubsection{Fit-1}\label{sec:resultfit1}

In this fit, as mentioned in the previous section, we do not consider the systematic errors or their correlations.
The best probable NP cases (scenarios), which are obtained after minimizing the $\chi^2_{NP}$ and using $w_i$ (eq. (\ref{omegai})),  
are listed in table \ref{tab:Result1}. Then using the formalism defined in section \ref{sec:goodfit}), we find the distribution 
of the residuals for all those fits and we check whether that distribution is accordant with a normal distribution with mean 
$0$ and variance $1$. As was mentioned and justified in section \ref{sec:goodfit}, we use Shapiro-Wilk's normality-test 
for this. Also, in order to check the normality of the residuals, we use the graphical method 
known as quantile-quantile ($Q-Q$) plot. In general, the $Q-Q$ plots are used to compare two probability ditributions. 
In fig. \ref{fig:hypotestnp}, we show the residual-distributions while comparing them with the reference Gaussian 
($\mu = 0$, $\sigma = 1$). The $p$-value obtained in the normality-test quantifies the probability of $H_0$ being true. 
In table \ref{tab:Result1}, the last column lists the $p$-values for the performed S-W test. 

Only those NP scenarios, which pass the normality test, are listed in table \ref{tab:Result2} with the best-fit 
values and $1 \sigma$ uncertainties of their parameters. Other than that, some cases are not shown in the Table, 
where the minimum, instead of being an isolated point, 
is actually a contour in the parameter-space. For such cases, we have plotted the best-fit contours in the parameter-space. 
These are shown in fig. \ref{fig:contours}. We have prepared these plots in terms of the goodness-of-fit contours for 
joint estimation of multiple NP parameters at a time. $1\sigma$ %, $2\sigma$, $3\sigma$,
and $4\sigma$ contours that are equivalent to $p$-values of $0.3173$ %, $0.0455$, $0.0027$,
and $0.0001$, correspond to confidence levels of $68.27\%$ %, $95.45\%$, $99.73\%$,
and $99.99\%$, respectively.

For our purpose, each confidence interval corresponds to a particular value of 
$X = \Delta\chi^2$ (i.e.\ $\chi^2 - \chi^2_{min}$) for a particular model with $d.o.f = N_{params}$, where the SM is 
considered to be the model with no free parameters. For cases up-to 3 parameters, errors on parameters can be 
estimated from the edges of the 2 or 3 dimensional contours as they properly reflect the correlation between 
the involved parameters.

From Table \ref{tab:Result1}, we note that all types of new interactions considered 
in our analysis can individually explain the data on $R(D)_{bin}$ published by \Babar. 
However, when it comes to the $q^2$-distribution of decay rate of $\bdasttau$, both \Babar~and Belle data 
independently allow a contribution from a new left or right-handed vector current effective operator (cases 1 and 2) 
as plausible explanation. Moreover, when the data ($q^2$-bins) from both the \Babar~and Belle are combined, the most 
likely scenarios are the cases with new right handed vector current, either alone or along with other new right or 
left handed scalar current effective operators. In addition to binned data, we have done the analysis by taking into account 
the Belle and LHCb measurements of the $q^2$ integrated $R(D^{(\ast)})$ (see Table \ref{tab:expinput} for numerical values). 
The outcome of these analyses are shown for datasets 6 and 7 in the table \ref{tab:Result1}. No scenario
passes the normality test for dataset-6. In dataset-7, the 
most likely scenarios are the new left or right handed scalar or vector current operators, though, across all the 
cases the reduced $\chi^2$s are $>1$.

Accross all the datasets discussed above, we note that wherever measurements of $R(D)$s are 
included in our fit the effective operators associated with the scalar current become relevant, either alone (less
preferable) or along with the right handed vector current operator. It could be considered as an indication that 
current data on $R(D)$ still allow a scalar current contribution as a possible explanation of the observed deviations.     
Also, across all the scenarios which qualify our predefined test criteria, a common NP explanation is case 2, 
i.e the presence of a new $(V + A)$ type interaction. Here, we can not distinguish whether the new contribution 
is a vector or a pseudo-vector or both. However, if we combine the information obtained from the parametric fit of the form
factors, it won't be wrong to conclude that the most favorable solution of the present data on the decay $\bdasttau$ 
could be obtained from the presence of a pseudo-vector current.

\subsubsection{Fit-2}\label{sec:resultfit2}

In this fit, as mentioned earlier, we consider the systematic error-sizes to be same as the statistical ones and 
assume 100\% correlation among them. The best cases according to their Akaike weights are listed in table \ref{tab:Result3}.
The results are obtained and analyzed in the same manner as for `Fit-1'. Here too, no fit-result for data-set `6' 
passes the normality criteria. Hence we drop that set from further analysis. The outcome of the analyses of the rest of 
the datasets are similar to the ones obtained in `Fit-1', i.e both the fits have almost identical conclusions.  
The only exception is that, here, the role of left handed vector current becoms equally important as the right handed vector 
current, i.e apart from a new $(V+A)$ type interaction, the presence of a new $(V-A)$ type interaction can also be considered 
as common NP explanation of the current data. The best fit values of the fitted parameters along with the corresponding 
errors are shown in table \ref{tab:Result3a}.

\section{Summary}

We look for possible new physics effects in the decays $\bdtau$ in the light of the recently available data from 
Belle, \Babar~and LHCb. At first, the form-factors, relevant in these decays, are fitted assuming the absence of any 
contribution coming from operators other than the SM. The fitted results are then compared with those obtained by 
HFAG from a fitting to the available data on $\bdell$. We note that the fit results of the parameter $R_2(1)$ largely 
disagree with each other, while the rest are more or less consistent with each other within errors. 
The effects are prominent in all the regions of the $q^2$ distribution of the form-factor $A_2(q^2)$, 
which is associated with a pseudo-vector current. Therefore, assuming the decays $\bdell$ are free from 
any new physics effects, such a difference in the $q^2$ distribution of $A_2$ (obtained from $\bdasttau$ and $\bdstell$) 
can be compensated by adding a contribution from new pseudo vector and/or pseudo tensor currents. 

In the next part of our analysis, we consider the new physics contributions in the decays $\bdtau$ which come from  
new vector, scalar or tensor type operators. In this case, we take the relevant form-factors as obtained using 
the fit results by HFAG. We define different 
possible NP scenarios which are obtained after combining contributions from the new operators in many different ways.  
Our goal is to select the best possible NP scenarios (new interactions) that can accommodate all the available data. 
In doing so, we use the AIC$_c$ in the analysis of the empirical data. 
Such procedures lead to more robust inferences in simultaneous comparative analysis of multiple competing scenarios. 
In order to check whether all the NP scenarios that are coming out of AIC$_c$ test can fit the data well or not, 
we have done Shapiro-Wilk's normality-test for each selected model. For a comparative study, we have also analyzed the data 
for selecting the best model using Schwarz-Bayesian Criterion (BIC). For our different 
datasets the best selected models are identical in both the selection criteria. 

Our analysis of the available data on $R(D^{\ast})$ from \Babar, Belle, and LHCb shows that the most plausible explanation 
of the data can be obtained from the presence of new effective oparators with left or right handed charged vector current. 
In addition, if we include $R(D)$ in our fit, apart from the vector currents the contributions from charged scalar currents 
might become relevant, either alone (though less preferable) or along with right handed vector current operators.   

Overall, our analysis of $\bdasttau$ shows that it is the 
contribution from a left or right-handed charged vector current effective operator, that, as well as accommodating all 
the available data, passes all the selection criteria for being the best possible NP scenario.

Here, we would like to point out that we have made use of the available data on the $q^2$ (bins) distributions of the decays
$\bdtau$, which have large errors. This, in turn, gives our fitted results large errors. 
Once the more precise data on the $q^2$ bins are made available, one may and should repeat the 
analysis to check sustainability of the above conclusions.

\section*{Acknowledgements}

We thank Gabriele Simi (Univ. of Padova) and Devdatta Majumder (Univ. of Kansas) for really helpful discussions on binned data and 
residual-distributions.

\end{document}